\documentclass[a4paper,fleqn,usenatbib]{mnras}

\usepackage{newtxtext,newtxmath}
\usepackage[T1]{fontenc}
\usepackage{ae,aecompl}

\usepackage{graphicx}	% Including figure files
\usepackage{amsmath}	% Advanced maths commands
\usepackage{amssymb}	% Extra maths symbols
\usepackage[para,online,flushleft]{threeparttable}
\usepackage{booktabs, caption, makecell}
\usepackage{lscape}
\usepackage{multirow}
\numberwithin{equation}{section}
\DeclareMathOperator\arctanh{arctanh}
\usepackage{xcolor} % package only to mark sections changed for revision
\definecolor{revision1}{rgb}{0.0,0.0,0.0}

\title[Optical follow-up of galaxy cluster candidates from \textit{Planck}]{Optical follow-up study of 32 high-redshift galaxy cluster candidates from \textit{Planck} with the William Herschel Telescope}

\author[Zohren et al.]{Hannah Zohren$^{1}$\thanks{E-mail: hzohren@astro.uni-bonn.de},
Tim Schrabback$^{1}$,
Remco F. J. van der Burg$^{2,3}$,
Monique Arnaud$^{3,4}$
\newauthor Jean-Baptiste Melin$^{3}$,
Jan Luca van den Busch$^{1,5}$,
Henk Hoekstra$^{6}$, 
and Matthias Klein$^{7,8}$
\\
$^{1}$Argelander-Institut f\"ur Astronomie, Rheinische Friedrich-Wilhelm Universit\"at Bonn, Auf dem H\"ugel 71, D-53121 Bonn, Germany\\
$^{2}$European Southern Observatory, Karl-Schwarzschild-Str. 2, D-85748, Garching, Germany\\
$^{3}$IRFU, CEA, Universit{\'e} Paris-Saclay, F-91191 Gif-sur-Yvette, France\\
$^{4}$Universit{\'e} Paris Diderot, AIM, Sorbonne Paris Cit{\'e}, CEA, CNRS, F-91191, Gif-sur-Yvette, France\\ 
$^{5}$Astronomisches Institut, Ruhr-Universit\"at Bochum, Universit\"atsstr. 150, 44780 Bochum, Germany\\
$^{6}$Leiden Observatory, Leiden University, PO Box 9513, 2300 RA Leiden, the Netherlands\\
$^{7}$Faculty of Physics, Ludwig-Maximilians-Universit\"at, Scheinerstr. 1, 81679 Munich, Germany\\
$^{8}$Max Planck Institute for Extraterrestrial Physics, Giessenbachstrasse, 85748 Garching, Germany
}

\date{Accepted XXX. Received YYY; in original form ZZZ}

\pubyear{2019}

\begin{document}
\label{firstpage}
\pagerange{\pageref{firstpage}--\pageref{lastpage}}
\maketitle

\begin{abstract}
\textcolor{revision1}{
The \textit{Planck} satellite has detected cluster candidates via the Sunyaev Zel'dovich (SZ) effect, but the optical follow-up required to confirm these candidates is still incomplete, especially at high redshifts and for SZ detections at low significance. In this work we present our analysis of optical observations obtained for 32 \textit{Planck} cluster candidates using ACAM on the 4.2-m William Herschel Telescope. These cluster candidates were preselected using SDSS, WISE, and Pan-STARRS images to likely represent distant clusters at redshifts $z \gtrsim 0.7$. We obtain photometric redshift and richness estimates for all of the cluster candidates from a red-sequence analysis of $r$-, $i$-, and $z$-band imaging data. In addition, long-slit observations allow us to measure the redshifts of a subset of the clusters spectroscopically. The optical richness is often lower than expected from the inferred SZ mass when compared to scaling relations previously calibrated at low redshifts. This likely indicates the impact of Eddington bias and projection effects or noise-induced detections, especially at low SZ-significance. Thus, optical follow-up not only provides redshift measurements, but also an important independent verification method. We find that 18 (7) of the candidates at redshifts $z > 0.5$ ($z > 0.8$) are at least half as rich as expected from scaling relations, thereby clearly confirming these candidates as massive clusters. While the complex selection function of our sample due to our preselection hampers its use for cosmological studies, we do provide a validation of massive high-redshift clusters particularly suitable for further astrophysical investigations. }
\end{abstract}


%%%%%%%%%%%%%%%%%%%%%%%%%%%%%%%%%%%%%%%%%%%%%%%%%%

%%%%%%%%%%%%%%%%% BODY OF PAPER %%%%%%%%%%%%%%%%%%

\section{Introduction}
\label{Introduction}

One of the core challenges in contemporary astrophysics is to explain the nature of dark matter and dark energy. Past efforts in understanding the parameters that govern our Universe have led to our fiducial Lambda-cold-dark-matter ($\Lambda$CDM) cosmological model, which includes a hierarchical structure formation, where dark energy takes the form of a spatially uniform and non-evolving energy density. In order to constrain the cosmological model from an observational point of view, galaxy clusters have proven to be valuable objects to study. They are the most massive gravitationally bound structures, which reside in the densest regions of the cosmic large-scale structure. Driven by gravity the large-scale structure emerged from small over-densities in the density field of the early Universe. Probing the growth of the densest fluctuations, the number of clusters as a function of mass and redshift sensitively depends on cosmological parameters \citep[e.g.][]{Allen2011}.

Samples of galaxy clusters form the foundation for such cosmological investigations. In order to compare their properties to theoretical predictions, they should ideally be selected based on their mass. Unfortunately, the mass is not directly observable. However, galaxy clusters are multi-component objects observable in various wavelength regimes. \textcolor{revision1}{They can be detected from their emission in the X-ray \citep[e.g.][]{Piffaretti2011,Pacaud2016}, in the optical and near-infrared \citep[e.g.][]{Rykoff2016}, via the Sunyaev Zel'dovich (SZ) effect \citep[e.g.][]{Bleem2015,PlanckCollab2016,Hilton2018}, and recently through their gravitational lensing signal \citep[e.g.][]{Miyazaki2018}. Scaling relations then allow to connect the cluster observables to their mass and make it possible to assemble samples of galaxy clusters with a known selection function \citep{Pratt2019}}. The challenge in this context is to carefully calibrate these relations to connect the observables and still account for intrinsic scatter \citep{Allen2011}.

The detection of galaxy clusters via the SZ effect provides cluster samples that are nearly mass limited. This is because the SZ effect, caused by an inverse Compton scatter of CMB photons by the hot electrons in the cluster plasma, is not subject to cosmic dimming \textcolor{revision1}{\citep{Carlstrom2002}}. Specifically, the \textit{Planck} SZ Survey provides the first all-sky SZ detected cluster catalogue including detections of massive clusters out to redshifts of $z \approx 1$. The full mission catalogue is called PSZ2 and was publicly released in 2016 \citep{PlanckCollab2016}. It contains SZ-detections down to a significance of signal-to-noise ratio $\mathrm{S/N} \geq 4.5$. The follow-up and verification process is ongoing \citep{Liu2015,PlanckCollab2016,VanDerBurg2016,Burenin2018,Amodeo2018,Barrena2018,Streblyanska2018,Boada2018}, but still incomplete, especially at high redshifts. The \textcolor{revision1}{primary goal} of this work is to help complete the follow-up of cluster candidates in the PSZ2 catalogue at high redshifts $z \gtrsim 0.7$ with the help of optical data from the William Herschel Telescope. In the redshift regime above $z \sim 0.7$ the PSZ2 catalogue has a completeness of about 80 per cent for massive clusters of $M_\mathrm{500c} \gtrsim 7.5 \times 10^{14}\, \mathrm{M}_\odot$. The completeness decreases, however, to 20 per cent for masses of $M_\mathrm{500c} \gtrsim 5 \times 10^{14}\, \mathrm{M}_\odot$ in that redshift regime \citep[fig. 26 in][]{PlanckCollab2016}. Considering cluster candidates at lower S/N threshold is a way to raise the completeness and reveal more massive high-$z$ clusters. The sample studied in this work therefore also includes low significance candidates detected from the \textit{Planck} SZ-maps via the Matched Multi-Filter 3 \citep{Melin2006,Melin2012} detection method with a SZ-significance down to $\mathrm{S/N} \gtrsim 3$. 
Since a lower S/N threshold also implies a lower reliability of the sources, confirmation using additional data is critically required. In this work we apply a preselection of cluster candidates based on optical and infrared data as suggested in \citet{VanDerBurg2016}. This helps to exclude those SZ sources that are likely spurious detections because they lack a counterpart in the optical and infrared data. As a result of these considerations, this work deals with the analysis of spectroscopic and photometric data of a sample of 32 cluster candidates, which originate either from the PSZ2 or from detection in the \textit{Planck} maps below the PSZ2 significance cut with the primary aim of confirming massive galaxy cluster candidates at high redshifts.

We structure this paper as follows. In Section \ref{Cluster_Candidate Sample} we present the \textit{Planck} SZ-Survey which builds the foundation for the cluster candidate sample that we follow-up optically in this work. We focus on the photometric observations in Section \ref{Photometric Observations} explaining the data reduction and the strategy to obtain redshift and richness estimates. We present the analysis of the spectroscopic observations in Section \ref{Spectroscopic Observations}. In Section \ref{Confirmation of Cluster Candidates} we discuss which cluster candidates are likely counterparts to the SZ-detections by comparing our richness estimates to the SZ-masses inferred from the SZ-signal. We give notes on individual cluster candidates and briefly discuss our results in Sections \ref{Notes on Individual Cluster Candidates} and \ref{Discussion}. Finally, we give a summary and conclusions of our work in Section \ref{Summary and Conclusions}.

Unless otherwise noted, we adopt a flat $\Lambda$CDM cosmology with $\Omega_{\mathrm{M}}=0.3$\,, $\Omega_{\Lambda}=0.7$\,, and $H_0 = 70\,\mathrm{km}\,\mathrm{s}^{-1}\mathrm{Mpc}^{-1}$ in this work, as approximately consistent with recent CMB results \citep[e.g.][]{Hinshaw2013,PlanckCollab2016CosmParam}. All magnitudes are given in the AB magnitude system.

\section{The \textit{Planck} Catalogue as Basis for the Cluster Candidate Sample}
\label{Cluster_Candidate Sample}

Most of the cluster candidates in our sample originate from the second \textit{Planck} catalogue of Sunyaev Zel'dovich (SZ) sources (PSZ2). This catalogue represents the largest SZ-selected sample of galaxy clusters to date and it is the deepest systematic all-sky survey of galaxy clusters \citep{PlanckCollab2016}. It includes 1653 detections in the 29 month full mission data, 1203 of which are confirmed with identified counterparts in external data sets and 1094 of which have redshift estimates. Clusters are included in the public PSZ2 catalogue down to a signal-to-noise ratio of $\mathrm{S/N}=4.5$, defined via three different detection methods: MMF1, MMF3 and PwS. We refer the reader to \citet{PlanckCollab2016} for a more detailed description. 
The parameter estimates are taken from the detection pipeline with the highest S/N ratio for a given detection.

The remainder of cluster candidates in our sample is assembled from SZ-detections in the \textit{Planck} maps with $\mathrm{S/N}\geq 3$ that are solely based on the MMF3 detection method. The masses and S/N ratios in the PSZ2 catalogue are by construction always larger than or equal to the corresponding values obtained from the MMF3 detection method.

We selected targets from the PSZ2 with the prospect of contributing to a complete follow-up of all targets down to $\mathrm{S/N}=4.5$. A complete follow-up is essential to understand the selection function including the completeness and purity of the PSZ2 catalogue in order to use it for cosmological studies. In particular, we focus on the high redshift regime at $z \gtrsim 0.7$ among the PSZ2 candidates that were still unconfirmed at the time of target selection. We inspect optical and NIR data from SDSS \citep{SDSSDR8}, Pan-STARRS \citep{Chambers2016PS1} and WISE 3.4\,$\mu$m \citep{Wright2010WISE} to help us identify the high redshift targets. Images in the $r$-, $i$- and $z$-band from SDSS and Pan-STARRS should display colours that are consistent with early-type galaxies at $z\gtrsim 0.7$. Here, particularly a missing counterpart in the $r$-band provides hints at a high-redshift candidate. In principle, a reliability of $\approx 90$ \textcolor{revision1}{per cent} is expected for cluster candidates from the PSZ2 catalogue \citep[see fig. 11 in][]{PlanckCollab2016}. We also expect a positional uncertainty of approximately 1.5~arcmin for the PSZ2 cluster candidates \citep{PlanckCollab2016}.

The \textit{Planck} maps can be exploited further by exploring the lower S/N regime for massive high redshift clusters suitable for astrophysical investigations. We additionally selected targets detected through the MMF3 detection method for our study with this purpose in mind. Here, we focused on the regime $3 < \mathrm{S/N} < 4.5$ which is not covered by the PSZ2 catalogue, to look for rich cluster candidates at redshifts above $z\gtrsim0.7$. 
Due to the decreasing reliability in the low S/N regime of the MMF3 detected clusters, an identification of likely cluster candidates from among the numerous detections down to $\mathrm{S/N}=3$, requires an adequate preselection. For this, we focused on the Pan-STARRS/SDSS $i$-band and the WISE 3.4\,$\mu$m band, looking for over-densities of red galaxies by eye.
These complex selection criteria make it hard to assess the reliability of our MMF3 cluster candidates and render the MMF3 sample unsuitable for cosmological studies. 
Our investigated sample finally includes a total of 32 cluster candidates, with 23 candidates from the PSZ2 catalogue and 9 candidates detected with the MMF3 method.
 
The PSZ2 catalogue includes an additional parameter $Q_\mathrm{neural}$, which is an indicator for the quality of a detection, i.e. $Q_{\mathrm{neural}}<0.4$ marks detections of low reliability \citep{PlanckCollab2016,Aghanim2015,Hurier2017}. This quantity is based on the spectral energy distribution for each detection over the different frequency bands, as assessed by a neural network. It is sensitive to IR induced spurious detections, but is not constructed to flag detections caused by noise, which are more likely to occur in the low S/N regime. We examine the use of $Q_\mathrm{neural}$ in this regime in Section \ref{Confirmation of Cluster Candidates}. 

Noise-induced detections or projection effects of multiple clusters contributing to the SZ signal can cause a discrepancy between optical and SZ measurements. Apart from that, we expect Eddington bias to play a significant role for our cluster candidate sample. This purely statistical type of bias leads to a distorted view of the underlying distribution of objects when a cut in significance is applied \citep{Eddington1913}. It can be comprehended from the following considerations: galaxy clusters follow a steep halo mass function \citep{Tinker2008} with numerous low-mass halos but only few high-mass halos. When these halos are detected in the \textit{Planck} maps they carry an additional (approximately Gaussian) noise contribution \citep{PlanckCollab2016}. Accordingly, it is expected that more low-mass clusters scatter over the SZ-significance threshold for detection than high-mass clusters scatter below the threshold. This implies that sources at low S/N are more likely to be up-scattered and hence their SZ-based mass will be overestimated. This causes a systematic bias which depends on the significance threshold and the redshift \citep{VanDerBurg2016}.

\section{Photometric Observations}
\label{Photometric Observations}

We use imaging data in the Sloan $r$-, $i$- and $z$-band obtained with the Auxiliary-port CAMera \cite[ACAM,][]{Benn2008ACAM} at the 4.2-m William Herschel Telescope (WHT) to optically follow-up the 32 cluster candidates in our selected sample. In imaging mode, ACAM has a circular field of view of 8.3~arcmin diameter with a pixel scale of $0\farcs25$/pix on a red-optimised chip with $2148 \times 2500$ pixels. The filters are especially useful for the red-sequence analysis, because they bridge the 4000~{\AA}-break in the targeted redshift regime (z $\gtrsim 0.7$). 

The observations were completed in four separate runs: two service mode observation runs (PI: Hoekstra)  
on 2015 December 12 and 2016 January 19, and two visitor mode runs (PI: Schrabback)  
with two nights on 2016 October 6 and 2016 October 7 and three nights from 2017 March 20 to 2017 March 22. We observed the clusters with a total integration time between 630~s and 1800~s per filter depending on the roughly estimated redshift of the cluster and the observing conditions of the night. 

\subsection{Data reduction and calibration}
\label{Sec:DataReductionImaging}

For the data reduction of the WHT imaging data we employ the {\tt GUI} version of the \textsc{THELI}\footnote{\url{https://astro.uni-bonn.de/~theli/gui/index.html}} pipeline \citep{Erben2005,Schirmer2013}. The reduction includes a bias subtraction, flat-field correction, and a subtraction of a background and a fringe model. For the background model we make use of the dither pattern that was applied between exposures. This allows us to distinguish between features at a fixed position on the CCD and sky-related signals. The astrometric solution is calculated in \textsc{THELI} with the help of the SDSS DR8 or the USNO-B1 reference catalogue. Finally, the images are co-added.

We decide to use aperture magnitudes for the colour measurement of the galaxies, because they are reasonably robust at the low S/N regime of faint galaxies. For reliable colours, we need to make sure to always consider the flux from the same intrinsic part of the galaxy in each band. The image quality in the co-added images varies depending on the night of observation and on the band that was used (see Table \ref{tab:PFSsizesLimmags}). On average we find a FWHM PSF size of $1\farcs21$ in the $r$-band, $1\farcs20$ in the $i$-band and $1\farcs14$ in the $z$-band. To enable robust photometric measurements, we therefore perform a PSF homogenisation of the co-added images using the software \textsc{PSFEx} \citep{Bertin2011PSFEx}. 
The PSF profile in our observations is best described by a Moffat profile: 

\begin{equation}
I(r)=I_0\left[1+\left(\frac{r}{\alpha}\right)^2\right]^{-\beta}\quad.
\label{eq:Moffat profile}
\end{equation}
We target a PSF profile with a 10 per cent larger FWHM than the largest PSF size measured in the $r$-, $i$- or $z$-band. Here,  $\alpha=\frac{\mathrm{FWHM}}{2\sqrt{2^{1/\beta}-1}}$ and we use $\beta = 2.5$. With this setup we make sure that no deconvolution is required because the targeted PSF will always be broader than the original PSF.

The PSF-homogenised images provide the basis for the colour measurements. We measure the colours in circular apertures of $2''$ diameter with the software \textsc{Source Extractor} \citep{Bertin1996Sextractor}. Here, we make use of the dual image mode, where we take the $i$-band unconvolved co-added image as a detection image and the PSF-homogenised $r$-, $i$- or $z$-band image as the measurement image. We check the result of the PSF-homogenisation by comparing the flux of stars in a fixed $2''$ aperture to the flux measured in a flexible elliptical aperture ("FLUX\_AUTO" in \textsc{Source Extractor}). In case of a successful PSF-homogenisation the average ratio of the two fluxes should be the same in all three bands. We assess the performance with the quantity $\Delta f_\mathrm{loss}$ which denotes the maximum difference between the average flux ratios in stars in the $r$-, $i$- and $z$-bands.

We perform a photometric calibration by matching the instrumental magnitudes from the ACAM instrument to the magnitudes in the Pan-STARRS (PS1) catalogue \citep{Chambers2016PS1} based on the stars in the observed fields. We chose this catalogue as a reference because of its depth and because its footprint covers all of our targets. We obtain the zeropoints and colour terms which account for slightly different filter curves in ACAM and PS1. For the colour terms we fit a linear relation to the PS1 colour and the ACAM colour in $r-i$ and $i-z$ and apply a 5$\sigma$ clipping to exclude outliers.  
In the following, we refer to calibrated total magnitudes (Kron magnitudes) as $m_r$, $m_i$ and $m_z$ and to calibrated colour measurements as $r-i$, $r-z$ and $i-z$. All given magnitudes are in the AB magnitude system. 

We characterise the quality of our data with $5\sigma$ limiting magnitudes defined as:
\begin{equation}
m_{\mathrm{aplim}}= \mathrm{ZP} -2.5\,\log_{10}5\sigma_{\mathrm{sky}}\quad,
\label{eq:MagLim}
\end{equation}
where $\mathrm{ZP}$ is the zero-point of the field and $\sigma_\mathrm{sky}$ is the standard deviation of the sky background measured in 1000 randomly placed apertures of $2''$ diameter that do not contain any detected source \citep{Klein2017}. We measure averaged $5\sigma$ limiting magnitudes of $m_{r\mathrm{,aplim}}=24.93$ in the $r$-band, $m_{i\mathrm{,aplim}}=24.54$ in the $i$-band and $m_{z\mathrm{,aplim}}=23.82$ in the $z$-band.

We also quantify the depth limit of our observations. To do so, we inject simulated galaxies into our images and define the 80 per cent detection limit of the respective observation as the magnitude at which we still recover 80 per cent of the injected sources. For these sources, we assume a S\'{e}rsic light profile with a constant S\'{e}rsic parameter of $n=4$ and give them a random half-light radius drawn from a uniform distribution of 1--3~kpc (which we convert into the corresponding angular diameter assuming a redshift of $z=0.7$). The resulting detection limit for our detection band ($i$-band) is called $m_\mathrm{i,totlim}$. An example of recovery fraction $N_\mathrm{detected}/N_\mathrm{injected}$ of sources as a function of the $i$-band magnitude $m_i$ is presented in figure \ref{fig:80limit-histo}. On average the detection limit is $m_{i\mathrm{,totlim}}=23.43$.

Additionally, we define a corresponding limiting redshift as the redshift, at which the detection limit $m_\mathrm{i,totlim}$ in the $i$-band
coincides with the limit $m_i^*(z) + 1.25$. Here, $m_i^*(z)$ is the redshift-dependent characteristic $i$-band magnitude of the stellar mass function as measured in \citet{Muzzin2013a} and \citet{Ilbert2013}. A redshift-dependent characteristic mass of quiescent galaxies in the redshift range of interest can be deduced, which is expressed as $\log M_\mathrm{star}^*/M_\odot = 10.95 - 0.167 \times z$. We infer a corresponding $i$-band magnitude as expected from a quiescent galaxy with stellar mass $M_\mathrm{star}^*$, which formed at redshift $z=3$. We adopt this magnitude as our redshift-dependent characteristic magnitude $m_i^*(z)$. The limiting redshift therefore indicates the redshift at which the faintest and still detectable galaxies have a magnitude of $m_\mathrm{i,totlim} = m_i^*(z) + 1.25$. On average our observations are limited at redshifts of 0.80. The limiting magnitudes, detection limits, and limiting redshifts of our observed fields are reported in Table \ref{tab:PFSsizesLimmags}. We base our photometric redshift analysis on a catalogue of galaxies in our observations, detected with the software \textsc{Source Extractor} \citep{Bertin1996Sextractor}. We include all objects with the internal flags \texttt{FLAG}=0, \texttt{FLAG}=1 and \texttt{FLAG}=2, to reduce the number of blend rejections.

As a final step, we apply an extinction correction to the colours and magnitudes of the galaxies. We base the extinction correction on the method described in \cite{Tonry2012PS1Photosystem}, who use the value of $E(B-V)$ by \cite{Schlafly2011}\footnote{\url{ https://irsa.ipac.caltech.edu/applications/DUST/}}.

\begin{table*}
	\begin{threeparttable}
%	\vspace{13mm}
	\centering
	\caption{Properties of the imaging data in the $r$-, $i$- and $z$-band from the WHT. }
	\begin{tabular}{c l c c c c c c c c c}
		\hline\hline
			\textcolor{revision1}{ID} & Name  & $r$-band IQ$^a$ & $m_{r\mathrm{,aplim}}$$^{b}$& $i$-band IQ$^a$ & $m_{i\mathrm{,aplim}}$$^{b}$ & $z$-band IQ$^a$ & $m_{z\mathrm{,aplim}}$$^{b}$ & $m_{i\mathrm{,totlim}}$$^{c}$ & limiting & $\Delta f_\mathrm{loss}$ $^{e}$\\
		 & & [$''$] & [$\mathrm{mag}_\mathrm{AB}$] & [$''$] & [$\mathrm{mag}_\mathrm{AB}$] & [$''$] & [$\mathrm{mag}_\mathrm{AB}$] & [$\mathrm{mag}_\mathrm{AB}$] & redshift$^{d}$ & [\%]\\
	
		\textcolor{revision1}{115} & PSZ2 G032.31+66.0 & 0.93 &	24.96 &	1.06 &	24.81 &	1.04 &	24.10 & 23.95 & 0.90 & 2.39\\
		\textcolor{revision1}{277} & PSZ2 G066.34+26.1 & 1.46 &	24.67 &	1.47 &	24.35 &	1.65 &	23.56 & 22.75 & 0.67 & 1.73\\
		\textcolor{revision1}{378} & PSZ2 G085.95+25.2 & 1.41 &	25.10 &	1.05 &	24.08 &	1.06 &	23.75 & 23.05 & 0.73 & 1.96\\
		\textcolor{revision1}{381} & PSZ2 G086.28+74.7 & 1.07 &	25.35 &	0.86 &	24.94 &	0.74 &	23.93 & 23.85 & 0.88 & 4.95\\
		\textcolor{revision1}{420} & PSZ2 G092.64+20.7 & 0.91 &	24.00 &	0.83 &	23.90 &	0.79 &	22.69 & 23.15 & 0.75 & 3.23\\
		\textcolor{revision1}{421} & PSZ2 G092.69+59.9 & 1.10 &	24.90 &	1.23 &	24.96 &	1.26 &	23.95 & 23.75 & 0.86 & 2.48\\
		\textcolor{revision1}{483} & PSZ2 G100.22+33.8 & 1.38 &	25.29 &	1.35 &	24.68 &	1.13 &	24.17 & 23.25 & 0.77 & 1.39\\
		\textcolor{revision1}{545} & PSZ2 G112.54+59.5 & 1.12 &	25.07 &	1.21 &	24.45 &	1.23 &	24.43 & 23.25 & 0.77 & 4.45\\
		\textcolor{revision1}{623} & PSZ2 G126.28+65.6 & 0.92 &	25.21 &	0.89 &	24.50 &	0.87 &	23.76 & 23.55 & 0.82 & 1.81\\
		\textcolor{revision1}{625} & PSZ2 G126.57+51.6 & 1.08 &	25.01 &	1.16 &	24.38 &	1.20 &	23.65 & 23.25 & 0.77 & 1.20\\
		\textcolor{revision1}{667} & PSZ2 G136.02--47.1 & 0.79 &	24.66 &	0.75 &	24.50 &	0.81 &	23.67 & 23.75 & 0.86 & 1.24\\
		\textcolor{revision1}{681} & PSZ2 G139.00+50.9 & 1.44 &	25.44 &	1.42 &	24.91 &	1.15 &	24.00 & 23.55 & 0.82 & 1.94\\
		\textcolor{revision1}{690} & PSZ2 G141.98+69.3 & 0.98 &	25.56 &	0.99 &	25.09 &	0.91 &	24.38 & 24.15 & 0.93 & 3.13\\
		\textcolor{revision1}{740} & PSZ2 G152.47+42.1 & 1.45 &	25.02 &	1.52 &	24.49 &	1.29 &	23.80 & 22.95 & 0.72 & 0.91\\
		\textcolor{revision1}{769} & PSZ2 G160.94+44.8 & 1.00 &	24.74 &	1.29 &	24.35 &	1.00 &	24.10 & 23.35 & 0.79 & 1.72\\
		\textcolor{revision1}{789} & PSZ2 G165.41+25.9 & 1.81 &	24.98 &	1.66 &	23.71 &	1.65 &	23.93 & 23.25 & 0.77 & 2.26\\
		\textcolor{revision1}{1074} & PSZ2 G237.68+57.8 & 1.46 &	25.19 &	1.49 &	24.77 &	1.53 &	24.04 & 23.15 & 0.75 & 0.22\\
		\textcolor{revision1}{1121} & PSZ2 G246.91+24.6 & 1.64 &	22.21 &	1.59 &	24.49 &	1.76 &	22.55 & 23.15 & 0.75 & 2.25\\
		\textcolor{revision1}{1441} & PSZ2 G305.76+44.7 & 1.52 &	25.07 &	1.41 &	23.92 &	1.45 &	23.53 & 22.65 & 0.66 & 0.99\\
		\textcolor{revision1}{1493} & PSZ2 G316.43+54.0 & 0.76 &	25.03 &	1.18 &	24.58 &	0.74 &	24.02 & 23.55 & 0.82 & 5.34\\
		\textcolor{revision1}{1512} & PSZ2 G321.30+50.6 & 1.28 &	24.95 &	1.30 &	24.61 &	1.20 &	23.79 & 23.25 & 0.77 & 1.28\\
		\textcolor{revision1}{1539} & PSZ2 G326.73+54.8 & 1.25 &	25.16 &	1.22 &	24.52 &	1.02 &	23.61 & 23.35 & 0.79 & 0.42\\
		\textcolor{revision1}{1606} & PSZ2 G343.46+52.6 & 1.20 &	25.06 &	1.18 &	24.80 &	1.15 &	24.16 & 23.45 & 0.81 & 0.24\\
		
		\textcolor{revision1}{-} & PLCK G55.00--37.0 & 0.99 &	24.91 &	1.02 &	24.38 &	0.93 &	23.47 & 23.55 & 0.82 & 1.73\\
		\textcolor{revision1}{-} & PLCK G58.14--72.7 & 1.15 &	25.14 &	1.00 &	24.39 &	0.94 &	23.86 & 23.65 & 0.84 & 2.83\\
		\textcolor{revision1}{-} & PLCK G82.51+29.8 & 1.48 &	25.20 &	1.32 &	24.80 &	1.27 &	23.99 & 23.55 & 0.82 & 1.07\\
		\textcolor{revision1}{-} & PLCK G98.08--46.4 & 1.70 &	25.26 &	1.68 &	24.84 &	1.29 &	24.42 & 23.45 & 0.81 & 1.81\\
		\textcolor{revision1}{-} & PLCK G122.62--31.9 & 1.31 &	25.02 &	1.19 &	24.64 &	1.10 &	23.77 & 23.55 & 0.82 & 0.47\\
		\textcolor{revision1}{-} & PLCK G150.77+17.1 & 1.16 &	24.80 &	1.19 &	24.43 &	1.33 &	23.62 & 23.35 & 0.79 & 0.67\\
		\textcolor{revision1}{-} & PLCK G164.82--47.4 & 1.27 &	24.79 &	1.30 &	24.44 &	1.19 &	23.75 & 23.35 & 0.79 & 1.10\\
		\textcolor{revision1}{-} & PLCK G174.14--27.5 & 1.01 &	25.32 &	0.92 &	24.76 &	0.93 &	23.94 & 23.85 & 0.88 & 1.86\\
		\textcolor{revision1}{-} & PLCK G184.49+21.1 & 0.77 &	24.79 &	0.67 &	24.68 &	0.76 &	23.93 & 23.95 & 0.90 & 3.74\\

		 \hline

	\end{tabular}
	\label{tab:PFSsizesLimmags}
		\begin{tablenotes}
		\textbf{\textit{Notes.}} $^a$ FWHM of the PSF (seeing). $^b$ $5\sigma$ limiting magnitudes as defined by Equation \ref{eq:MagLim}. $^c$ Detection limit at which 80 per cent of the simulated, injected galaxies are still recovered in the source detection in the $i$-band with \textsc{Source Extractor}. $^d$ Limiting redshifts of the observations defined as the redshift at which $m_{i,\mathrm{totlim}} = m_i^*(z) + 1.25$. $^e$ Maximum difference between the average flux ratios (comparing $2''$ aperture flux and FLUX\_AUTO) in stars in the $r$-, $i$- and $z$-bands after the PSF-homogenisation (see Section \ref{Sec:DataReductionImaging}).\\
		\textcolor{revision1}{We indicate the PSZ2 ID of the candidates in column 1 and the full name in column 2 \citep{PlanckCollab2016}}. In case this is not available, we give the candidates generic names starting with "PLCK" followed by a notation of the galactic coordinates. 
		\end{tablenotes}
	
	\end{threeparttable}

\end{table*}

\subsection{Red-sequence models and redshift estimates}

We aim to extract redshift and richness information about the galaxy clusters from the available optical data. For this, we make use of the fact that early-type galaxies, which are the dominant population in massive galaxy clusters, follow a tight correlation between colour and magnitude with a typically very small intrinsic scatter of $< 0.1\, \mathrm{mag}$ \citep{Bower1992}. These galaxies are host to stellar populations that have evolved passively since $2 < z < 5$ \citep[e.g.][]{Bower1992,Lin2006}. This so called red-sequence is characterised by its slope and intercept, which depend on the redshift \citep{Gladders1998}. For the galaxy clusters detected with \textit{Planck} at redshifts up to $z\sim 1$, there are enough red-sequence galaxies in the clusters to present an excess to the background field galaxies. This allows us to estimate the redshift of the cluster by comparing the colours of the galaxies with the colours of empirical red-sequence models, which predict the colour of red-sequence galaxies as a function of their magnitude and redshift. We construct the empirical red-sequence models analogously to the work by \citet{VanDerBurg2016} on the basis of the deep 30-band photometric data of the COSMOS/UltraVISTA field \citep{Muzzin2013b}. This provides us with a catalogue of galaxies down to faint magnitudes and with high quality photometric redshifts allowing us to constrain the red-sequence models over the full magnitude range of interest. 
We select quiescent galaxies based on their rest-frame $U-V$ and $V-J$ colours up to redshifts of 1.2 and down to a magnitude of $m_i = 24.0$. These galaxies have similar properties as the cluster red-sequence galaxies. 
The $r$-, $i$-, and $z$-band magnitudes in the UltraVISTA catalogue were obtained with the Subaru filters, which do not match the filters of PS1. Therefore, we transform the UltraVISTA colours to the PS1 photometric system. As a result, a set of quiescent galaxies is available with COSMOS/UltraVISTA redshifts and colours and total magnitudes corresponding to the PS1 system.

\begin{figure}
	\centering
	\includegraphics[width=1.0\columnwidth]{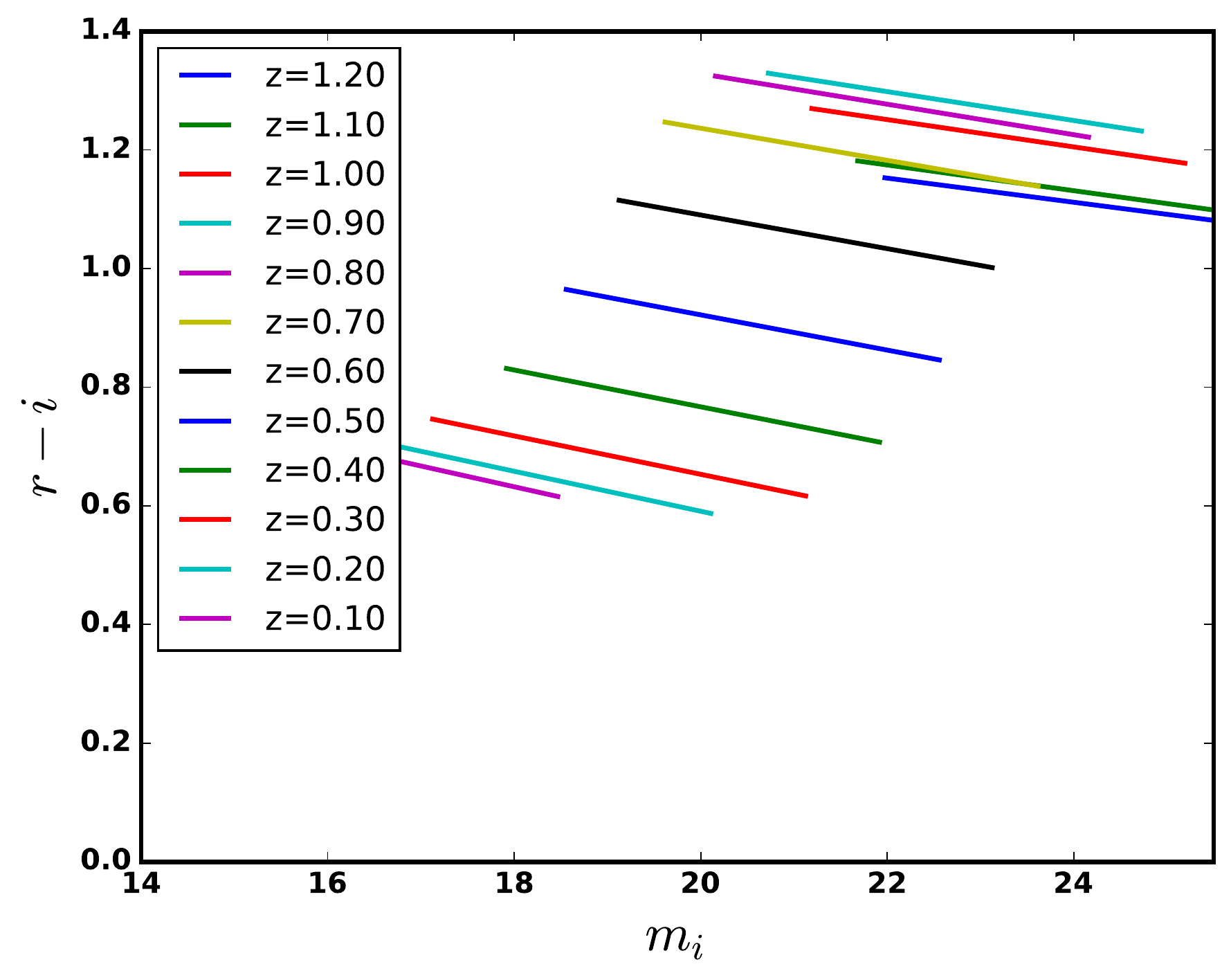}
	\includegraphics[width=1.0\columnwidth]{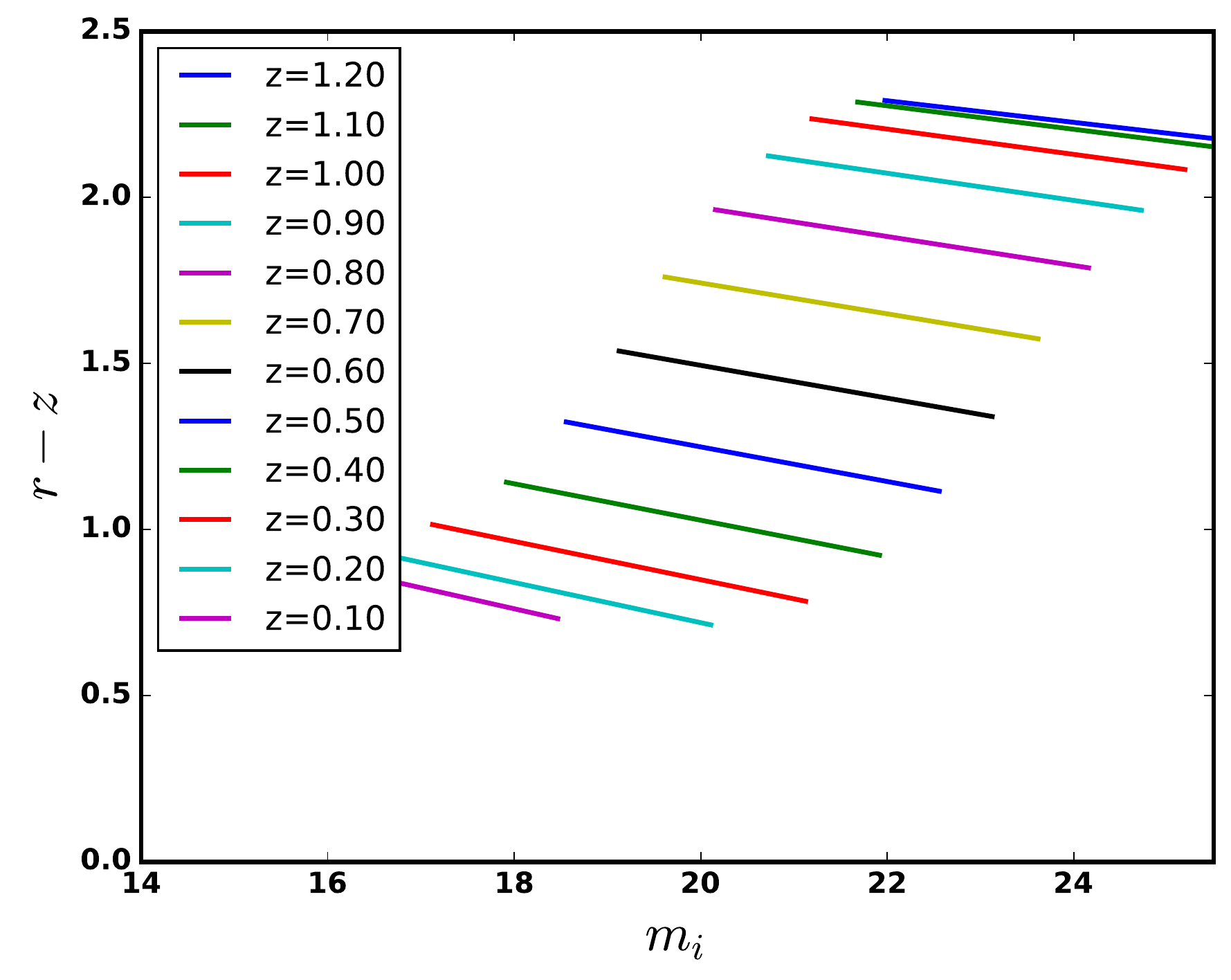}
	\includegraphics[width=1.0\columnwidth]{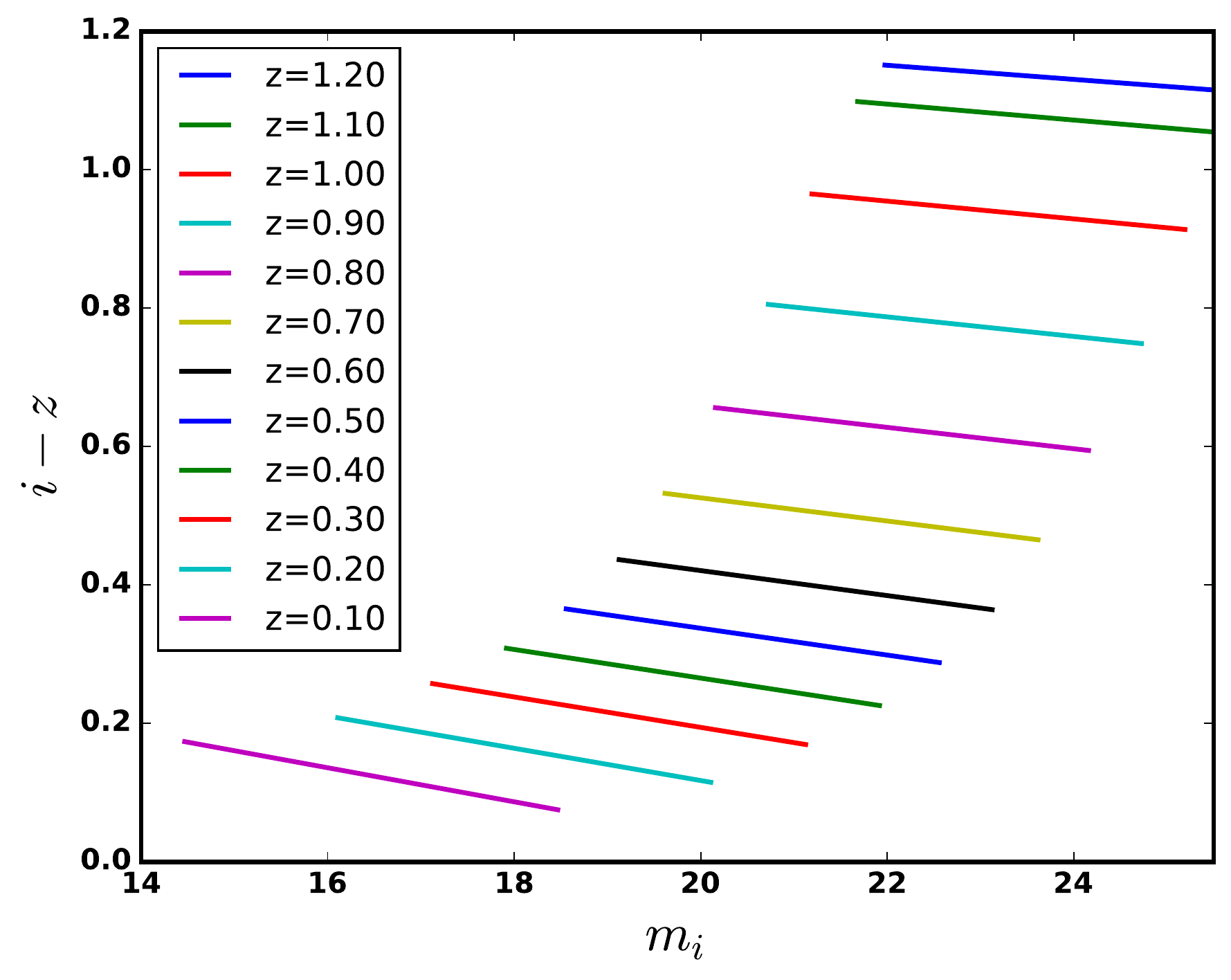}
	\caption{Empirical red-sequence models for three colours (recalibrated, see Section \ref{Sec:RScalib}). \textit{Top}: $r-i$ vs $m_i$, highest sensitivity in regime $0.3 \lesssim z \lesssim 0.7$, \textit{Middle}: $r-z$ vs $m_i$, highest sensitivity in regime $0.4 \lesssim z \lesssim 0.9$, \textit{Bottom}: $i-z$ vs $m_i$, highest sensitivity in regime $0.6 \lesssim z \lesssim 1.1$. Each line covers the range $m_i^*-2.0 \leq m_i \leq m_i^*+2.0$.   }
	\label{fig:RS-models}
\end{figure}

As described in \citet{VanDerBurg2016}, we then divide the galaxies into redshift bins with width 0.04 and step size 0.01 and fit a linear relation to the colours ($r-i$, $r-z$ or $i-z$) as a function of the total $i$-band magnitude $m_i$. This provides us with a slope, intercept (at magnitude $m_i = 22.0$) and scatter for each redshift step. We show the models for the three available colour combinations in Fig. \ref{fig:RS-models}. The models have the highest sensitivity at redshifts where the two involved filters enclose the 4000~{\AA}-break. Hence, the ($r-i$) vs $m_i$, ($r-z$) vs $m_i$, and ($i-z$) vs $m_i$ model is most sensitive to redshifts of ($0.3 \lesssim z \lesssim 0.7$), ($0.4 \lesssim z \lesssim 0.9$), and ($0.6 \lesssim z \lesssim 1.1$), respectively.

For our task to estimate the redshifts empirically with the help of our red-sequence models we conduct several steps, which are largely based on works by \citet{Klein2017}, \citet{Klein2019} and \citet{VanDerBurg2016}. 
Our basic strategy is to count how many galaxies in an 0.5~Mpc radius around the cluster centre agree with the red-sequence models at different redshift steps. This way, we obtain a histogram of counted galaxies versus redshift, which exposes an over-density of galaxies at the cluster redshift because of the red-sequence galaxies in the cluster. To enhance this over-density, we apply certain filters and weighting techniques to the galaxies. We describe these below.

\subsubsection{Magnitude cuts and colour weights}
\label{Sec:MagCutsColourWeights}
In a first step, we identify a candidate for the brightest cluster galaxy (BCG) of the cluster candidate from a colour image in the $r$-, $i$- and $z$-band. Considering the galaxies in a 0.5~Mpc radius around the BCG \citep[a value adapted from][]{Buddendiek2015},  
we then count all galaxies $p$ as a match to the red-sequence models at a given redshift step if they fulfil all of these criteria:

\begin{enumerate}
	\item The colour of galaxy $p$ agrees with all three red-sequence models at the same time within three times the standard deviation of the respective red-sequence model
	\begin{equation}
	 \Delta c_{p,k}=|c_{p,k}-\langle c(z,m_i)\rangle_{k}|<3\sigma_{c_{k}}(z)\quad. 
	\end{equation}
	Here, the colour combinations are described by the index $k\in {1,2,3}$ with $(c_{1},c_{2},c_{3})=(r-i,i-z,r-z)$, $c_{p,k}$ is the measured colour $k$ of galaxy $p$ and $\langle c(z,m_i)\rangle _{k}$ is the model colour $k$ at redshift $z$ with a scatter of the model $\sigma_{c_{k}}(z)$. We also use galaxies if they only fall in the $3\sigma_{c_{k}}(z)$ range of the red-sequence models taking their $1\sigma$ photometric errors into account.
	\item The galaxy is brighter than the detection limit $m_i < m_{i\mathrm{,totlim}}$.
	\item The galaxy is brighter than $m_i < m_i^*(z)+1.25$, where $m_i^*(z)$ is the characteristic magnitude at the respective redshift.
\end{enumerate}

We then weight each galaxy by its colour. At each redshift step we assign a weight $w_{p}(z)$ to the galaxy depending on how close the galaxy's colour is to the colour predicted by the red-sequence models at that step \citep{Klein2019}: 

\begin{equation}
	w_{p}(z)=\frac{\text{\ensuremath{\prod}}_{k=1}^{3}G(\Delta c_{p,k}\,,\sigma_{c_{k}}(z))}{N(\sigma_{c_{1}}(z),\sigma_{c_{2}}(z),\sigma_{c_{3}}(z))}\quad.
	\label{eq:ColourWeight}
\end{equation}
$G(\Delta c_{p,k},\sigma_{c_{k}}(z))$ is the value of a normalised Gaussian function at colour offset $\Delta c_{p,k}$.
The normalisation $N$ is defined as 

\begin{equation}
	N(\sigma_{c_{1}}(z),\sigma_{c_{2}}(z),\sigma_{c_{3}}(z))=\text{\ensuremath{\prod}}_{k=1}^{3}G(0,\sigma_{c_{k}}(z))\quad.
\end{equation} 

\subsubsection{Completeness correction}
\label{Sec:CompletenessCorrection}
In the third criterion of our list, we apply a magnitude cut-off to the galaxies that we take into consideration. However, it is possible that the detection limit $m_{i\mathrm{,totlim}}$ (second criterion), which represents a limit to the completeness of the detections, is brighter than the limit $m_i < m_i^*(z)+1.25$. In that case, we have to account for the galaxies that we miss due to the limited depth of our data. Following \citet{Klein2017}, we estimate the amount of galaxies, which we expect to miss, by extrapolating the Schechter function down to our magnitude cut-off. The Schechter function $S(m,m_i^*,\alpha)$ is defined via

\begin{equation}
\begin{aligned}[b]
S(m,m_i^*,\alpha)\mathrm \, {d}m = & \;0.4\,\ln(10)\,\Phi^*10^{-0.4(m-m_i^*)\cdot(\alpha+1)}\\
& \cdot\exp{[-10^{-0.4(m-m_i^*)}]}\mathrm {d}m \quad.
\end{aligned}
\label{eq:SchechterFunction}
\end{equation}
We chose a value for the faint-end slope of $\alpha = -1.0$ adapted from \citet{Klein2017}. This results in a completeness correction factor

\begin{equation}
	c_{\mathrm{cmp}}=\frac{\int_{-\infty}^{m_i^{*}+1.25}S(m,m^{*},\alpha)\mathrm{d}m}{\int_{-\infty}^{m_{i,\mathrm{totlim}}}S(m,m^{*},\alpha)\mathrm{d}m}\quad
\end{equation}
\citep{Klein2017}. We only apply this correction factor in case that the detection limit $m_{i\mathrm{,totlim}}$ is brighter than $m_i^*(z)+1.25$. Otherwise, we set $c_{\mathrm{cmp}}=1.0$.

In addition, it is possible to correct for the magnitude-dependent fraction of retrieved versus injected galaxies from the estimation of the 80 per cent depth of the data (see Section \ref{Sec:DataReductionImaging} and Appendix \ref{Sec:AppA80percentLimit}). However, we only apply this type of correction when we estimate the cluster richness (see Section \ref{Sec:Rich+MassEstimates}).

\subsubsection{Radial weights}
\label{Sec:RadialWeights}

The galaxies in a cluster are typically more abundant towards the centre. To include this information in our analysis, we follow an approach by \citet{Klein2019}. We weight galaxies according to their distance from the cluster centre (characterised by the BCG position) with the help of a Navarro Frenk White (NFW) profile \citep{Navarro1997}. Although originally intended to describe the distribution of dark matter in $N$-body simulations, it also provides a good description of the number density profile of cluster galaxies \citep[e.g.][]{Lin2004,Hansen2005}. The surface density of galaxies can be expressed as 

\begin{equation}
	\Sigma (R) \propto \frac{1}{(R/R_\mathrm{S})^2-1}f(R/R_\mathrm{S})
\end{equation}
\citep{Bartelmann1996}. Here, we set $R_\mathrm{S}=0.15h^{-1}\mbox{ Mpc}$ \citep{Rykoff2012} as the characteristic scale radius and

\begin{equation}
f(x) = 
	\begin{cases}
		1-\frac{2}{\sqrt{x^2-1}}\arctan{\sqrt{\frac{x-1}{x+1}}} &  (x > 1)\\
		1-\frac{2}{\sqrt{1-x^2}} \arctanh {\sqrt{\frac{1-x}{x+1}}} &  (x < 1) \quad.
	\end{cases}
\end{equation}
Below the minimum radius of $0.1h^{-1}\mbox{ Mpc}$ we set
the radial weight to be constant, to avoid a singularity for $R=0$ \citep{Rykoff2012}. The profile is truncated at a cutoff radius $R_C=0.5\mbox{ Mpc}$ for the redshift estimates, because we consider all galaxies out to this radius for the estimate. Accordingly, we normalise the profile with the help of a correction term $C_\mathrm{rad}$ as

\begin{equation}
	1=C_{\mathrm{rad}}\int_{0}^{R_\mathrm{C}}{\mathrm{d}R\ 2\pi R_p\Sigma(R_p)}
\end{equation}
\citep{Klein2019}. Thus, the radial weight for a galaxy $p$ at distance $R_p$ from the estimated centre is
\begin{equation}
	n_p(z)=C_{\mathrm{rad}}(z)2\pi R_p\Sigma(R_p)\quad.
	\label{eq:RadialWeight}
\end{equation}

\subsubsection{Masking and statistical background estimate}
\label{Sec:Masking+BGEstimate}
As a next step, we account for the contribution of field galaxies that do match the models but do not actually belong to the cluster itself. Since the field of view of ACAM is too small to estimate the local background contribution from the available images, we make use of the UltraVISTA catalogue from \cite{Muzzin2013b}, which we matched to the PS1 photometric system. This field covers around 1.62 deg$^2$ corresponding to more than 100 times the field of view of ACAM. The deep photometric data in a field of this size provide a good basis for an estimate of the background contribution from field galaxies. However, this approach cannot account for a variation of this contribution over large spatial scales.

In order to mimic similar conditions as for the observations with the ACAM instrument, we add Gaussian noise  to the flux of the galaxies with a standard deviation equal to the one from the sky background $\sigma_\mathrm{sky}$. Next, we count the number of galaxies that agree with the available red-sequence models at the different redshift steps. Here, we apply the same criteria and weighting steps (see Sections \ref{Sec:MagCutsColourWeights}, \ref{Sec:CompletenessCorrection}, \ref{Sec:RadialWeights}) as during the analysis of the galaxies in the actual co-adds.
We estimate the contribution of background galaxies this way in 75 ACAM-sized, non-overlapping apertures covering the UltraVISTA field. For each galaxy $q$ in an aperture we calculate the quantities $w_q(z)$ and $n_q(z)$ analogously to Equations \ref{eq:ColourWeight} and \ref{eq:RadialWeight}. We average the results from 75 apertures, calculate the standard deviation and normalise the values to an area of 1~arcmin$^2$, so that we can subtract our statistical background estimate from the weighted number of galaxies in differently sized areas in the co-added image.

Additionally, we mask bright foreground objects in the field of view that potentially cover galaxies in the cluster. We subtract the area covered by these masks from the total area within a 0.5~Mpc radius.

\subsubsection{Iteration of the redshift estimate}

\begin{figure}

	\includegraphics[width=1.0\columnwidth]{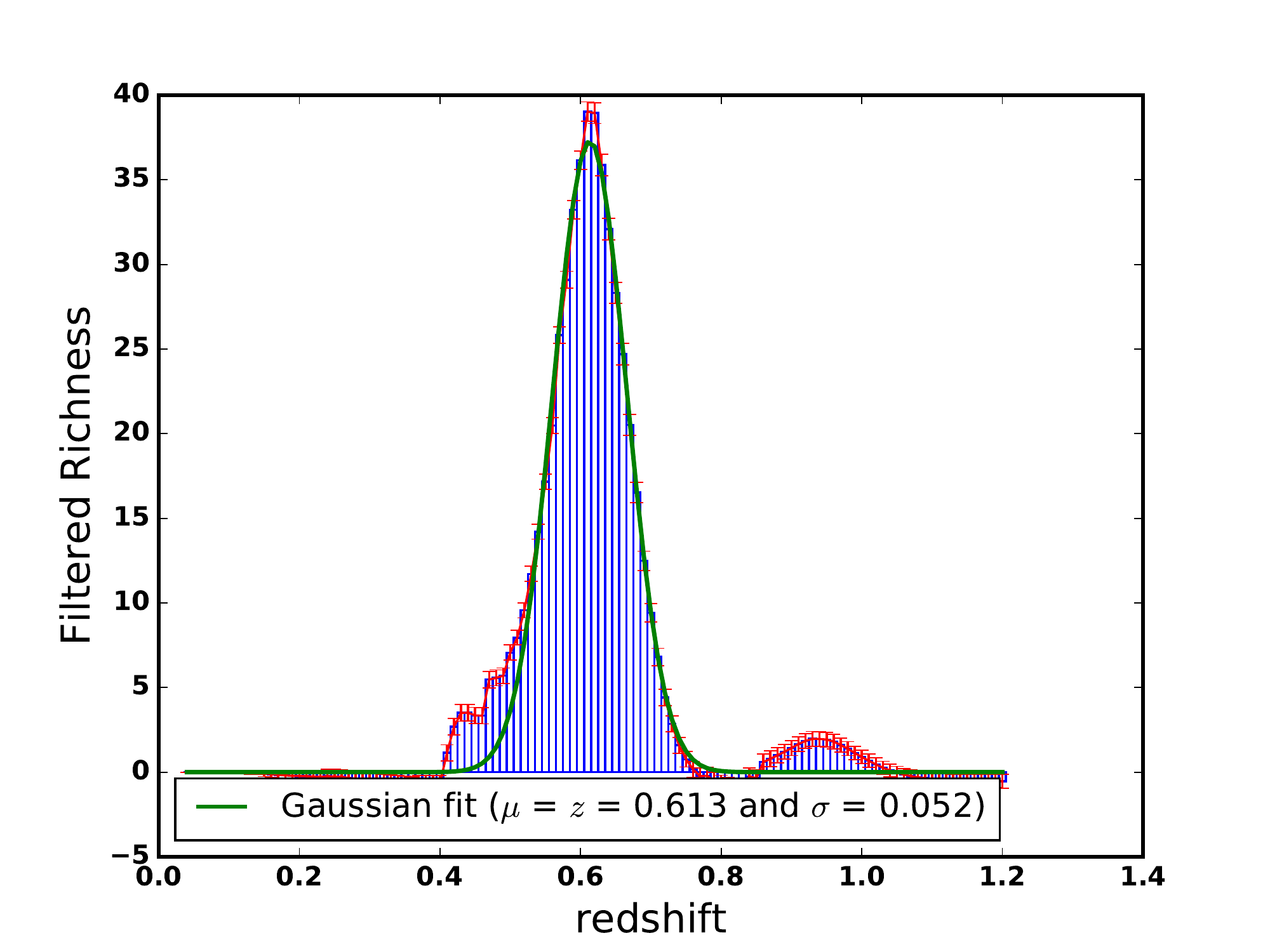}
	\includegraphics[width=1.0\columnwidth]{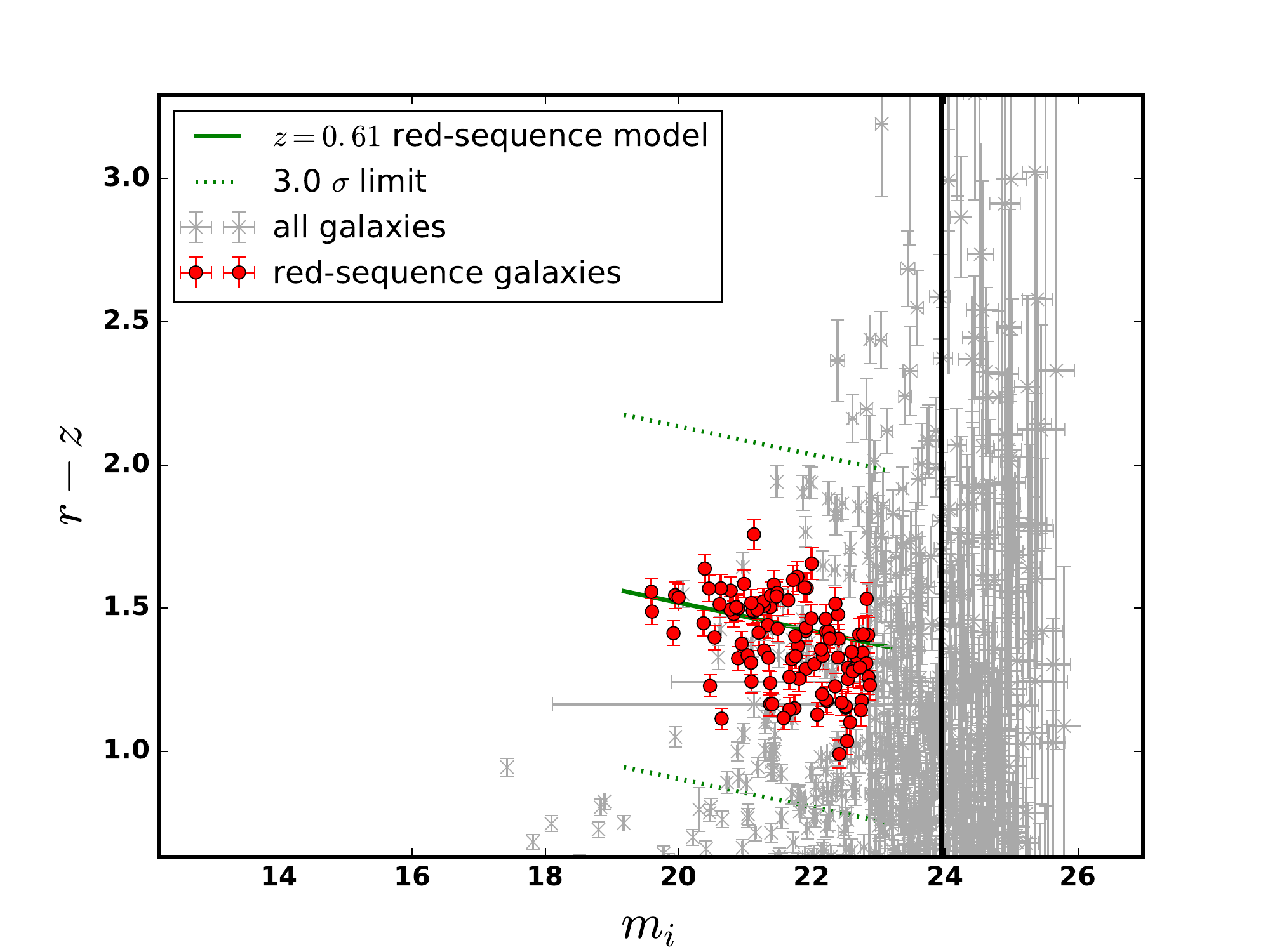}
	\caption{\textit{Top:} Filtered richness $\lambda_{\mathrm{MCMF}}$ versus redshift with Gaussian fit in green for cluster candidate PSZ2 G032.31+66.07. The blue bars display $\lambda_\mathrm{MCMF}$ at the different redshift steps. The red error bars display the uncertainties that emerge from the background subtraction. \textit{Bottom:} Diagram of $(r-z)$ colour versus $i$-band magnitude for cluster candidate PSZ2 G032.31+66.07. The grey points mark all galaxies observed in the image. The red points are the red-sequence galaxies within $R_\mathrm{C}$ consistent with the (recalibrated) red-sequence model at $z=0.61$, that is indicated by a green, solid line. The green, dotted lines mark the $\pm3\sigma$ band around that model. The red dots only display red-sequence galaxies that are brighter than the limit $m_i < m_i^* + 1.25$. The vertical, black line represents the detection limit $m_{i\mathrm{,totlim}}$ of this observation. }
	\label{fig:RShistogram+CMD}
\end{figure}

Evaluating the weighting and masking schemes described in the previous sections at each redshift step, provides us with the so called filtered richness \citep{Klein2017,Klein2019} 

\begin{equation}
\begin{aligned}[b]
	\lambda_{\mathrm{MCMF}}(z)	= &	\,c_{\mathrm{cmp}}(z)\cdot\Sigma w_{p}(z)n_{p}(z)\\
	& -c_{\mathrm{cmp}}(z)\cdot A_{\mathrm{cl}}(z)\cdot\Sigma w_{q}(z)n_{q}(z)\quad.
\end{aligned}
\end{equation}
Here, $A_\mathrm{cl}$ is the area within a 0.5~Mpc radius (with the masked regions excluded). The index $p$ denotes galaxies from the observations and the index $q$ indicates galaxies from the estimate of the (average) background from the UltraVISTA field. The filtered richness quantifies how many galaxies are consistent with the red-sequence models at different redshift steps. For an initial redshift estimate, we take the BCG as the centre of the cluster and identify the most prominent over-density in the distribution of filtered richness $\lambda_{\mathrm{MCMF}}(z)$ as a function of redshift. We fit a Gaussian function to this over-density and take the peak position of the Gaussian as our initial redshift estimate.  At this point, we set $n_{p}(z) = n_{q}(z) = 1$. We do this because we want to account for different cluster morphologies, where the BCG is not necessarily always right in the centre for a cluster. To obtain an estimate of the centre of the galaxy over-density, we then identify all galaxies in the field of view that agree with the initial redshift estimate and pick that galaxy as a new centre that has the maximum number of neighbours within a 0.5~Mpc radius. Afterwards, we repeat the redshift estimate analogously now around the new centre and including the radial weights $n_{p}(z)$ and $n_{q}(z)$. An overview of the redshift estimation procedure is given in Fig. \ref{fig:RShistogram+CMD}. We obtain statistical errors for the estimated redshifts by bootstrapping the catalogue of galaxies in the respective observation. Here, we assemble a new catalogue by drawing galaxies from the original catalogue at random until we have a catalogue of the same length again, where it is possible that some galaxies enter the new catalogue multiple times and others do not enter it at all. We create 1000 new catalogues and re-estimate the redshift a thousand times per cluster candidate. To account for variation in the background of field galaxies, we only pick one position in the UltraVISTA catalogue per bootstrap step for the background subtraction. Thus, the statistical errors also mirror the variation of the background. The symmetric 68 per cent uncertainty can be calculated via
 
 \begin{equation}
	 \sigma_\mathrm{sym}^2 = \frac{\sum_{i=1}^{B}\left(z(\bold{x}_{i}^{*}) - \bar{z}(\bold{x}^*)\right)^2}{B-1} \quad.
	\label{eq:SymBootError}
 \end{equation}
Here, $B=1000$ is the number of bootstrap iterations, $z$ is the measured quantity, i.e. the redshift, $\bold{x}$ represents the original catalogue of galaxies and then $\bold{x}_{i}^{*}$ is the $i$-th bootstrapped version of this catalogue. Finally, $\bar{z}(\bold{x}^*)$ is the average of all redshifts that result from the bootstrap iterations. We find an average uncertainty of $\sigma_{z_\mathrm{phot}} = 0.062$ for the photometric redshift measurements.

\subsubsection{Calibration of the redshift estimate}
\label{Sec:RScalib}

\begin{figure}
	\centering

\includegraphics[width=1.0\columnwidth]{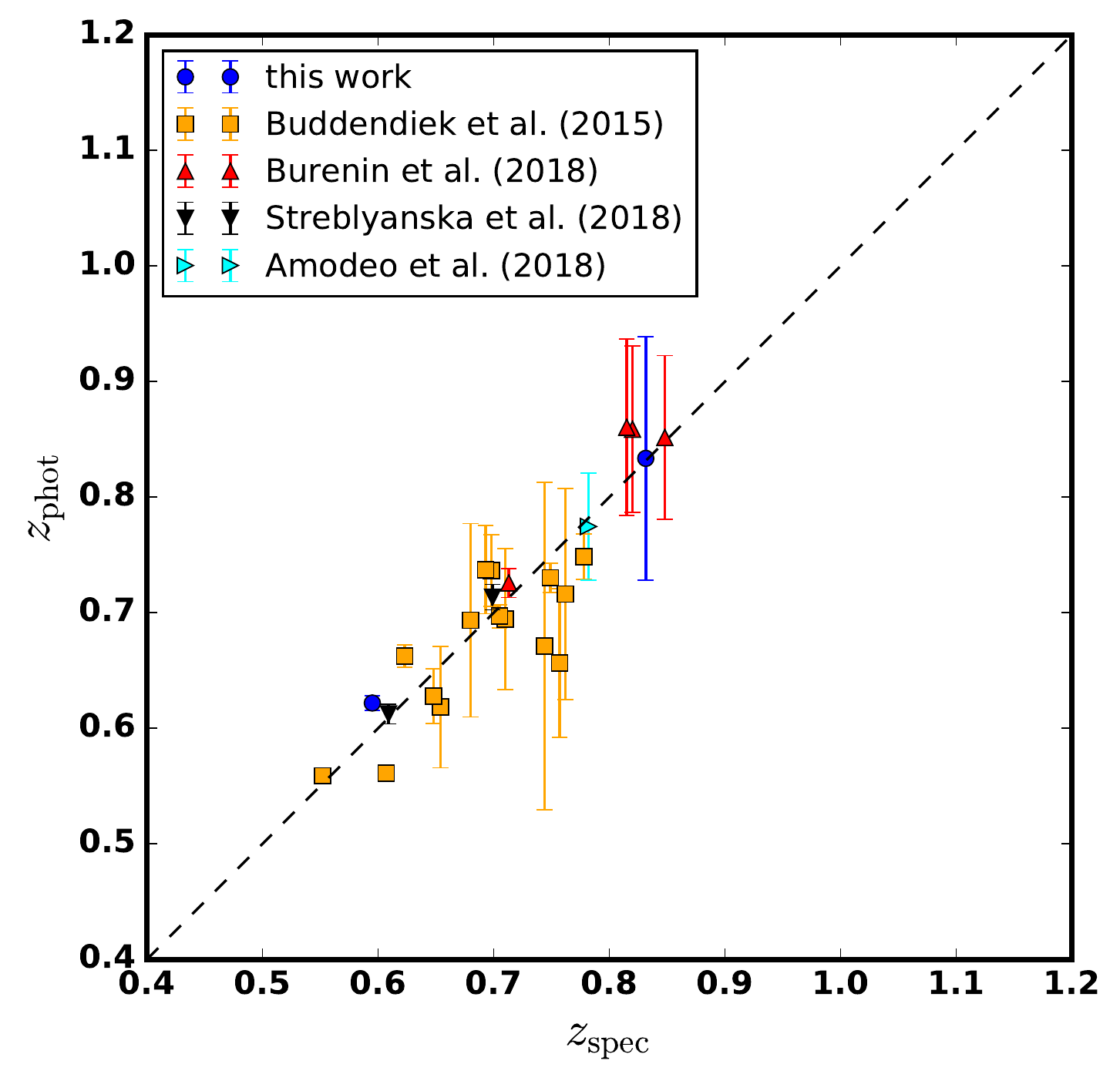}

	\caption{Comparison of red-sequence redshifts (obtained through the analysis steps described in this work) after recalibration of the red-sequence models and spectroscopic redshifts from this work, \citet{Buddendiek2015}, \citet{Burenin2018}, \citet{Streblyanska2018} and \citet{Amodeo2018}.
	 The error bars show the statistical 68 per cent errors, which result from a bootstrapping according to equation \ref{eq:SymBootError}. We find a scatter of $\sigma_z = 0.021$. 
	 }
	\label{fig:Spec-Photo-z-comparison}
\end{figure}

Since our red-sequence models are based on galaxies in the field, which may differ from the cluster member galaxies in age or metallicity and thus in broadband colour, we perform a correction to our best-fit photometric redshifts based on a direct comparison with spectroscopically determined redshifts.
Apart from our own spectroscopic redshift estimates (see Section \ref{Spectroscopic Observations}) for two clusters with a sufficiently well-defined red-sequence, we include results from several other authors. \citet{Burenin2018} measured spectroscopic redshifts with the BTA 6-m telescope using the instruments SCORPIO and SCORPIO-2 for five of the clusters in our sample, namely PSZ2 G092.69+59.92, PSZ2 G126.28+65.62, PSZ2 G126.57+51.6, PSZ2 G237.68+57.83 and PSZ2 G343.46+52.65. \citet{Amodeo2018} provide a spectroscopic redshift for PSZ2 G085.95+25.23 with Keck/LRIS spectroscopy. \citet{Streblyanska2018} list spectroscopic redshifts originating from the SDSS DR12 spectroscopic information for PSZ2 G032.31+66.07 and PSZ2 G086.28+74.76.  Additionally, we resort to data by \citet{Buddendiek2015}, who measured spectroscopic and photometric redshifts with the help of observations also obtained at the WHT using ACAM. \citet{Buddendiek2015} investigate a sample originating from the \textit{ROSAT} All Sky Survey. It therefore does not relate to the PSZ2 follow-up pursued here and we only use this sample for calibration purposes of the photometric redshifts. We estimate red-sequence based redshifts using their $r$-, $i$- and $z$-band observations of 15 galaxy clusters and compare these photometric redshifts to the corresponding spectroscopic redshift results found by \citet{Buddendiek2015}.
From the overall comparison of spectroscopic and photometric redshifts we find that we overestimate the redshift by a median offset of $\Delta z = (z_\mathrm{phot} - z_\mathrm{spec})_\mathrm{median} = 0.104 \pm 0.045$. We decide to recalibrate our red-sequence models in an iterative manner. For this purpose we compare the colours of the red-sequence models at the photometric redshift and at the spectroscopic redshift. We then adjust the intercepts of our three red-sequence models in a redshift dependent way. For this, we fit a line to the colour offset as a function of redshift and modify the intercepts of the red-sequence models accordingly. Subsequently, we re-estimate the photometric redshifts. We repeat this process until we minimise the scatter between photometric and spectroscopic redshifts given by 
\begin{equation}
\sigma_{z} = \sqrt{\frac{1}{N}\sum \left(\frac{z_\mathrm{spec}-z_\mathrm{phot}}{1+z_\mathrm{spec}}\right)^2}\quad.
\end{equation}
Here, $N$ is the number of galaxy clusters with available spectroscopic redshifts and $z_\mathrm{spec}$ and $z_\mathrm{phot}$ are the spectroscopic and red-sequence redshifts of the clusters, respectively. Table \ref{tab:ResultsPSZ2} gives an overview of the calibrated red-sequence redshifts of our complete cluster sample. In Fig. \ref{fig:Spec-Photo-z-comparison}, we plot the calibrated photometric redshift versus the corresponding spectroscopic redshift of all clusters used for the recalibration. From this we can see that the systematic bias has been removed. The remaining scatter is $\sigma_z = 0.021$ as compared to the average uncertainty of $\sigma_{z_\mathrm{phot}} = 0.062$ for the photometric redshift measurements.

\subsection{Richness and mass estimates}
\label{Sec:Rich+MassEstimates}

We want to relate the results from our optical data to the SZ-based results (in particular the mass $M_\mathrm{500c,SZ}$) inferred from the \textit{Planck} measurements. In this work, we make use of the richness-mass scaling relation established by \citet{Rozo2015}, which connects the richness estimated from optical data of the SDSS DR8 \texttt{redMaPPer} catalogue \citep{Rykoff2014} to the SZ-mass from the PSZ1 catalogue \citep{PlanckCollab2014XXIX}, the progenitor of the PSZ2 catalogue. This scaling relation was already used for comparison of optical observations to SZ-observations e.g. by \citet{PlanckCollab2016} and \citet{VanDerBurg2016}. Additionally, this scaling relation is suitable for our work in contrast to other richness-mass scaling relations because it directly relates the mass estimate inferred from the SZ-signal (in particular from PSZ1, which is very similar to PSZ2) to a richness, that is inferred from a red-sequence analysis. Both of these quantities are available for the cluster candidates in our sample. It also extends to relatively high redshifts ($z\leq0.5$) compared to other richness-mass scaling relations and uses a relatively large sample size of 191 cluster to base their results on.

The scaling relation by \citet{Rozo2015} is based on the richness as defined in \citet{Rykoff2014}. Therefore, we aim to estimate the richness in a similar way. The richness is estimated by counting all the galaxies that agree with the redshift estimate above a certain magnitude threshold and within some cut-off radius $R_\mathrm{C}$. However, the cut-off radius itself depends on the richness as well. In particular, \citet{Rykoff2014} assume a power-law relation between the richness $\lambda$ and the cut-off radius $R_{\mathrm{C}}$ of the following shape

\begin{equation}
\lambda(R_{\mathrm{C}}) = 100 \left(\frac{R_{\mathrm{C}}}{R_0}\right)^{1/\beta}\quad.
\label{eq:RichnessRadiusRykoff}
\end{equation}
The cut-off radius $R_\mathrm{C}$ characterises the circular area around the cluster centre within which the galaxies contributing to the richness are counted. It is, however, not comparable to common radii describing an over-density like $R_{500\mathrm{c}}$ (the radius within which the density is 500 times higher than the critical density of the Universe at a given redshift). In the above equation, we have $R_0=1.0\,h^{-1}\,\mathrm{Mpc}$ and $\beta=0.2$ according to Equation (4) in \citet{Rykoff2014}.  \citet{Rozo2009} determine the optimal choice of the parameters $\beta$ and $R_0$ empirically by minimising the scatter in the relation between richness and X-ray luminosity of their investigated galaxy cluster sample.
In their work \citet{Rykoff2012} report that the optimal richness measurements are obtained when considering galaxies with luminosities $L\geq 0.2 L^{*}$, which translates to apparent magnitudes as $m_i\leq m_i^*(z)+1.75$. Thus, the richness $\lambda$ can be calculated as
\begin{equation}
\lambda(R_{\mathrm{C}})=\left[N(R_{\mathrm{C}})-N_{\mathrm{BG}}(R_{\mathrm{C}})\right]\cdot c_{\mathrm{cmp}}(z)/0.95\quad,
\label{eq:RichnessRadiusMeasured}
\end{equation}
where $N$ is the number of galaxies in a circle with radius $R_{\mathrm{C}}$ in the co-added frame with a magnitude $m_i < m_i^*(z)+1.75$ and $m_i<m_{i\mathrm{,totlim}}$ that agree within $2\sigma_{c_{k}}(z)$ with the red-sequence model of the estimated redshift. We count the galaxies correcting for the magnitude-dependent fraction of retrieved versus injected galaxies from the estimation of the 80 per cent depth of the data (see Section \ref{Sec:DataReductionImaging} and Appendix \ref{Sec:AppA80percentLimit}). $N_{\mathrm{BG}}$ is the corresponding number of background galaxies. The completeness correction is applied in case $m_{i\mathrm{,totlim}}<m_i^*(z)+1.75$. Additionally, we divide the result by 0.95 in order to account for galaxies with a larger scatter than $2\sigma_{c_{k}}(z)$. Assuming Gaussian scatter, their number is expected to be 5 per cent of the total.

Since it is not clear what the cut-off radius is beforehand, we estimate the richness within a range of different values of $R_{\mathrm{C}}$. We then compare the richness $\lambda(R_{\mathrm{C}})$ we obtain this way to the richness we expect from Equation (\ref{eq:RichnessRadiusRykoff}). Thus, the richness estimate is the unique point where the two values coincide. We estimate the uncertainty of the richness in a purely statistical way. This means that we consider the Poisson errors on the number counts of galaxies in the observation and on the background counts. Any uncertainty introduced by the redshift uncertainty is therefore not included. It has to be noted however that the richness is closely linked to the redshift estimate since it builds the basis for the richness estimate and can also have an impact on a possible completeness correction.

The $\lambda-M_\mathrm{500c,SZ}$ scaling relation is described by
\begin{equation}
\langle \ln \lambda | M_\mathrm{500c,SZ}\rangle = a + \alpha \ln{\left(\frac{M_\mathrm{500c,SZ}}{5.23\times 10^{14}\,\mathrm{M}_\odot}\right)}\quad,
\label{eq:RichnessMassRozo}
\end{equation}
with $a=4.572\pm 0.021$ and $\alpha=0.965\pm 0.067$, and an intrinsic scatter in richness of $\sigma_{\ln \lambda | M_\mathrm{500c,SZ}}=0.266\pm0.017$ \citep{Rozo2015}.
Assuming that there is no redshift-dependent evolution of this relation, we apply this scaling relation to our sample in order to infer a corresponding mass (we call this mass $M_\mathrm{500c,\lambda}$) from the richness we measured. Given that the red-sequence assembles over time, it is possible for the normalisation of the richness-mass scaling relation to change with redshift \citep{VanUitert2016}. However, there are also studies indicating only little evolution of the scaling relation out to redshifts of $z=0.6$ \citep{Andreon2014} or even $z=0.8$ \citep{Saro2015}.

Based on the obtained redshift estimates and central positions, it is also possible to estimate the mass based on the SZ signal in the \textit{Planck} maps, as detailed in \ref{Confirmation of Cluster Candidates}. It refers to the mass within a radius of $R_{500c}$. 
The measured redshifts, richnesses and masses are summarised in Table \ref{tab:ResultsPSZ2}.

\begin{landscape}
\begin{table}
	\begin{threeparttable}
		\caption{Results of the photometric analysis of the cluster sample studied here.}
		\begin{tabular}{ c c c l c c c c c c c | c c c c}

		\hline\hline
		&&\textcolor{revision1}{ID} &Name & RA & Dec & $z_\mathrm{spec}$ & $z_\mathrm{RS}$ & $\lambda$  & $M_{\mathrm{500c,}\lambda}$ & $c_\mathrm{cmp}$ & $M_\mathrm{500c,SZ}$ $^{a}$& $ S/N $  & $D_{\mathrm{blind}}$ $^{b}$ & $Q_\mathrm{neural}$ $^{c}$\\
		&& & & [$^\circ$] & [$^\circ$] &  &  &  & [$10^{14}\mathrm{M}\odot$] & & [$10^{14}\mathrm{M}\odot$] & &  [arcmin] & \\ 
		
		%----------------- M_lambda/MSZ > 0.5-------------------------------------------------------------
		
		\parbox[t]{2mm}{\multirow{19}{*}{\rotatebox[origin=c]{90}{ \textcolor{revision1}{$\frac{M_{\mathrm{500c,}\lambda}}{M_\mathrm{500c,SZ}}$}}}}& \parbox[t]{2mm}{\multirow{19}{*}{\rotatebox[origin=c]{90}{ \textcolor{revision1}{$> 0.5$}}}} & \textcolor{revision1}{115}  & PSZ2 G032.31+66.0 & 219.348 &	24.407 &  0.609$^{g}$	&$0.61\pm 0.01$ & $71\pm 13$ &	$3.81\pm 0.72$ & 1.00 &	$5.64^{+0.77} _{-0.84}$ &	5.14 & 3.18 & 0.98 \\		
		&& \textcolor{revision1}{277}  &PSZ2 G066.34+26.1 & 270.268 &	39.876 & -	&$0.62\pm 0.02$ & $79\pm 14$ &	$4.22\pm 0.78$ & 1.16 &	$4.81^{+0.67} _{-0.74}$ &	5.63 & 2.71 & 0.96 \\
		&& \textcolor{revision1}{378}  &PSZ2 G085.95+25.2 $^\star$ & 277.648 &	56.892 & $0.782\pm0.003^{f}$	&$0.77\pm 0.05$ & $160\pm 21$ &	$8.86\pm 1.22$ & 1.66 &	$4.97^{+0.58} _{-0.63}$ & 5.55 & 1.96 & 0.98 \\		
%		&& \textcolor{revision1}{378-B}  &\textcolor{revision1}{PSZ2 G085.95+25.2} $^\star$ & \textcolor{revision1}{277.599} &	\textcolor{revision1}{56.885} & \textcolor{revision1}{$0.782\pm0.003^{f}$}	&\textcolor{revision1}{$0.74\pm 0.03$} & \textcolor{revision1}{$148\pm18$} &	\textcolor{revision1}{$8.15\pm 1.05$} & \textcolor{revision1}{1.45} &	\textcolor{revision1}{???} & \textcolor{revision1}{5.55} & \textcolor{revision1}{1.08} & \textcolor{revision1}{0.98} \\		
		
		&& \textcolor{revision1}{420}  &PSZ2 G092.64+20.7 & 289.196 &	61.665 & \textcolor{revision1}{0.545$^{h}$}	&$0.58\pm 0.02$ & $36\pm 11$ &	$1.87\pm 0.61$ & 1.00 &	$4.45^{+0.46} _{-0.49}$ &	5.12 & 1.10  & 0.92\\
		
		&& \textcolor{revision1}{483}  &PSZ2 G100.22+33.8 & 258.417 &	69.355 & \textcolor{revision1}{0.598$^{h}$}	&$0.56\pm 0.06$ & $35\pm 9$ &	$1.78\pm 0.49$ & 1.00 &	$4.04^{+0.51} _{-0.55}$ &	5.69 & 1.13 & 0.99 \\
		
		&& \textcolor{revision1}{623}  &PSZ2 G126.28+65.6 & 190.599 &	51.443 & 0.820$^{e}$	&$0.83\pm 0.07$ & $79\pm 16$ &	$4.21\pm 0.87$ & 1.46 &	$5.00^{+0.67} _{-0.71}$ &	4.77 & 2.51 & 0.92 \\		
		&& \textcolor{revision1}{625}  &PSZ2 G126.57+51.6 & 187.444 &	65.354 & 0.815$^{e}$	&$0.80\pm 0.08$ & $81\pm 16$ &	$4.35\pm 0.91$ & 1.60 &	$5.82^{+0.56} _{-0.60}$ &	6.35 & 0.53 & 0.91  \\
		
		&& \textcolor{revision1}{690}  &PSZ2 G141.98+69.3 $^\star$ & 183.109 &	46.395 & 0.714$^{g}$	&$0.71\pm 0.03$ & $40\pm 8$ &	$2.08\pm 0.45$ & 1.00 &	$2.74^{+0.96} _{-1.33}$ &	4.71 & 7.96 & 0.84 \\
		
		&& \textcolor{revision1}{1074}  &PSZ2 G237.68+57.8 $^\dagger$ $^\star$ & 163.336 &	10.877 & $0.894\pm0.007^{d}$	&$0.97\pm 0.05$ & $148\pm 27$ &	$8.15\pm 1.55$ & 2.94 &	$5.38^{+0.76} _{-0.81}$ &	5.36 & 4.48 & 0.94 \\
		
		&& \textcolor{revision1}{1493}  &PSZ2 G316.43+54.0 & 200.820 &	-7.998 & -	&$0.53\pm 0.02$ & $72\pm 13$ &	$3.82\pm 0.74$ & 1.00 &	$5.82^{+0.75} _{-0.82}$ &	5.18 & 0.87 & 0.73\\		
		&& \textcolor{revision1}{1512}  &PSZ2 G321.30+50.6 $^\star$ & 204.611 &	-10.550 & -	&$0.79\pm 0.04$ & $131\pm 19$ &	$7.18\pm 1.07$ & 1.53 &	$5.28^{+0.94} _{-1.08}$ &	4.63 & 1.08 & 0.96  \\	
%		&& \textcolor{revision1}{1512-B}  &\textcolor{revision1}{PSZ2 G321.30+50.6} $^\star$ & \textcolor{revision1}{204.661} &	\textcolor{revision1}{-10.566} & \textcolor{revision1}{-}	&\textcolor{revision1}{$0.68\pm 0.04$} & \textcolor{revision1}{$37\pm10$} &	\textcolor{revision1}{$1.90\pm0.55$} & \textcolor{revision1}{1.02} &	\textcolor{revision1}{???} &	\textcolor{revision1}{4.63} & \textcolor{revision1}{4.06} & \textcolor{revision1}{0.96}  \\	
		&& \textcolor{revision1}{1539}  &PSZ2 G326.73+54.8 & 206.320 &	-5.526 & -	&$0.60\pm 0.02$ & $76\pm 14$ &	$4.05\pm 0.76$ & 1.00 &	$5.98^{+0.81} _{-0.89}$ &	5.92 & 3.57 & 1.00 \\
		
		&& \textcolor{revision1}{1606}  &PSZ2 G343.46+52.6 & 216.094 &	-2.731 & 0.713$^{e}$	&$0.72\pm 0.01$ & $115\pm 15$ &	$6.28\pm 0.87$ & 1.02 &	$6.26^{+0.89} _{-0.98}$ &	4.90 & 0.99 & 0.96 \\
		
		&& \textcolor{revision1}{-}  &PLCK G58.14--72.7 $^\dagger$ $^\star$ & 356.394 &	-18.803 & $0.938\pm0.003^{d}$	&$1.03\pm 0.10$ &  $60\pm 16$ &	$3.15\pm 0.90$ & 2.23 &	$3.90^{+0.88} _{-1.08}$ &	4.20 & 2.64 & 0.99 \\		
		&&\textcolor{revision1}{-}  &PLCK G82.51+29.8 & 268.725 &	54.478 & -	&$0.86\pm 0.10$ & $25\pm 11$ &	$1.25\pm 0.57$ & 1.67 &	$2.83^{+0.96} _{-1.34}$ &	3.51 & 4.49 & 0.71 \\		
		&&\textcolor{revision1}{-}  &PLCK G98.08--46.4 $^\dagger$ $^\star$ & 355.524 &	13.023 & $0.983\pm0.005^{d}$	&$1.06\pm 0.04$ & $53\pm 21$ &	$2.80\pm 1.17$ & 3.47 &	$2.63^{+1.20} _{-1.20}$ &	3.34 & 4.38 & 0.98 \\
		&&\textcolor{revision1}{-}  &PLCK G174.14--27.5 & 59.050 &	16.448 & $0.834\pm0.005^{d}$	&$0.84\pm 0.11$ & $15\pm 8$ &	$0.73\pm 0.44$ & 1.22 &	$2.30^{+1.55} _{-1.55}$ &	3.81 & 4.44 & -  \\
		&&\textcolor{revision1}{-}  &PLCK G184.49+21.1 & 111.081 &	34.045 & $0.596\pm0.007^{d}$	&$0.62\pm 0.01$ & $121\pm 15$ &	$6.64\pm 0.89$ & 1.00 &	$5.88^{+0.80} _{-0.87}$ &	4.26 & 0.21 & 0.99 \\

		\hline %----------------- M_lambda/MSZ > 0.25-------------------------------------------------------------
		
		\parbox[t]{2mm}{\multirow{6}{*}{\rotatebox[origin=c]{90}{ \textcolor{revision1}{$\frac{M_{\mathrm{500c,}\lambda}}{M_\mathrm{500c,SZ}}$}}}}& \parbox[t]{2mm}{\multirow{6}{*}{\rotatebox[origin=c]{90}{ \textcolor{revision1}{$> 0.25$}}}} &\textcolor{revision1}{381}  &PSZ2 G086.28+74.7 & 204.480 &	38.900 & 0.699$^{g}$	&$0.71\pm 0.01$ & $36\pm 10$ &	$1.87\pm 0.55$ & 1.00 &	$5.47^{+0.69} _{-0.75}$ &	5.07 & 1.78 & 0.96\\	
		&&\textcolor{revision1}{421}  &PSZ2 G092.69+59.9 $^\star$ & 216.646 &	51.268 & 0.848$^{e}$	&$0.86\pm 0.07$ & $20\pm 10$ &	$0.99\pm 0.51$ & 1.42 &	$3.38^{+0.78} _{-0.95}$ &	4.90 & 1.16 & 0.96 \\		
		&&\textcolor{revision1}{1121}  &PSZ2 G246.91+24.6 & 141.495 &	-15.162 & -	&$0.60\pm 0.14$ & $35\pm 9$ &	$1.81\pm 0.50$ & 1.00 &	$5.67^{+0.71} _{-0.77}$ &	4.80 & 2.13 & 0.97 \\	
		&&\textcolor{revision1}{-}  &PLCK G55.00--37.0 & 325.138 &	-0.432 & -	&$0.62\pm 0.14$ & $23\pm 9$ &	$1.19\pm 0.50$ & 1.00 &	$5.30^{+0.78} _{-0.85}$ &	3.85 & 0.58 & 0.90 \\		
		&&\textcolor{revision1}{-}  &PLCK G122.62--31.8 & 12.583 &	30.990 & -	&$0.91\pm 0.06$ & $28\pm 13$ &	$1.42\pm 0.69$ & 2.15 &	$4.99^{+0.93} _{-1.07}$ &	3.32 & 1.96 & 0.81 \\		
		&&\textcolor{revision1}{-}  &PLCK G150.77+17.1 & 87.435 &	62.406 & -	&$0.59\pm 0.05$ & $32\pm 10$ &	$1.63\pm 0.54$ & 1.00 &	$4.94^{+0.93} _{-1.05}$ &	3.6 & 2.30  & 0.98 \\
		
		\hline %----------------- M_lambda/MSZ < 0.25-------------------------------------------------------------	
		
		\parbox[t]{2mm}{\multirow{4}{*}{\rotatebox[origin=c]{90}{ \textcolor{revision1}{$\frac{M_{\mathrm{500c,}\lambda}}{M_\mathrm{500c,SZ}}$}}}}& \parbox[t]{2mm}{\multirow{4}{*}{\rotatebox[origin=c]{90}{ \textcolor{revision1}{$< 0.25$}}}}&\textcolor{revision1}{545}  &PSZ2 G112.54+59.5 $^\star$ & 202.464 &	56.797 & -	&$0.83\pm 0.06$ & $4.4\pm 5.4$ &	$0.21\pm 0.27$ & 1.89 &	$5.32^{+0.60} _{-0.64}$ &	5.10 & 1.14 & 0.90 \\		
		&&\textcolor{revision1}{667}  &PSZ2 G136.02--47.1 $^\star$ & 22.008 &	14.756 & -	&$0.61\pm 0.07$ & $5.8\pm 6.6$ &	$0.28\pm 0.34$ & 1.00 &	$5.65^{+0.95} _{-1.07}$ &	4.64  & 1.98 & 0.56 \\		
		&&\textcolor{revision1}{681}  &PSZ2 G139.00+50.9 & 170.075 &	63.248 & -	&$0.71\pm 0.10$ &  $12\pm 7$ &	$0.59\pm 0.38$ & 1.00 &	$4.61^{+0.72} _{-0.81}$ &	4.98 & 1.90 & 0.78  \\
%		&&\textcolor{revision1}{740}  &PSZ2 G152.47+42.1 $\diamond$ & 142.409  &  61.660 & - &	- & - &	- & - &	- &	4.81 & 3.34 & 1.00 \\
%		&&\textcolor{revision1}{769}  &PSZ2 G160.94+44.8 $^\star$ & 143.418 &	54.992 & -	&$0.74\pm 0.10$ & $2.3\pm 5.2$ &	$0.10\pm 0.26$ & 1.16 &	-  &	4.98 & 3.83 & 0.06 \\	
%		&&\textcolor{revision1}{789}  &PSZ2 G165.41+25.9 $^\star$ & 110.990 &	52.160 & -	&$0.67\pm 0.11$ & $-4.8\pm 3.6$ &	-  & 1.00 &	$4.04^{+0.97} _{-1.15}$ &	4.51 & 0.35 & 0.99 \\	
		&&\textcolor{revision1}{1441}  &PSZ2 G305.76+44.7 & 195.003 &	-18.022 & -	&$0.76\pm 0.13$ & $20\pm 11$ &	$0.99\pm 0.59$ & 2.29 &	$7.17^{+0.71} _{-0.76}$ &	5.72 & 1.73 & 0.97 \\	
%		&&\textcolor{revision1}{-}  &PLCK G164.80-47.4 $\diamond$ & 39.918  &  6.450 &  - &	- & - &	- & - &	- &	4.24  & 1.51 & 0.99 \\
		
		\hline %----------------- M_lambda/MSZ < unknown -------------------------------------------------------------
			
		\parbox[t]{2mm}{\multirow{4}{*}{\rotatebox[origin=c]{90}{ \textcolor{revision1}{$\frac{M_{\mathrm{500c,}\lambda}}{M_\mathrm{500c,SZ}}$}}}}& \parbox[t]{2mm}{\multirow{4}{*}{\rotatebox[origin=c]{90}{\textcolor{revision1}{unknown}}}}&\textcolor{revision1}{740}  &PSZ2 G152.47+42.1 $\diamond$ & 142.409  &  61.660 & - &	- & - &	- & - &	- &	4.81 & 3.34 & 1.00 \\
		&&\textcolor{revision1}{769}  &PSZ2 G160.94+44.8 $^\star$ & 143.418 &	54.992 & -	&$0.74\pm 0.10$ & $2.3\pm 5.2$ &	$0.10\pm 0.26$ & 1.16 &	-  &	4.98 & 3.83 & 0.06 \\
		&&\textcolor{revision1}{789}  &PSZ2 G165.41+25.9 $^\star$ & 110.990 &	52.160 & -	&$0.67\pm 0.11$ & $-4.8\pm 3.6$ &	-  & 1.00 &	$4.04^{+0.97} _{-1.15}$ &	4.51 & 0.35 & 0.99 \\	
		&&\textcolor{revision1}{-}  &PLCK G164.80-47.4 $\diamond$ & 39.918  &  6.450 &  - &	- & - &	- & - &	- &	4.24  & 1.51 & 0.99 \\
			\hline
			
		\end{tabular}
		\label{tab:ResultsPSZ2}

		\begin{tablenotes}
			\textbf{\textit{Notes.}} All the values to the left of the solid line result from the analysis of the optical data, values to the right are based on the SZ-signal from the \textit{Planck} maps. We sort the cluster candidates into three categories of validation: conservative confirmation with $M_{\mathrm{500c,}\lambda}/M_{\mathrm{500c,SZ}}>0.5$, loose confirmation with $M_{\mathrm{500c,}\lambda}/M_{\mathrm{500c,SZ}}>0.25$ and unconfirmed with $M_{\mathrm{500c,}\lambda}/M_{\mathrm{500c,SZ}}<0.25$ (each within the $1\sigma$ uncertainties). \textcolor{revision1}{Cluster where results for $M_{\mathrm{500c,}\lambda}$ or $M_\mathrm{500c,SZ}$ could not be measured are categorised as "unknown".}\\
			$^{a}$ The SZ-mass is inferred from the \textit{Planck} maps based on the photometric redshift and position information from the optical data.
			$^{b}$ Distance from optical centre to position of SZ blind detection. 
			$^{c}$ The values of $Q_\mathrm{neural}$ for the MMF3 detected candidates (identifier "PLCK" and $\mathrm{S/N} < 4.5$) were computed by G. Hurier (private communication) following the method described in \citet{Aghanim2015} and \citet{Hurier2017}. 
			Spectroscopic redshifts $z_\mathrm{spec}$ are indicated as published in $d$) this work (BCG candidate) with a detailed description in Section \ref{Spectroscopic Observations}, $e$) \citet{Burenin2018}, $f$) \citet{Amodeo2018}, $g$) \citet{Streblyanska2018}, \textcolor{revision1} {$h$) \citet{Streblyanska2019}}.\\
			$^\dagger$ For these candidates we found very high completeness correction factors of $c_\mathrm{cmp} > 3.5$ when using the photometric redshift. We therefore resort to the spectroscopic redshifts for the indicated richness $\lambda$ and richness-based mass $M_{\mathrm{500c,}\lambda}$. In Table \ref{tab:AlternativeRichMassResults}, we list similar results based on the spectroscopic redshifts when available for the other cluster candidates.\\
			$\diamond$ For these candidates the analysis did not lead to a conclusive redshift result because the red-sequence models do not have enough constraining power. The reason is that we do not have a sufficiently deep detection limit in these cases. We only report the position of the potential brightest cluster galaxy and available quantities related to the SZ-based measurements.\\
			$^\star$ More remarks on these candidates in Section \ref{Notes on Individual Cluster Candidates}.\\
		\end{tablenotes}

	\end{threeparttable}
\end{table}
\end{landscape}

\section{Spectroscopic Observations}
\label{Spectroscopic Observations}

For nine out of 32 cluster candidates in our sample, we performed spectroscopic observations with the long-slit of the ACAM instrument at the WHT during the visitor mode observation runs in October 2016 and March 2017. We selected these targets based on a preliminary reduction of the $r$-, $i$- and $z$-band images obtained in the same night. In case of a galaxy over-density at a likely high redshift and with a visible BCG candidate, we conducted three spectroscopic exposures of 1200~s each per cluster. We positioned the slit such that it covered the BCG candidate and at least one other candidate cluster galaxy. We used a V400 grating and the GG495A order-sorting filter in combination with a slit width of $1\farcs0$. Thus, we obtained observations at 3.3 {\AA}/pixel with a resolution of $R\approx 450$ at $\lambda=6000$~{\AA} covering a wavelength range between 4950 and 9500 {\AA}. This agrees well with the range of expected emission and absorption features in the targeted redshift regime.

\subsection{Data reduction}

The data reduction of the spectroscopic observations includes a bias subtraction and flat-fielding. We remove cosmic rays with the help of the algorithm LA-cosmic \citep{VanDokkum2001} for all of the spectroscopic frames. We then co-add three frames of the same target, respectively, and employ \textsc{IRAF} for the further reduction of the co-added spectroscopic frames. We extract spectra using the task \texttt{apall}, which follows the "Optimal Extraction Algorithm for CCD Spectroscopy" by \citet{Horne1986}. This includes a background subtraction and trace fitting of the position of the spectrum in the spatial direction of the frame. For the background subtraction a polynomial is fitted to the sky background along the spatial direction at each wavelength step and subsequently subtracted to remove the contribution of the night sky to the spectrum. The tasks \texttt{identify} and \texttt{dispcor} allow us to perform a wavelength calibration with the help of well-known sky emission lines\footnote{\url{ http://www.astrossp.unam.mx/~resast/standards/NightSky/skylines.html}}. 
Finally, we perform a flux calibration using the tasks \texttt{sensfunc} and \texttt{calibrate}. Our primary goal here is not to get an absolute flux calibration but to find the correct relative fluxes over the range of the spectrum. We use standard star observations of the star BD+332642 (observed by us on 2017 March 21, slit width: $0\farcs75$, exposure time: 150 seconds). They enable us to estimate the sensitivity function (\texttt{sensfunc}) which connects the measured spectra (e.g. of the standard stars) to reference spectra of the same object. When applied to the wavelength-calibrated spectra (\texttt{calibrate}), the sensitivity function allows us to obtain the flux-calibrated spectra.

\subsection{Spectroscopic redshifts}

\begin{figure}
	\centering
	\includegraphics[width=1.0\columnwidth]{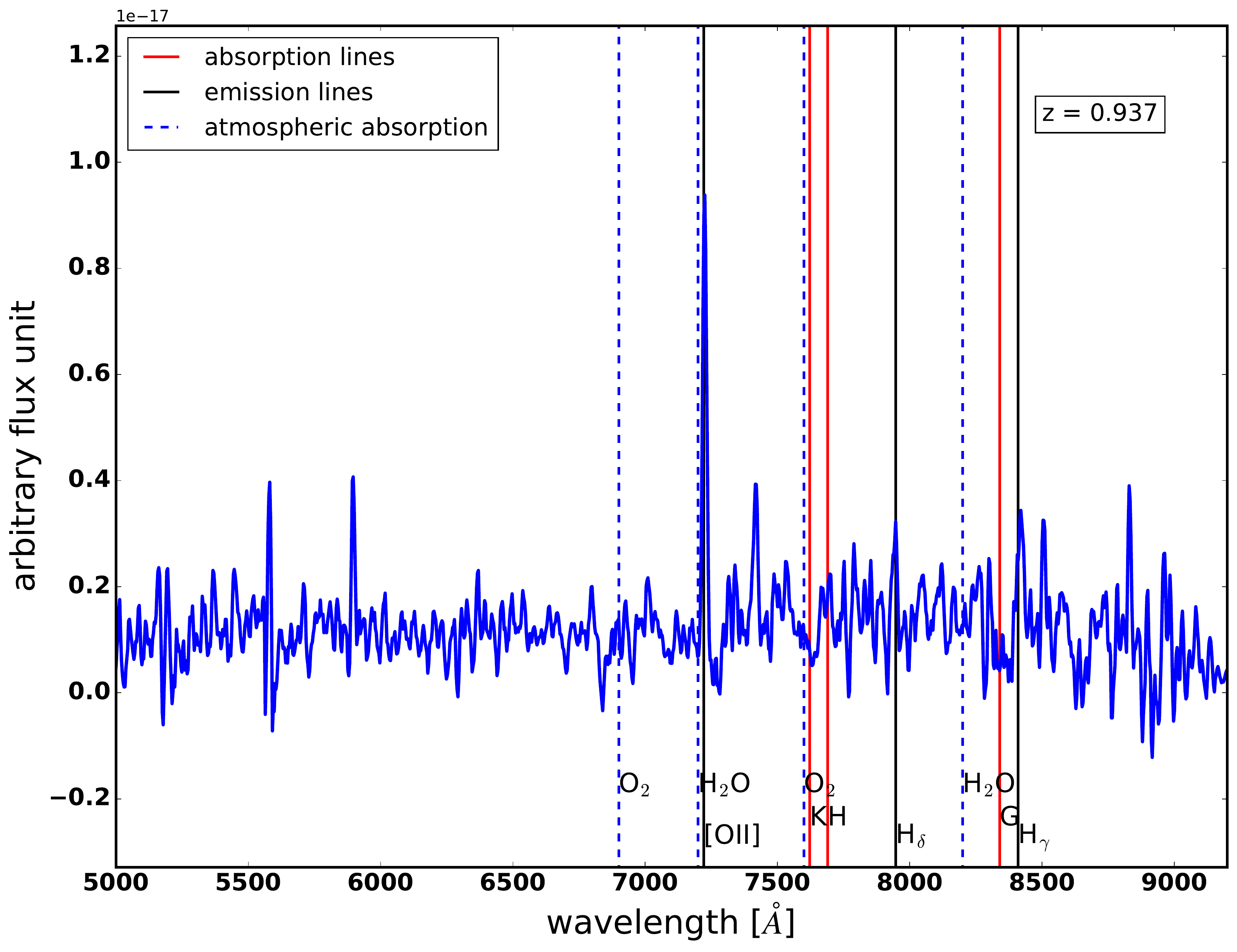}
	\caption{Spectrum of the BCG candidate in cluster PLCK G58.14--72.7. The vertical, dashed blue lines indicate where atmospheric absorption lines caused by water and oxygen molecules are to be expected. The vertical, solid black lines represent the position of emission lines. The vertical, solid red lines represent the position of absorption lines that may occur especially in elliptical galaxies. The original wavelength at emission/absorption is shifted to the redshift indicated in the upper right corner of the plot. The [OII] line is the only feature we clearly identified for this spectrum.} 
	\label{fig:MMF3-18209Spectrum}
\end{figure}

We examine the fully extracted and calibrated spectra of the candidate BCGs and other cluster member galaxies for prominent emission and absorption line features in order to obtain a rough spectroscopic redshift estimate. The Calcium H and K absorption lines at about 3935~{\AA} and 3970~{\AA} and the 4000~{\AA}-break are the main features that we can identify in the majority of the galaxies\footnote{We use the lines listed at the SDSS homepage as a reference \url{http://classic.sdss.org/dr6/algorithms/linestable.html}.}. Additionally, we have one galaxy (in PLCK G58.14--72.7) with a prominent emission line (see Fig. \ref{fig:MMF3-18209Spectrum}).
We perform a cross-correlation with the absorption line template \texttt{spDR2-023}\footnote{\url{http://classic.sdss.org/dr5/algorithms/spectemplates/}} and the emission line template \texttt{femtemp97}\footnote{\url{http://tdc-www.harvard.edu/iraf/rvsao/Templates/}} following these steps:
\begin{enumerate}
	\item We fit a polynomial of order 7 to the spectrum and the template, to subtract the continuum contribution. 
	\item We mask wavelength regimes that include known intense sky lines and atmospheric lines to minimise the impact of residuals from the sky background subtraction and to prevent the cross-correlation with atmospheric features. The most common molecules to cause atmospheric absorption are $\mbox{H}_{2}\mbox{O}$ (at 7200 {\AA}) and $\mbox{O}_{2}$ (at 6900~{\AA}  and 7600~{\AA}).
	\item We perform the cross-correlation with the function \texttt{crosscorrRV()} from the \texttt{PyAstronomy.pyasl} package in python. While the spectroscopic data are relatively shallow, we were able to fine-tune the redshift using the photo-$z$ information as a prior. The cross-correlation gives us a spectroscopic redshift estimate.
	\item We estimate the uncertainty of the spectroscopic redshift result by fitting a Gaussian to the correlation peak and by taking the half width at half maximum as the uncertainty.
\end{enumerate} 

The spectroscopic redshift estimates from the cross-correlation analysis are in close agreement with our estimates by eye. In particular, we can identify the emission line from the BCG candidate of PLCK G58.14--72.7 (see Fig. \ref{fig:MMF3-18209Spectrum}) as the [OII] emission line with a rest frame wavelength of 3727~{\AA} based on our photometric redshift result and an estimate by eye upon inspection of the spectrum. Additionally, when two galaxies falling on the same (long)slit have both a reliable redshift measurement, we also find a good consistency between those two measurements. We report spectroscopic redshifts for five cluster candidates in Table \ref{tab:ResultsSpectroscopy}. The spectra from galaxies in the remaining cluster candidates were unfortunately too noisy to extract a reasonable redshift by eye or with the cross-correlation technique. This was, however, fully expected given the relatively short spectroscopic integration times used in this study.

\begin{table*}
	\begin{threeparttable}
		\caption{Redshift results for the spectroscopic sub-sample.}
		\begin{tabular}{ c l c c c c l }
			\hline\hline 
			\textcolor{revision1}{ID} & Name & RA & Dec & $z_{\mathrm{spec}}$ & $f_{\mathrm{peak}}^\star$ & features \\ 
			\hline
			\textcolor{revision1}{1074} &PSZ2 G237.68+57.83  & $163.3244$ & $10.8770$ & $z_1 = 0.894\pm0.007$ & 1.12 & Ca H+K lines, 4000~{\AA}-break\\ 
			&& $163.3107$ & $10.8834$ & $z_2 = 0.878\pm0.006$ & 1.84 & Ca H+K lines, 4000~{\AA}-break\\ 
			\textcolor{revision1}{-} &PLCK G58.14--72.7  & $356.3987$ & $-18.8019$ & $z_1 = 0.938\pm0.003$ & 2.98 & [OII] emission line \\ 
			&& $356.3982$ & $-18.8020$ & $z_2 = 0.927\pm0.004$ & 2.37 & Ca H+K lines, 4000~{\AA}-break\\
			\textcolor{revision1}{-} &PLCK G98.08--46.4 & $355.5238$ & $13.0233$ & $z_1 = 0.983\pm0.005$ & 1.74 & Ca H+K lines, 4000~{\AA}-break\\  
			&& $355.5201$ & $13.0139$ & $z_2 = 0.988\pm0.008$ & 1.38 & Ca H+K lines, 4000~{\AA}-break\\ 
			\textcolor{revision1}{-} &PLCK G174.14--27.5  & $59.0335$ & $16.4454$ & $z_1 = 0.834\pm0.005$ & 1.19 & Ca H+K lines, 4000~{\AA}-break\\
			&& $59.0371$ & $16.4443$ & $z_2 = 0.829\pm0.003$ & 2.31 & Ca H+K lines, 4000~{\AA}-break \\
			\textcolor{revision1}{-} &PLCK G184.49+21.1  & $111.0799$ & $34.0573$ & $z_1 = 0.596\pm0.007$ & 1.81 & Ca H+K lines, 4000~{\AA}-break, Mg-line\\
			&& $111.0749$ & $34.0465$ & $z_2 = 0.594\pm0.007$ & 1.85 & Ca H+K lines, 4000~{\AA}-break\\ 
			
			\hline 
		\end{tabular}
		
		\begin{tablenotes}
			\textbf{\textit{Notes.}} Results of the spectroscopic analysis. $z_1$ and $z_2$ correspond to the redshifts of the first (= BCG candidate) and second galaxy spectrum on the slit. \\ In case of the clusters PSZ2 G112.54+59.53, PLCK G82.51+29.8, PLCK G122.62--31.9 and PLCK G164.82--47.4 the extracted spectra were not pronounced enough to obtain a reliable redshift estimate.\\
			$^\star$ $f_{\mathrm{peak}}$ is the ratio between the cross-correlation peak and the next weaker peak in the cross-correlation function. It characterises how reliable the redshift result is.
		\end{tablenotes}
		
		\label{tab:ResultsSpectroscopy}
	\end{threeparttable}
\end{table*}

\section{Confirmation of Cluster Candidates}
\label{Confirmation of Cluster Candidates}

To verify or invalidate the cluster candidates in our sample in a quantitative manner we resort to the richness-mass relation by \citet{Rozo2015} which allows us to asses if our measured richness suggests a halo that can account for the SZ-signal in the \textit{Planck} maps.

We derive the \textit{Planck} SZ mass proxy $M_\mathrm{500c,SZ}$ following mainly Section 7.2 in  \citet{PlanckCollab2014XXIX}. To do so, we filter the \textit{Planck} maps at the position of the optical cluster centre from our analysis with the Matched Multi-Filter 3 \citep[MMF3, ][]{Melin2006} varying the angular cluster size $\theta_s$ between 0.8 and 32~arcmin. Here, the six frequency channels are linearly combined, and the filtering takes into account the cluster pressure profile and thermal SZ spectrum as prior knowledge \citep{PlanckCollab2014XXIX}. For each cluster size we estimate the SZ flux, $Y_{500}$, within the radius $R_{500}$ from the filtered maps. We can break the size-flux degeneracy with an X-ray scaling relation \cite[see also fig. 16 in][]{PlanckCollab2016}. Since it is redshift dependent, we use our photometric redshift results as prior information to then obtain the SZ mass proxy $M_\mathrm{500c,SZ}$.

\begin{figure}
	\includegraphics[width=1.0\columnwidth]{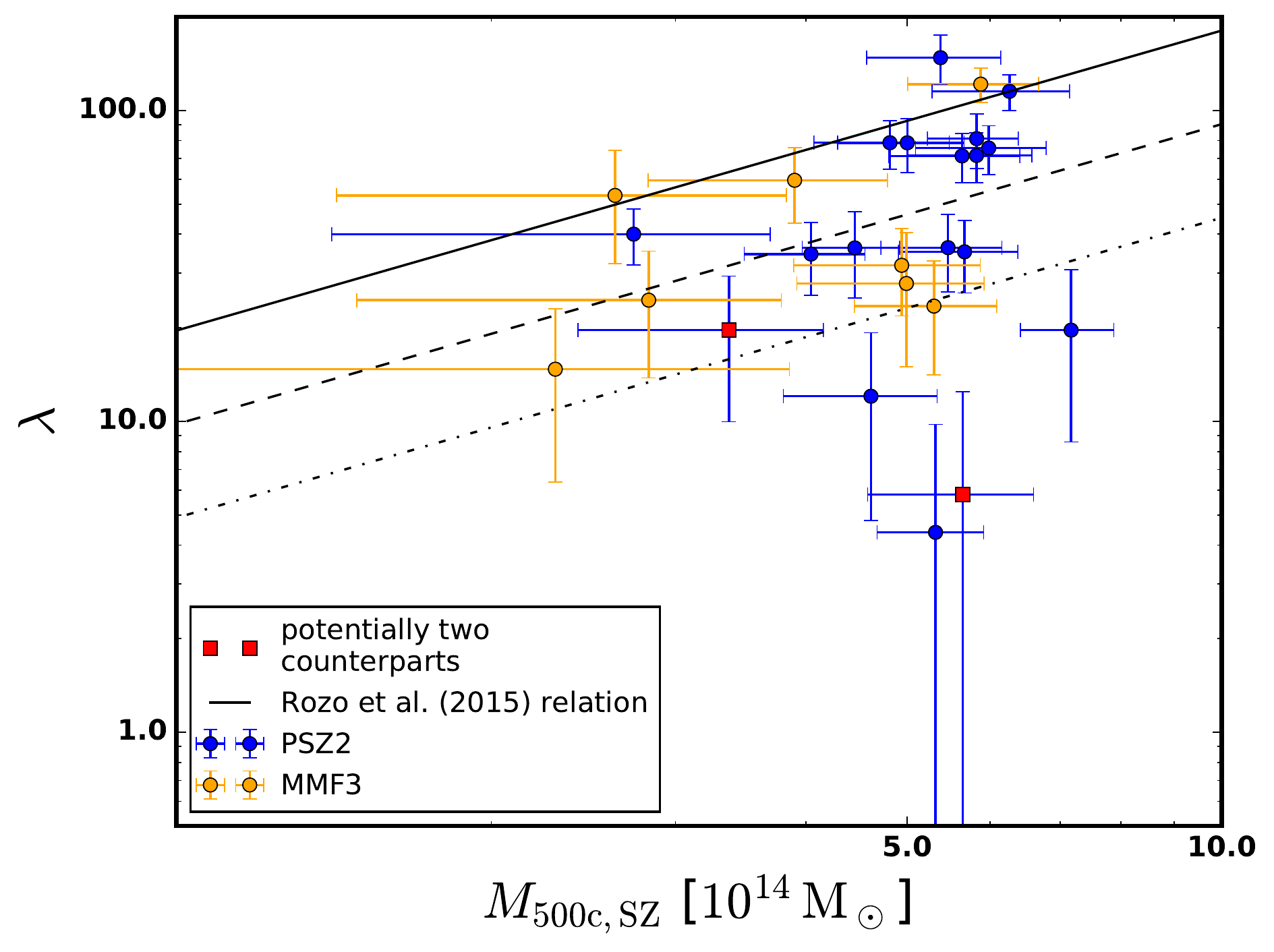}
	\caption{\textcolor{revision1}{Comparison of the richness $\lambda$ obtained from our optical data and the SZ-based mass $M_\mathrm{500c,SZ}$ from the \textit{Planck} measurements. The solid line marks the richness-mass relation from \citet{Rozo2015}. The dashed (dash-dotted) line marks, where the richness is 50 per cent (25 per cent) of what is expected from the scaling relation. Red squares mark PSZ2 cluster candidates with two potential optical counterparts. In these cases, the SZ-based mass $M_\mathrm{500c,SZ}$ should be seen as an upper limit.}}
	\label{fig:ResultsRichnessMassRelation}
\end{figure}

We compare our estimates of $\lambda$ and $M_\mathrm{500c,SZ}$ in Fig. \ref{fig:ResultsRichnessMassRelation}, finding that the majority of clusters are close to the scaling relation by \citet{Rozo2015} or at least approximately half as rich as expected from the relation. 
This suggests that the richness-mass relation, which was established for systems measured at higher S/N and lower redshifts, may indeed be applicable even under the assumption that there is no redshift evolution and given slight differences in the richness definition. We also find a fraction of candidates that are clearly below the scaling relation.

Indeed we even expect a fraction of cluster candidates to lie notably below the richness-mass relation because we investigate a sample in the low S/N regime. Even with our preselection it is possible that spurious detections or projection effects occur. Even more importantly, the Eddington bias starts to play a significant role at low S/N. As shown in \citet{VanDerBurg2016}, this effect can even lead to an overestimation of the measured SZ-mass by a factor of 2 with respect to the real SZ-mass for clusters at $\mathrm{S/N} \lesssim 4.0 $ and $z \gtrsim 0.6$.

In order to distinguish between confirmed and invalidated cluster candidates, it is necessary to find criteria to base this classification on. 
We expect that  the measured SZ signal is predominantly caused by inverse Compton scattering of CMB photons instead of noise in the \textit{Planck} maps. \citet{VanDerBurg2016} assume that this is the case when the richness-based mass amounts to 50 per cent or more of the SZ-based mass within the error bars.

\begin{figure}
	\includegraphics[width=1.0\columnwidth]{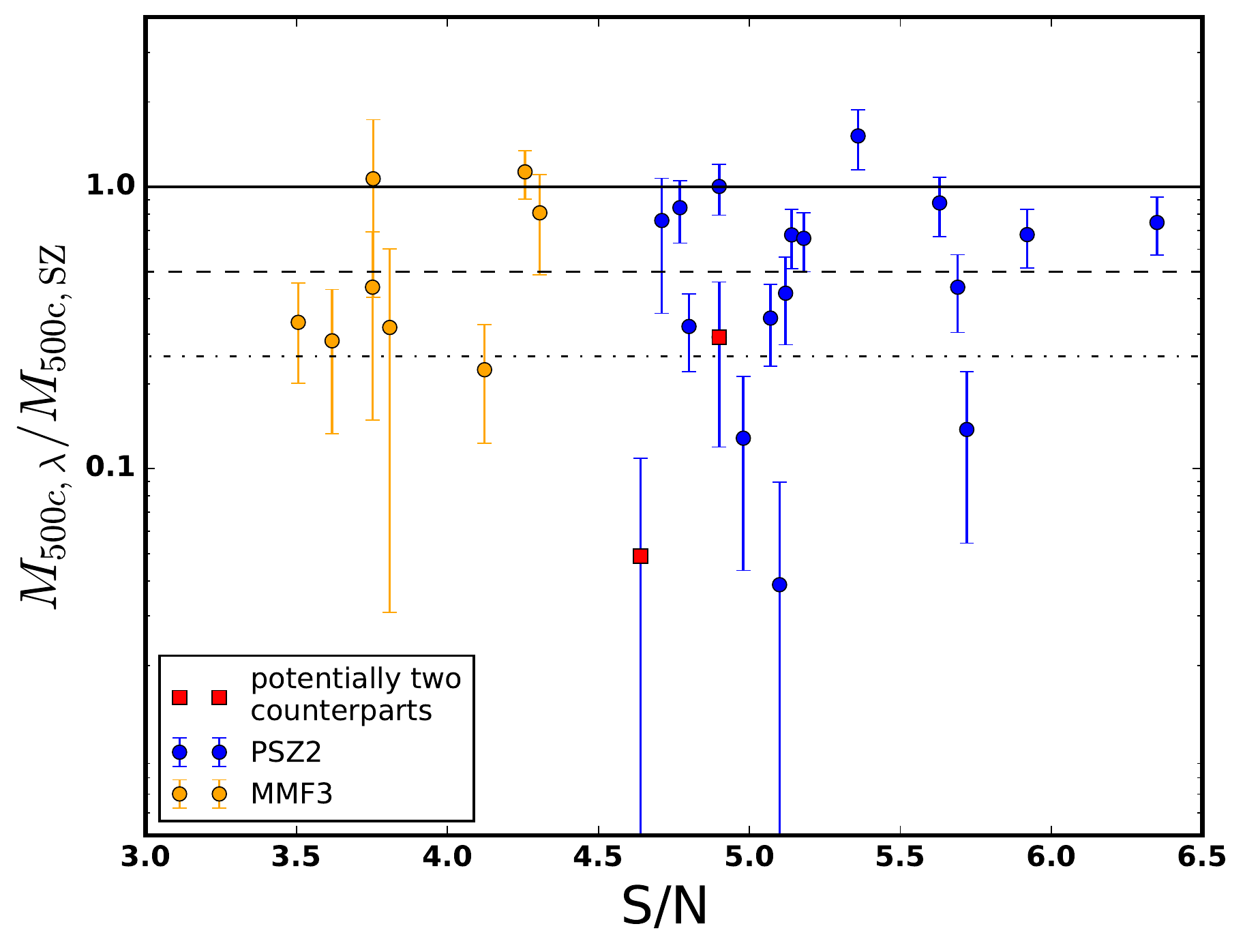}
	\includegraphics[width=1.0\columnwidth]{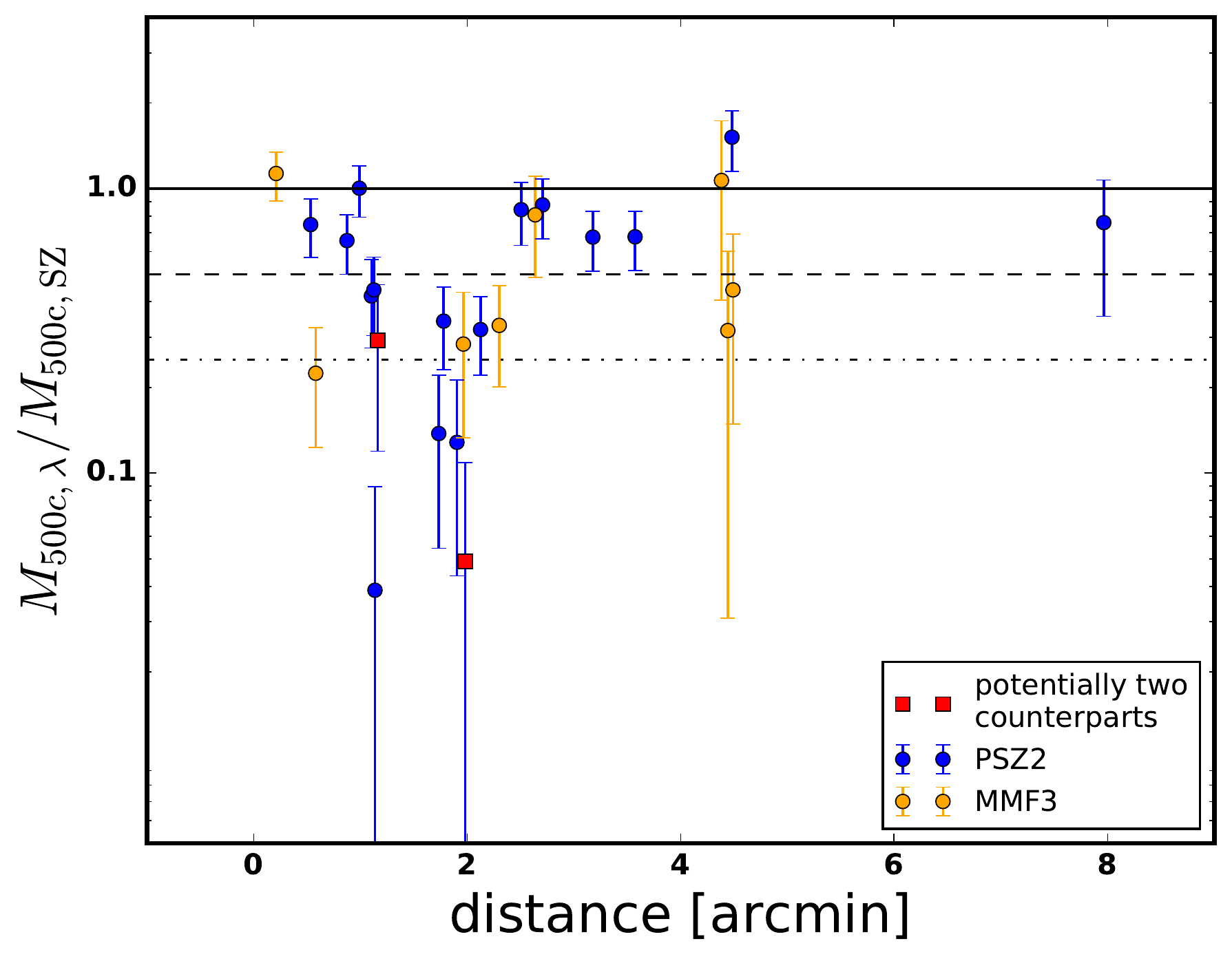}
	\includegraphics[width=1.0\columnwidth]{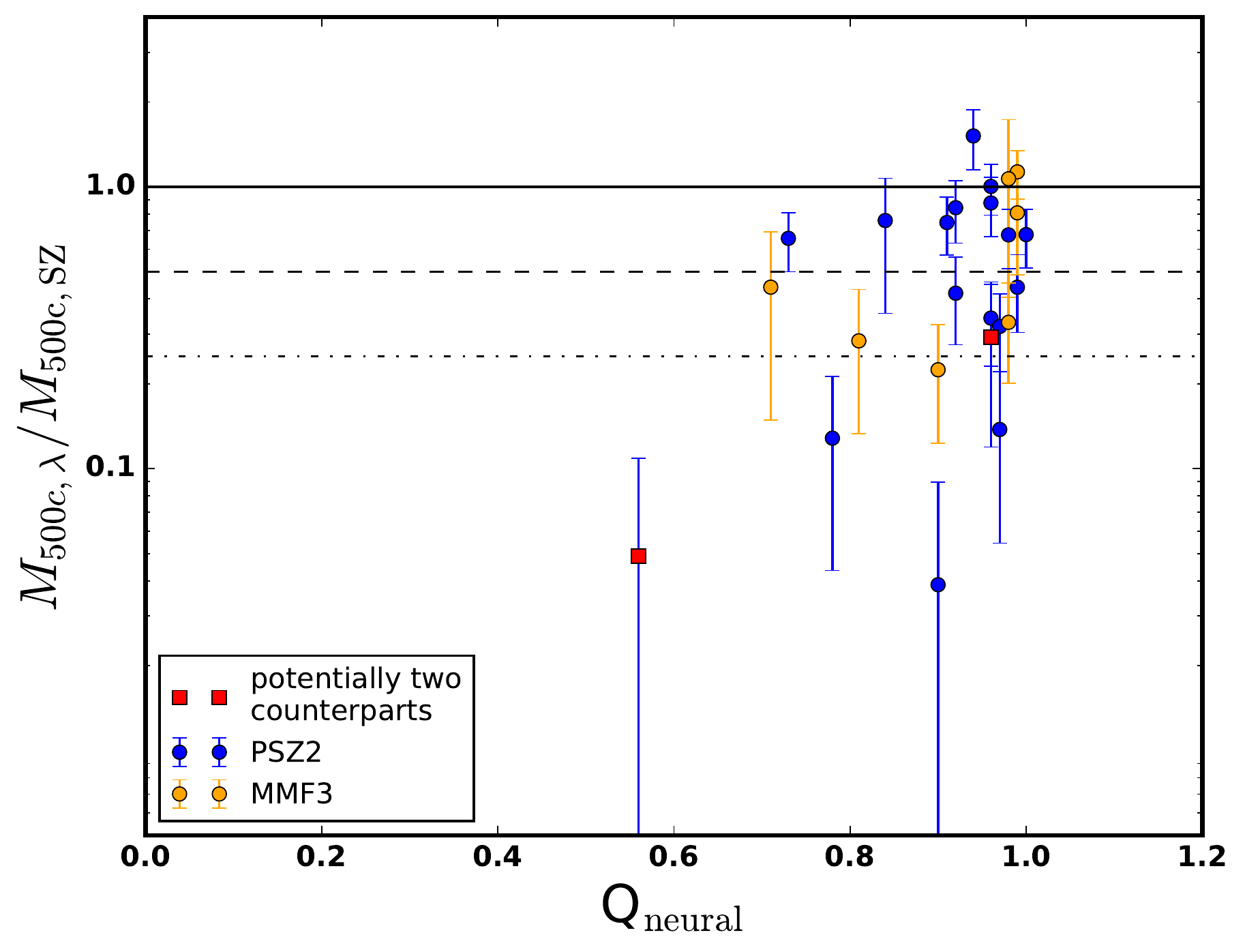}
	
	\caption{\textcolor{revision1}{Comparison of the mass ratio $M_{\mathrm{500c,}\lambda}/M_\mathrm{500c,SZ}$ versus S/N (\textit{top}) of the \textit{Planck} SZ detection, versus the distance between the optical centre and the SZ peak coordinates of the blind detection (\textit{middle}) and versus $Q_\mathrm{neural}$ (\textit{bottom}). We only plot the candidates that have the corresponding information available in Table \ref{tab:ResultsPSZ2}. The solid, dashed and dash-dotted line mark, where $M_{\mathrm{500c,}\lambda}/M_\mathrm{500c,SZ}=1.0$, $M_{\mathrm{500c,}\lambda}/M_\mathrm{500c,SZ}=0.5$, and $M_{\mathrm{500c,}\lambda}/M_\mathrm{500c,SZ}=0.25$, respectively. Red squares mark PSZ2 cluster candidates with two potential optical counterparts. In these cases, the mass ratio $M_{\mathrm{500c,}\lambda}/M_\mathrm{500c,SZ}$ should be seen as a lower limit.}
		}
	\label{fig:SN+distance+Q_NEURAL}
\end{figure}

From the sample in this work this criterion is met for 18 candidates (13 are in the PSZ2 catalogue, 5 were detected with the MMF3 method at lower S/N). However, there is a continuous transition between a SZ signal which is noise dominated and one which is dominated by the presence of a cluster. 
Especially in the light of the multiple effects that can lead to a discrepancy between richness-based mass and SZ-based mass (assumption of no redshift evolution of the scaling relation, potentially fewer galaxies on the red-sequence at higher redshifts, 25 per cent scatter in the richness-mass relation, Eddington bias especially at low S/N, projection effects with multiple clusters contributing to the SZ signal) the validation criterion of $M_{\mathrm{500c,}\lambda}/M_\mathrm{500c,SZ}\geq 0.5$ might be too strict. Inspecting the distribution of candidates in the richness-mass-plane once again (see Fig. \ref{fig:ResultsRichnessMassRelation}), we see that there is a bulk of candidates above a limit of $M_{\mathrm{500c,}\lambda}/M_\mathrm{500c,SZ}\geq 0.25$. Also inspecting the colour images, they are likely valid counterparts to the SZ-signal. For 24 (16 are in the PSZ2 catalogue, 8 were detected with the MMF3 method) out of 32 candidates the richness-based mass makes up at least 25 per cent of the SZ-based mass within the $1\sigma$ uncertainties and we conclude that they are likely optical counterparts to the SZ-signal. We distinguish a conservative cluster confirmation criterion with $M_{\mathrm{500c,}\lambda}/M_\mathrm{500c,SZ}\geq 0.5$ and a loose cluster confirmation criterion with $M_{\mathrm{500c,}\lambda}/M_\mathrm{500c,SZ}\geq 0.25$.
\textcolor{revision1}{It has to be noted that the mass ratio $M_{\mathrm{500c,}\lambda}/M_\mathrm{500c,SZ}$ as estimated here is not entirely applicable for clusters with multiple counterparts. The richness-based mass is inferred with the help of the richness-mass scaling relation by \citet{Rozo2015}. Therefore, it only relates to one particular optical counterpart, assuming that it fully accounts for the SZ-signal. However, our estimates of the SZ-based mass are based on the SZ-signal, possibly with contributions from several counterparts. In these cases our reported mass ratio is biased low. Our sample includes two candidates with two potential counterparts (PSZ2 G092.69+59.92 and PSZ2 G136.02-47.15; see Section \ref{Notes on Individual Cluster Candidates} for details). The mass ratios of these cluster candidates can be seen as a lower limit.}
Inspecting the additional available information in Table \ref{tab:ResultsPSZ2} beyond the  richness-mass relation, we notice the following:

Fig. \ref{fig:SN+distance+Q_NEURAL} displays that the fraction of loosely confirmed clusters is higher among the low S/N noise sources in our sample (MMF3 with $\mathrm{S/N} < 4.5$) than among the high S/N sources (PSZ2 with $\mathrm{S/N} > 4.5$). This is not necessarily to be expected since spurious detections are more likely to occur at lower S/N.
However, we had a larger number of candidates to choose from considering the MMF3 detection method down to $\mathrm{S/N} > 3$. This means we could identify particularly those candidates which appeared very rich in the WISE, PS1 and/or SDSS data. It is possible that this preselection counterbalances the effect of Eddington bias. An additional explanation for the high fraction of confirmed MMF3 cluster candidates could be the fact that these low S/N candidates have a higher positional uncertainty, lower masses and there are a lot more candidates available than in case of the PSZ2 candidates. Consequently, the probability for super-positions by chance is increased. We check if richness $\lambda$ and SZ-mass $M_\mathrm{500c,SZ}$ are related with the help of Spearman's rank correlation coefficient $r_\mathrm{s}$. We find $r_\mathrm{s,MMF3} = 0.36$ for the MMF3 targets and $r_\mathrm{s,PSZ2} = 0.23$ considering all PSZ2 targets available richness measurements or $r_\mathrm{s,PSZ2,0.25} = 0.35$ considering only the PSZ2 targets with $M_{\mathrm{500c,}\lambda}/M_\mathrm{500c,SZ}\geq 0.25$. We conclude that there is a tendency towards a positive correlation between richness and SZ-mass. There is a comparable correlation between loosely confirmed PSZ2 and MMF3 targets.

\textcolor{revision1}{With regard to the PSZ2 cluster candidates, we find that 16 out of 20 PSZ2 candidates (with available estimates for $M_{\mathrm{500c,}\lambda}$ and $M_\mathrm{500c,SZ}$) are above the limit $M_{\mathrm{500c,}\lambda}/M_\mathrm{500c,SZ}\geq 0.25$. This corresponds to 80 per cent}.
Considering the numerous different follow-up studies for the PSZ2 catalogue with confirmed as well as invalidated cluster candidates it is hard to quantify the expected fraction of confirmed clusters in our sample. We can, however, get a rough estimate based on the work by \citet{PlanckCollab2016}. They report 1653 detections in PSZ2 with and expected reliability of about 90 per cent. Thus, 165 candidates are expected to be invalidated. \citet{PlanckCollab2016} report 1203 confirmed cluster candidates. Consequently, there should be 165 candidates to be invalidated among the remaining $1653 - 1203 = 450$ cluster candidates. This corresponds to an expected fraction of confirmed clusters of about 63 per cent if one were to randomly select candidates from the remaining 450 of unconfirmed candidates. \textcolor{revision1}{This is slightly lower than the fraction of confirmed clusters that we find for the PSZ2 targets presented in this work. Our fraction is however comparable to the fraction of confirmed PSZ2 cluster candidates reported in \cite{VanDerBurg2016}}.

%This is comparable to the fraction of confirmed PSZ2 cluster candidates reported in \cite{VanDerBurg2016} and also comparable to the fraction of confirmed clusters that we find for the PSZ2 targets presented in this work.  

Fig. \ref{fig:SN+distance+Q_NEURAL} illustrates that we find no confirmed clusters with values of $Q_\mathrm{neural} < 0.7$. Additionally, the fraction of validated clusters and the ratio $M_{\mathrm{500c,}\lambda}/M_\mathrm{500c,SZ}$ both generally increase with increasing $Q_\mathrm{neural}$. There are, however, also cluster candidates with a high value of $Q_\mathrm{neural} > 0.7$ that do not fulfil our validation criteria. It is reasonable that this quantity becomes less reliable in the low S/N regime ($\mathrm{S/N} < 5$), since $Q_\mathrm{neural}$ is not suited to flag noise-induced detections \citep{PlanckCollab2016}.

Fig. \ref{fig:SN+distance+Q_NEURAL} shows that all except one of the loosely confirmed clusters in our sample are within a distance of $\lesssim 5$~arcmin from the \textit{Planck} SZ position. This corresponds to a cluster confirmation criterion used in several previous works \citep{Barrena2018,Streblyanska2018,Boada2018}. On average the distance is 2.4~arcmin, which is well within the size of the \textit{Planck} beam. We also find that 68 per cent of the confirmed clusters with $M_{\mathrm{500c,}\lambda}/M_\mathrm{500c,SZ}\geq 0.25$ are found within a distance of 2.6~arcmin from the respective detection in the \textit{Planck} maps. This is less than 2 times larger than the positional uncertainty of $\approx 1.5$~arcmin of the PSZ2 union catalogue \citep{PlanckCollab2016}. Additionally, this is in good agreement with values found for confirmed cluster candidates from PSZ2 by \citet{Streblyanska2018} with 68 per cent of confirmed clusters within 3.1~arcmin and from the first \textit{Planck} data release PSZ1 by \citet{Barrena2018} with 68 per cent of confirmed clusters within 2.8~arcmin.
From Fig. \ref{fig:SN+distance+Q_NEURAL} we see that the optical positions of the PSZ2 targets tend to be closer to the position of the SZ detection than the optical positions of the MMF3 targets. This agrees with the fact that the positional uncertainty of the PSZ2 targets is smaller because they have a higher S/N.

\section{Notes on Individual Cluster Candidates}
\label{Notes on Individual Cluster Candidates}

Table \ref{tab:ResultsPSZ2} summarises the estimated properties of the investigated sample and we present colour images of the confirmed cluster candidates in \textcolor{revision1}{appendix \ref{ColourImages}}. In this section we discuss some cluster candidates that are worth mentioning for example due to high redshifts, multiple counter parts or for special treatment in our analysis.

\subsubsection*{PSZ2 G085.95+25.23}
% (PSZ2-378)

\textcolor{revision1}{This cluster candidate likely has a complex structure. We find an optical counterpart around the position $\mathrm{RA} = 277.648^\circ$, $\mathrm{Dec} = 56.892^\circ$ at a photometric redshift of $z_\mathrm{phot} = 0.77\pm0.05$. We can also measure the redshift from a second position at $\mathrm{RA} = 277.599^\circ$, $\mathrm{Dec} = 56.885^\circ$ with an estimated photometric redshift of $z_\mathrm{phot}=0.74\pm0.03$ with a richness of $\lambda = 148\pm18$\footnote{Note that this richness is related to the richness measured at the first position because the measurements are close both in position and in redshift.}. Both redshifts agree within their uncertainties. We report the results for the first position (with higher richness) in Table \ref{tab:ResultsPSZ2}. The redshift of this optical counterpart is in good agreement with the spectroscopic redshift of $z_\mathrm{spec} = 0.782\pm0.003$ found by \citet{Amodeo2018} (see Table \ref{tab:ExternalResults}). Our richer (less rich) component has a distance of 1.98~arcmin (1.06~arcmin) to the optical cluster centre reported by \citet{Amodeo2018}. }

\subsubsection*{PSZ2 G092.69+59.92}
% (PSZ2-421)
According to \citet{Burenin2018}, this cluster candidate consists of two optical counterparts in projection. They estimate a spectroscopic redshift of $z_\mathrm{spec} = 0.848$ for a counterpart at a position which coincides with our own observations. They mention that a redshift of $z=0.463$ was reported in the \textit{redMaPPer} cluster survey \citep{Rykoff2014}. This agrees with the spectroscopic redshift of $z_\mathrm{spec}=0.461$ which \citet{Streblyanska2018} mention in their work. However, the geometric centre of their optical cluster does not lie within the field of view of our observation.  Since our photometric redshift estimate and position are in better agreement with the results by \cite{Burenin2018} we only make use of their spectroscopic redshift estimate for our red-sequence calibration. \textcolor{revision1}{We note that the richness $\lambda$ of the cluster candidate measured at a photometric redshift of $z_\mathrm{phot} = 0.86\pm0.07$ is lower than expected given the measured SZ signal. This is not indicative of a halo that may solely be responsible for the measured SZ signal, but it is a hint that an additional counterpart (at $z=0.46$) may be contributing to the SZ signal.}

\subsubsection*{PSZ2 G112.54+59.53}
% (PSZ2-545)
In the colour image there are only a few cluster members visible and we find that some of them are fainter than the 80 per cent depth of our data. Accordingly, there is only a very weak peak in the histogram of filtered richness versus redshift resulting in a redshift estimate of $z_\mathrm{phot} = 0.83\pm0.06$.

Since the density of galaxies at the same redshift is so low, the step of finding the galaxy which maximises the richness is not reasonable. We therefore only report the redshift, richness and mass results at the position of the BCG identified by eye. The richness is not high enough to fulfil the confirmation criterion. However, deeper data might reveal more of the potential cluster galaxies and provide a more robust estimate of richness and mass. 

\subsubsection*{PSZ2 G136.02--47.15}
% (PSZ2-667)
\textcolor{revision1}{\citet{Streblyanska2018} list a spectroscopic redshift of $z_\mathrm{spec}=0.465$ from SDSS DR12 for this cluster candidate. They categorise this counterpart as "potentially associated", i.e. this candidate does not meet their richness/distance requirements for cluster confirmation. The possible counterpart we discuss in this paper is closer in sky position to the SZ detection, but has a richness-based mass that is also lower than expected for the SZ detection. Even though both sources (at $z=0.47$ and $z_\mathrm{phot}=0.61\pm0.07$ measured in this work) may have contributed to the measured SZ signal, it may be a largely noise-induced detection.
We find that the given geometric centre from \citet{Streblyanska2018} does not lie within the field of view of our observation of this target. We therefore conclude that we do not investigate the same optical counterpart as \citet{Streblyanska2018}. For this reason we do not use the spectroscopic redshift estimate for the recalibration of the red-sequence models.}

\subsubsection*{PSZ2 G141.98+69.31}
% (PSZ2-690)
\citet{Streblyanska2018} identify an optical counterpart as potentially associated with PSZ2 G141.98+69.31 and report a spectroscopic redshift of $z_\mathrm{spec} = 0.714$. Since the given geometrical centre is about 3.5~arcmin from the centre we find in this work, we do not include this spectroscopic redshift for the recalibration of the red-sequence models. Our photometric redshift estimate of $z_\mathrm{phot} = 0.71\pm0.03$ is however close to the spectroscopic redshift from \citet{Streblyanska2018}. We additionally find a quite large offset between the optical centre and the SZ detection of 7.96~arcmin. The similar redshift results and offsets between optical and SZ position could be a hint at the presence of large-scale structure.

\subsubsection*{PSZ2 G160.94+44.8}
% (PSZ2-769)
This candidate likely corresponds to a false detection when carefully reinspecting the \textit{Planck} maps. This is supported by the fact that PSZ2 G160.94+44.8 has a quality flag of $Q_\mathrm{neural} = 0.06$. Additionally, we find a large distance of more than 10~arcmin between the SZ-peak and the optical position when we re-sample the \textit{Planck} maps at the optical position for the SZ-mass measurement. This gives a second strong argument for a false \textit{Planck} detection. 

\subsubsection*{PSZ2 G165.41+25.93}
% (PSZ2-789)
Upon the inspection of the colour image an optical counterpart to the SZ-signal is hardly identifiable. Accordingly, there is only a very weak peak in the histogram of filtered richness versus redshift resulting in a redshift estimate of $z_\mathrm{phot} = 0.67\pm0.03$. At this redshift, member galaxies of the cluster should be detectable in our imaging. We therefore believe that the optical counterpart is a small galaxy group at most.

Since the density of galaxies at the same redshift is so low, our pipeline is not able to reliably refine the cluster centre through identifying the galaxy which maximises the richness. We therefore, only report the redshift, richness and mass results at the position of the BCG identified by eye.

\subsubsection*{PSZ2 G237.68+57.83}
% (PSZ2-1074)
From our red-sequence analysis we obtain a high photometric redshift estimate of $z_\mathrm{phot} = 0.97\pm0.05$. This is close to the highest redshifts for which clusters can still be detected in the \textit{Planck} maps. However, our red-sequence models loose constraining power in this redshift regime. Additionally, we have a 80 per cent detection limit of $m_{i\mathrm{,totlim}}=23.15$. This photometric redshift and detection limit cause a high completeness correction factor of $c_\mathrm{cmp}>3.5$.
We therefore resort to our spectroscopic redshift result for the estimate of the optical richness and inferred mass. The spectroscopic redshift of the BCG candidate of $z_\mathrm{spec}=0.894$ is in good agreement with the spectroscopic redshift estimate of $z_\mathrm{spec}=0.892$ reported by \citet{Burenin2018}. Due to the problems with the photometric redshift measurement we do not include this target in the recalibration of the red-sequence models.

\subsubsection*{PSZ2 G305.76+44.79}
%  (PSZ2-1441)
Our observations of this cluster candidate provide only a shallow 80 per cent detection limit of $m_{i\mathrm{,totlim}}=22.65$, which means we have a limiting redshift of 0.66. We measure a photometric redshift of $z_\mathrm{phot} = 0.76\pm0.13$. However, this is a noisy measurement which is based only on a few galaxies. Therefore, our pipeline cannot reliably refine the cluster centre and we only report the redshift, richness and mass results at the position of the BCG identified by eye. The richness is not high enough to fulfil the confirmation criterion. However, deeper data might reveal more of the potential cluster galaxies and provide a more robust estimate of richness and mass.

\subsubsection*{PSZ2 G321.30+50.63}
%  (PSZ2-1512)
\textcolor{revision1}{Similar to cluster candidate PSZ2 G085.95+25.23, this cluster candidate likely has a complex structure. We measure a photometric redshift of $z_\mathrm{phot} = 0.79\pm0.04$ around the position \textcolor{revision1}{$\mathrm{RA} = 204.611^\circ$, $\mathrm{Dec} = -10.550^\circ$}. Additionally, we find $z_\mathrm{phot} = 0.68\pm0.07$ around the position $\mathrm{RA} = 204.661^\circ$, $\mathrm{Dec} = 10.566^\circ$. Here we measure a richness of $\lambda = 37\pm10$. The redshifts agree only within around $2\sigma$. This could however also be connected to the broad photo-$z$ peak in the distribution of filtered richness versus redshift. We report the results for the first, richer component in Table \ref{tab:ResultsPSZ2}.}

\subsubsection*{PLCK G58.14--72.7}
% (MMF3-18209)
Similar to PSZ2 G237.68+57.83, we measure a high photometric redshift of $z_\mathrm{phot} = 1.03\pm0.10$. This would require a high completeness correction factor of $c_\mathrm{cmp}>3.5$.
Thus, we do not include this cluster in the recalibration of the red-sequence models. Additionally, we use the spectroscopic redshift of the BCG candidate for our optical richness and mass estimates.

Concerning the spectroscopic redshifts estimated via cross-correlation with absorption and emission line templates we noticed that the $\mathrm{O}_2$ atmospheric absorption line at 7600~{\AA} coincides with the position of the 4000~{\AA}-break (see Fig. \ref{fig:MMF3-18209Spectrum}). The emission peak of the [OII] line is however prominent enough so that we are confident of our spectroscopic redshift result of $z_\mathrm{{spec}}=0.938\pm0.003$ for the BCG and  $z_\mathrm{{spec}}=0.927\pm0.004$ for the second galaxy on the slit.

\subsubsection*{PLCK G98.08--46.4}
% (MMF3-14073)
For this cluster candidate we measure a very high photometric redshift of $z_\mathrm{phot} = 1.06\pm 0.04$. This is close to the highest redshifts for which clusters can still be detected in the \textit{Planck} maps. Given the high redshift and an intermediate depth of the data we have to apply a large completeness correction factor of $c_\mathrm{cmp}>3.5$. In addition, the completeness curve describing how many of the injected sources are retrieved drops rather sharply in comparison to other observations. This could also lead to an overestimate of the richness. 
Since we have measured a spectroscopic redshift of $z_\mathrm{spec} = 0.983$ for the BCG candidate as well, we report the richness given this redshift estimate in Table \ref{tab:ResultsPSZ2}. We do however not use the spectroscopic redshift result for the calibration of the red-sequence models.

\section{Discussion}
\label{Discussion}

\textit{Planck} is the only all-sky SZ survey presently available. The public PSZ2 catalogue provides galaxy cluster candidates down to a SZ-significance threshold of $\mathrm{S/N} = 4.5$, forming an excellent basis for cosmological investigations. In this context, all candidates must be systematically followed up to understand the selection function and to fully exploit \textit{Planck}'s potential. For this we provide a noticeable contribution in this work. We pursue slightly different and partly complementary methods and strategies for the confirmation of cluster candidates in comparison to similar follow-up studies of the PSZ2 catalogue. 

For example \citet{Streblyanska2018} inspect PSZ2 cluster candidates in the Compton $y$-maps and in SDSS data. Their confirmation criteria are based on the richness and distance of the optical centre to the position of the PSZ2-detection. Additionally, they provide photometric and partially spectroscopic redshift information. Due to the limited depth of the SDSS data, they estimate the richness taking into account galaxies close to the photometric redshift estimate with an $r$-band magnitude in the range of ($r_\mathrm{BCG}$, $r_\mathrm{BCG} + 2.5$) in an 0.5~Mpc radius around the cluster centre. The limited depth allows them to provide this richness estimate only for clusters with $z<0.6$. \citet{Streblyanska2018} require a richness of $>5$ and a distance of the optical centre to the position of the SZ-detection of $<5'$ for a cluster to be confirmed. This way, they confirm 37 clusters from the PSZ2 catalogue and find 17 clusters as "potentially associated" with a PSZ2 detection. Our sample overlaps with the one from \citet{Streblyanska2018} for 9 cluster candidates. Our redshift estimates are in good agreement with their results. In addition, we provide richness-based masses obtained using the scaling relation by \citet{Rozo2015} and SZ-mass results which we inferred from the \textit{Planck} maps based on the optical positions and photometric redshifts.

\citet{Boada2018} apply confirmation criteria which are very similar to \citet{Streblyanska2018}. They use optical data from the Kitt Peak National Observatory 4m Mayall telescope to provide photometric redshifts, richnesses and the distance from the optical centre to the PSZ2-detection for their sample of PSZ2 cluster candidates. They focus on the regime of high SZ-significance with $\mathrm{S/N}>5$ confirming clusters in a low to intermediate redshift range of $0.13 < z < 0.74$. Our study does not overlap with their sample but complements this with a sample of \textit{Planck} cluster candidates at intermediate to high redshifts ($0.5 \lesssim z \lesssim 1.0$) and extending down to low SZ-significance $\mathrm{S/N} > 3$. 
 
The works by \citet{Amodeo2018} and \citet{Burenin2018} report precise spectroscopic redshifts for PSZ2 clusters, which are in part also included in our sample. Their spectroscopic information helps us with recalibrating our red-sequence models. As a result, we obtain photometric redshifts which allow us to estimate the richness and richness-based mass from our observations. Additionally, we can infer the SZ-based mass and therefore supplement the spectroscopic redshift information from \citet{Amodeo2018} and \citet{Burenin2018}. 

Finally, our strategy of cluster confirmation is very similar to \citet{VanDerBurg2016} who confirm 16 clusters detected with \textit{Planck}. Based on optical imaging with MegaCam at CFHT they estimate photometric redshifts and richnesses, and infer SZ-based masses. They illustrate the benefit of preselecting cluster candidates from optical and NIR-data in order to reliably uncover massive clusters at low SZ detection significance. In addition, they point out the advantage of a secondary mass proxy from optical data because it is independent of Eddington bias. These strategies have proven equally useful for our follow-up study. The reliability is of the order of \textcolor{revision1}{80 per cent} for our PSZ2 cluster candidates. This roughly meets our expectations (see Section \ref{Confirmation of Cluster Candidates}) and is comparable to the findings of \cite{VanDerBurg2016}. In particular, our selection was effective for the low S/N targets detected with the MMF3 method. We achieve a high reliability where only one of these targets remains unconfirmed. 
There is of course the possibility of an increased number of chance super-positions between MMF3 SZ detections and our selected optical counterparts for example due to the large number of candidates in the low S/N regime. However, we do not find that the optical positions of the MMF3 targets are particularly closer to the SZ detections than it is the case for the PSZ2 candidates. Additionally, the Spearman rank correlation coefficients of the richness and the SZ-mass show a very similar tendency towards a positive correlation both for the MMF3 and for the PSZ2 candidates. This gives us reason to believe that we did select cluster candidates in the optical which are indeed related to the SZ-detection in the \textit{Planck} maps.

Moreover, the fact that our sample is not purely SZ-selected implies that we have a complex selection function which is hard to quantify because it requires a careful modelling of the optical properties of the clusters. The consequential uncertainties in the selection function can easily present a problem for the use of such samples for precision cosmology. With our approach we do however increase the sample's purity especially towards lower SZ significance. It allows us to efficiently exploit the potential of the \textit{Planck} SZ-survey to lower S/N to find more massive, high-redshift clusters. Indeed, our study extends the number of massive, high-redshift clusters including some of the highest redshift clusters detected in \textit{Planck} so far. These could be of particular interest for further astrophysical investigations in the context of clusters as laboratories for the interaction and evolution of galaxies, potential mergers or AGN feedback. Therefore, combining a SZ-significance cut with deeper auxiliary data might be the best option to obtain representative massive and high-redshift samples based on \textit{Planck} as the only all-sky SZ survey presently available. 

\section{Summary and Conclusions}
\label{Summary and Conclusions}

We present a photometric (spectroscopic) follow-up analysis of a sample of 32 (9) galaxy cluster candidates observed with the 4.2-m William Herschel Telescope. The sample is selected from detections in the \textit{Planck} maps (PSZ2 catalogue and MMF3 detection method) while the candidates were preselected to likely have counterparts at redshifts above $z\gtrsim 0.7$ based on SDSS, Pan-STARRS and WISE 3.4 $\mu$m data. From a red-sequence based analysis of galaxy over-densities in the $r$-, $i$- and $z$-band imaging data we are able to obtain photometric redshift and richness estimates for the cluster candidates. 
We recalibrate our red-sequence models based on a comparison of our initial photometric redshift estimates with spectroscopic redshift estimates.

In combination with the richness-mass relation by \citet{Rozo2015}, the richness estimate provides us with a quantitative measurement to assess if the detected cluster candidate is likely massive/rich enough to account for the detected SZ signal.
We consider a cluster candidate to be a likely counterpart of a SZ detection in our sample according to quantitative and qualitative criteria. We require 1) that we can identify an over-density of cluster galaxies of the same colour in the colour images and 2) that the richness-based mass is high enough to account for the SZ signal considering the cluster should be at least 50 per cent as massive as expected from the SZ signal. We confirm 18 clusters out of 32 cluster candidates from the \textit{Planck} maps in our sample. The confirmed clusters cover a redshift range of $0.5 \lesssim z \lesssim 1.0$ and a richness-based mass range of $1.0\times 10^{14} \lesssim M_{\mathrm{500c,}\lambda}/\mathrm{M}_\odot \lesssim 8.0\times 10^{14}$.

There are, however, multiple effects that can lower the ratio of richness-based mass to SZ-based mass. Among them are projection effects, the assumption of no redshift evolution of the richness-SZ-mass scaling relation, the 25 per cent scatter in the richness-SZ-mass relation, and Eddington bias especially at low S/N. Applying a less strict criterion where the cluster should be 25 per cent as massive as expected from the SZ signal, we find 24 optical clusters which are likely counterparts to the SZ-signal.

Investigating candidates down to $\mathrm{S/N} \geq 3$ in the \textit{Planck} maps, we find that the SZ-based mass proxy is clearly influenced by Eddington bias. In this context, the richness-based mass represents a valuable quantity  because it is independent of the SZ-signal and hence not subject to Eddington bias.

While our sample has a complex selection function due to the preselection with auxiliary optical and infrared data, and is therefore not suitable for precision cosmology, our approach is still efficient in uncovering rich, high-$z$ clusters in the low S/N regime from the \textit{Planck} all-sky survey.

\section*{Acknowledgements}

\small{HZ is a member of and received financial support for this research from the International Max Planck Research School (IMPRS) for Astronomy and Astrophysics at the Universities of Bonn and Cologne.
HZ acknowledges support from the German Federal Ministry of Economics and Technology (BMWi) provided through DLR under project 50 OR 1803.
This article is based on observations obtained with the William Herschel Telescope, operated on the island of La Palma by the Isaac Newton Group of Telescopes in the Spanish Observatorio del Roque de los Muchachos of the Instituto de Astrof\'isica de Canarias. In particular, we thank the ING staff for their support at the WHT telescope. We thank  Alya Amira Azman, Beatriz Hern\'{a}ndez Mart\'{i}n, Manali Jeste, Nils Linz, Vyoma Muralidhara, Silvia N\"{o}sel, Fatimah Raihan and Felix Siebeneicker for their support in the observation run at the WHT in October 2016. We thank Beatriz Hern\'{a}ndez Mart\'{i}n, Fatimah Raihan and Nils Weissgerber for useful discussions and sharing scripts helping us in the photometric calibration of this work.  We thank the Instituto de Astrof\'isica de Canarias (IAC) cluster validation group, formed by Jos\'e Alberto Rubi\~no, Alina Streblyanska, Rafael Barrena, Alejandro Aguado-Barahona and Antonio Ferragamo, for discussing some PSZ2 counterparts reported here, which were also observed as part of the LP15 validation programme of PSZ2 sources at Canary Island observatories.
We thank Peter Schneider for useful discussions.
We thank Axel Buddendiek for his help with the RS calibration and for helpful discussions concerning the spectroscopic redshift analysis.
We thank Guillaume Hurier for providing the $Q_\mathrm{neural}$ values for low S/N candidates. 
The research leading to these results has received funding from the European Community's Seventh Framework Programme (FP7/2013-2016) under grant agreement number 312430 (OPTICON). This project has received funding from the European Union's Horizon 2020 research and innovation programme under grant agreement No 730890. This material reflects only the authors views and the Commission is not liable for any use that may be made of the information contained therein.
The research leading to these results has received funding from the European Research  Council  under  the  European Union Seventh Framework Programme (FP72007-2013) ERC grant agreement no 340519.
This research has made use of the NASA/ IPAC Infrared Science Archive, which is operated by the Jet Propulsion Laboratory, California Institute of Technology, under contract with the National Aeronautics and Space Administration.
The Pan-STARRS1 Surveys (PS1) and the PS1 public science archive have been made possible through contributions by the Institute for Astronomy, the University of Hawaii, the Pan-STARRS Project Office, the Max-Planck Society and its participating institutes, the Max Planck Institute for Astronomy, Heidelberg and the Max Planck Institute for Extraterrestrial Physics, Garching, The Johns Hopkins University, Durham University, the University of Edinburgh, the Queen's University Belfast, the Harvard-Smithsonian Center for Astrophysics, the Las Cumbres Observatory Global Telescope Network Incorporated, the National Central University of Taiwan, the Space Telescope Science Institute, the National Aeronautics and Space Administration under Grant No. NNX08AR22G issued through the Planetary Science Division of the NASA Science Mission Directorate, the National Science Foundation Grant No. AST-1238877, the University of Maryland, Eotvos Lorand University (ELTE), the Los Alamos National Laboratory, and the Gordon and Betty Moore Foundation.
This publication makes use of data products from the Wide-field Infrared Survey Explorer, which is a joint project of the University of California, Los Angeles, and the Jet Propulsion Laboratory/California Institute of Technology, funded by the National Aeronautics and Space Administration.
Funding for SDSS-III has been provided by the Alfred P. Sloan Foundation, the Participating Institutions, the National Science Foundation, and the U.S. Department of Energy Office of Science. The SDSS-III web site is http://www.sdss3.org/.
SDSS-III is managed by the Astrophysical Research Consortium for the Participating Institutions of the SDSS-III Collaboration including the University of Arizona, the Brazilian Participation Group, Brookhaven National Laboratory, Carnegie Mellon University, University of Florida, the French Participation Group, the German Participation Group, Harvard University, the Instituto de Astrofisica de Canarias, the Michigan State/Notre Dame/JINA Participation Group, Johns Hopkins University, Lawrence Berkeley National Laboratory, Max Planck Institute for Astrophysics, Max Planck Institute for Extraterrestrial Physics, New Mexico State University, New York University, Ohio State University, Pennsylvania State University, University of Portsmouth, Princeton University, the Spanish Participation Group, University of Tokyo, University of Utah, Vanderbilt University, University of Virginia, University of Washington, and Yale University.}

%%%%%%%%%%%%%%%%%%%%%%%%%%%%%%%%%%%%%%%%%%%%%%%%%%



\bibliographystyle{mnras}
\bibliography{references} % if your bibtex file is called references.bib

%\begin{thebibliography}{99}
%\end{thebibliography}

%%%%%%%%%%%%%%%%%%%%%%%%%%%%%%%%%%%%%%%%%%%%%%%%%%

%%%%%%%%%%%%%%%%% APPENDICES %%%%%%%%%%%%%%%%%%%%%

\appendix

\section{80 per cent depth limits}
\label{Sec:AppA80percentLimit}

In this section we show an example of the distribution of the recovery fraction $N_\mathrm{detected}/N_\mathrm{injected}$ of sources as a function of the $i$-band magnitude $m_i$. We define the 80 per cent depth limits $m_{i\mathrm{,totlim}}$ in the $i$-band as the faintest magnitude where we could still recover 80 per cent or more of the injected galaxies.

\begin{figure}
	\centering
	\includegraphics[width=1.0\columnwidth]{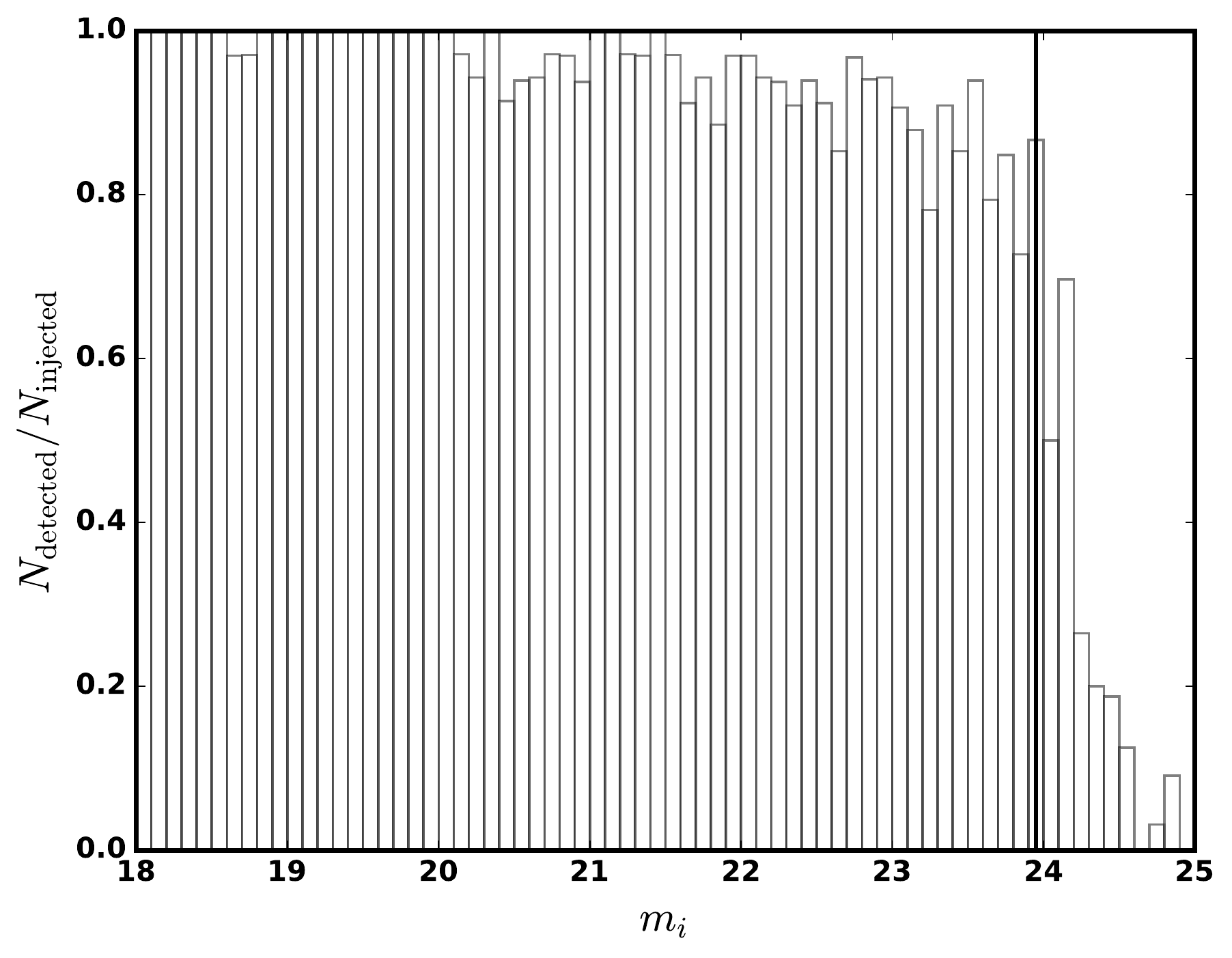}
	\caption{Distribution of the recovery fraction $N_\mathrm{detected}/N_\mathrm{injected}$ of sources as a function of the $i$-band magnitude $m_i$ for cluster candidate G032.31+66.07. The black line indicates the 80 per cent depth limit $m_{i\mathrm{,totlim}}$ for this particular observation.}
	\label{fig:80limit-histo}
\end{figure}

\section{External information about the candidates}

In this section we summarise results from the literature where targets from our sample have been investigated. This concerns spectroscopic redshifts and optical cluster centres.

\begin{table*}
	\begin{threeparttable}
			\hspace*{-25mm}
		\caption{Results from independent follow-up studies of cluster candidates from our sample.}
		\begin{tabular}{ c l c c c l}
			\hline\hline
			\textcolor{revision1}{ID} &Name & RA & Dec & $z_\mathrm{spec}$ & comments  \\
			 & & [$^\circ$] & [$^\circ$] &   &   \\ 
			\textcolor{revision1}{115} &PSZ2 G032.31+66.07 & 219.3540  &  24.3986 &	0.609 & confirmed cluster from \citet{Streblyanska2018}, \\
					& & & & & spec-$z$ from 6 galaxies  \\
			\textcolor{revision1}{277} &PSZ2 G066.34+26.14  & 270.2772  &  39.8685 &	-  & confirmed cluster from \citet{Streblyanska2018}\\
			\textcolor{revision1}{378} &PSZ2 G085.95+25.23  & 277.6164 & 56.8823 & $0.782\pm 0.003$ & from \cite{Amodeo2018} \\
			\textcolor{revision1}{381} &PSZ2 G086.28+74.76  & 204.4745  &  38.9019  &	0.699 & confirmed cluster from \citet{Streblyanska2018},\\
					& & & & &  spec-$z$ from 1 galaxy \\
			\textcolor{revision1}{420} &\textcolor{revision1}{PSZ2 G092.64+20.78}  & \textcolor{revision1}{289.1893}  &  \textcolor{revision1}{61.6782} & \textcolor{revision1}{0.545} & \textcolor{revision1}{confirmed cluster from \citet{Streblyanska2019}}\\
					& & & & & \textcolor{revision1}{spec-$z$ from 39 galaxies}  \\
			\textcolor{revision1}{421} &PSZ2 G092.69+59.92  & 216.5355  &  51.2373  & 0.461 & potentially associated cluster from \citet{Streblyanska2018}, \\
					& & & & & spec-$z$ from 3 galaxies, optical pos. outside the FOV of \\
					& & & & & our observation \\
					& & 216.6504 &  51.2642 & 0.848 & from \cite{Burenin2018}, spec-$z$ from 2 galaxies,\\
					& & & & & other close by cluster at $z=0.463$ mentioned \\
			\textcolor{revision1}{483} &\textcolor{revision1}{PSZ2 G100.22+33.81}  & \textcolor{revision1}{258.4232}  &  \textcolor{revision1}{69.3626} & \textcolor{revision1}{0.598} & \textcolor{revision1}{confirmed cluster from \citet{Streblyanska2019}}\\
					& & & & & \textcolor{revision1}{spec-$z$ from 18 galaxies}  \\			
			\textcolor{revision1}{623} &PSZ2 G126.28+65.62  & 190.5975 & 51.4394  &	0.820 & from \cite{Burenin2018}, spec-$z$ from 7 galaxies \\
			\textcolor{revision1}{625} &PSZ2 G126.57+51.61 & 187.4492  &  65.3536 & 0.815 & from \cite{Burenin2018}, spec-$z$ from 1 galaxy \\
			\textcolor{revision1}{667} &PSZ2 G136.02--47.15  & 22.0984 & 14.6871 & 0.465 &  potentially associated cluster from \citet{Streblyanska2018}, \\
					& & & & &  spec-$z$ from 1 galaxy, optical pos. outside the FOV of \\
					& & & & &  our observation \\
			\textcolor{revision1}{681} &PSZ2 G139.00+50.92  & 170.0707  &  63.2500  &	- & confirmed cluster from \citet{Streblyanska2018} \\
			\textcolor{revision1}{690} &PSZ2 G141.98+69.31  & 183.1693  &  46.3564 &  0.714 &  potentially associated cluster from \citet{Streblyanska2018},\\
					& & & & &  spec-$z$ from 2 galaxies \\
			\textcolor{revision1}{1074} &PSZ2 G237.68+57.83  & 163.3242  &  10.8770  & - & 	confirmed cluster from \citet{Streblyanska2018} \\
			         & & 163.3179  &  10.8794 & 0.892 & from \cite{Burenin2018}, spec-$z$ from 1 galaxy\\
			\textcolor{revision1}{1606} &PSZ2 G343.46+52.65 & 216.0963 & -2.7178 & 0.713 & from \cite{Burenin2018}, spec-$z$ from 3 galaxies  \\
			
		\end{tabular}

		\begin{tablenotes}
			\textcolor{revision1}{Column 1: ID of cluster in PSZ2,} column 2: name of the cluster, column 3 and 4: optical centre from the respective work, column 5: spectroscopic redshift if available, column 6: comments on the origin of the results.
			
		\end{tablenotes}
		
		\label{tab:ExternalResults}
	\end{threeparttable}
\end{table*}

\section{Alternative Richness and Mass Results at the Spectroscopic Redshifts}
\label{Sec:AlternativeRichness+Mass}

In this section we present our richness and mass results inferred from the optical data when using the spectroscopic redshifts (when available) instead of the photometric redshifts.

\begin{table*}
	\begin{threeparttable}
		\hspace*{-25mm}
		\caption{Alternative richness and mass results at the spectroscopic redshifts.}
		\begin{tabular}{ c l c c c c | c}
			\hline\hline
			\textcolor{revision1}{ID} &Name  & $z_\mathrm{spec}$ & $\lambda$ & $M_{\mathrm{500c,}\lambda}$ & $c_\mathrm{cmp}$ & $M_\mathrm{500c,SZ}$ \\
			& & & & [$10^{14}\mathrm{M}\odot$] & & [$10^{14}\mathrm{M}\odot$] \\ 
			
			\textcolor{revision1}{115} &PSZ2 G032.31+66.07  & 0.609$^d$ & $71\pm13$ & $3.81\pm0.72$ & 1.00 & $5.70^{+0.76}_{-0.84}$ \\
			\textcolor{revision1}{378} &PSZ2 G085.95+25.23  & $0.782\pm 0.003^c$ & $167\pm22$ & $9.24\pm1.27$ & 1.74 & $5.41^{+0.55}_{-0.59}$ \\
			\textcolor{revision1}{381} &PSZ2 G086.28+74.76  & 0.699$^d$ & $36\pm10$ & $1.88\pm0.55$ & 1.00 & $5.32^{+0.70}_{-0.76}$ \\
			\textcolor{revision1}{420}  &\textcolor{revision1}{PSZ2 G092.64+20.7}   & \textcolor{revision1}{0.545$^{e}$}	 & \textcolor{revision1}{$32\pm   11$} &	\textcolor{revision1}{$1.64\pm 0.57$} & \textcolor{revision1}{1.00} &	\textcolor{revision1}{$4.40^{+0.45}_{-0.49}$} \\
			\textcolor{revision1}{421} &PSZ2 G092.69+59.92  & 0.848$^b$ & $18\pm9.$ & $0.91\pm0.50$ & 1.36 & $3.39^{+0.77}_{-0.95}$ \\
			\textcolor{revision1}{483}  &\textcolor{revision1}{PSZ2 G100.22+33.8} & \textcolor{revision1}{0.598$^{e}$} & \textcolor{revision1}{$37\pm10$} &	\textcolor{revision1}{$1.94\pm 0.51$} & \textcolor{revision1}{1.00} &	\textcolor{revision1}{$4.09^{+0.51}_{-0.56}$} \\			
			\textcolor{revision1}{623} &PSZ2 G126.28+65.62  &	0.820$^b$ & $76\pm15$ & $4.04\pm0.84$ & 1.38 & $5.06^{+0.66}_{-0.71}$ \\
			\textcolor{revision1}{625} &PSZ2 G126.57+51.61  & 0.815$^b$ & $83\pm17$ & $4.45\pm0.94$ & 1.68 & $5.83^{+0.56}_{-0.60}$ \\
			\textcolor{revision1}{690} &PSZ2 G141.98+69.31  &  0.714$^d$ & $40\pm8$ & $2.08\pm0.45$ & 1.00 & $4.05^{+0.79}_{-0.90}$ \\
			\textcolor{revision1}{1074} &PSZ2 G237.68+57.83  & $0.894\pm0.007^a$ & $148\pm27$ & $8.15\pm1.55$ & 2.94 & $5.47^{+0.75} _{-0.80}$  \\
			\textcolor{revision1}{1606} &PSZ2 G343.46+52.65  & 0.713$^b$ & $115\pm15$ & $6.29\pm0.86$ & 1.00 & $6.35^{+0.87}_{-0.95}$ \\
			
			\textcolor{revision1}{-} &PLCK G58.14--72.7  & $0.938\pm0.003^a$ & $60\pm16$ & $3.15\pm0.90$ & 2.23 & $4.13^{+0.85} _{-1.01}$ \\
			\textcolor{revision1}{-} &PLCK G98.08--46.4  & $0.983\pm0.005^a$ & $53\pm21$ & $2.80\pm1.17$ & 3.47 & $3.24^{+1.06} _{-1.46}$ \\
			\textcolor{revision1}{-} &PLCK G174.14--27.5  & $0.834\pm0.005^a$ & $15\pm8$ & $0.75\pm0.43$ & 1.17 & $3.09^{+1.33}_{-2.15}$ \\
			\textcolor{revision1}{-} &PLCK G184.49+21.1  & $0.596\pm0.007^a$ & $118\pm15$ & $6.41\pm0.87$ & 1.00 & $5.80^{+0.79}_{-0.87}$ \\

		\end{tabular}
		
		\begin{tablenotes}
			\textcolor{revision1}{Column 1: ID of the cluster in PSZ2,} column 2: name of the cluster, column 3: spectroscopic redshift: $z_\mathrm{spec}$ a) from this work (BCG candidate), b) from \citet{Burenin2018}, c) from \citet{Amodeo2018}, d) from \citet{Streblyanska2018}, \textcolor{revision1}{e) from \citet{Streblyanska2019},} column 4 and 5: richness and inferred mass based on $z_\mathrm{spec}$, column 6: completeness correction factor at $z_\mathrm{spec}$, column 7: SZ-mass using $z_\mathrm{spec}$ and the optical position of the BCG in this work or the optical centre indicated in the literature.
			\\
			
		\end{tablenotes}
		
		\label{tab:AlternativeRichMassResults}
	\end{threeparttable}
\end{table*}

\section{Colour Images of Cluster Candidate Sample}
\label{ColourImages}

Here, we provide colour images of all clusters investigated in this work. All images have north up and east left orientation and the same colour scale. The images display roughly 4 arcmin by 3.4 arcmin field.

\begin{figure*}
	\begin{minipage}{0.32\textwidth}
		\includegraphics[width=\linewidth]{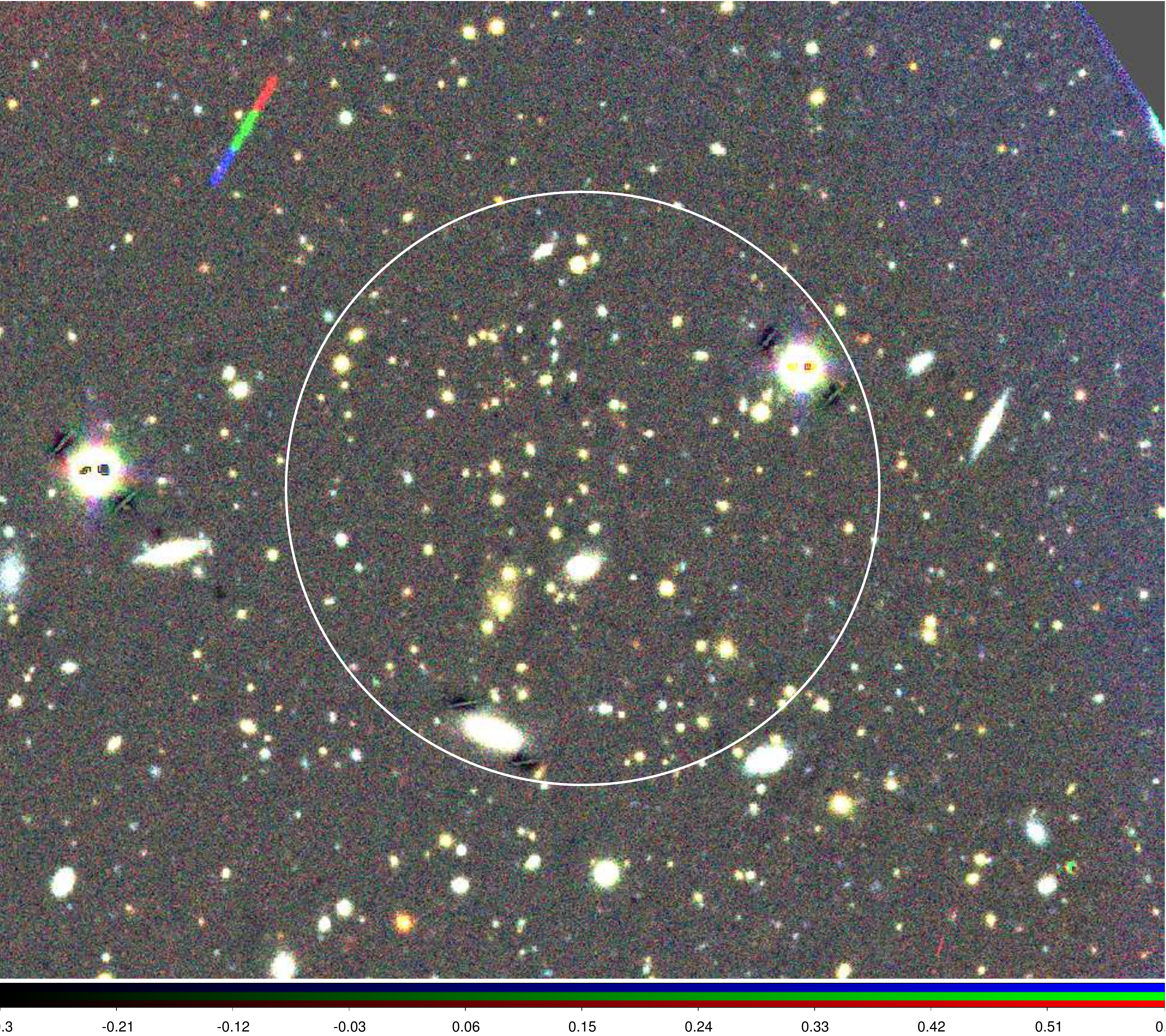}
		(a) PSZ2 G032.31+66.07
	\end{minipage}
	\hspace{0.1cm} % note: no blank line here
	\begin{minipage}{0.32\textwidth}
		\includegraphics[width=\linewidth]{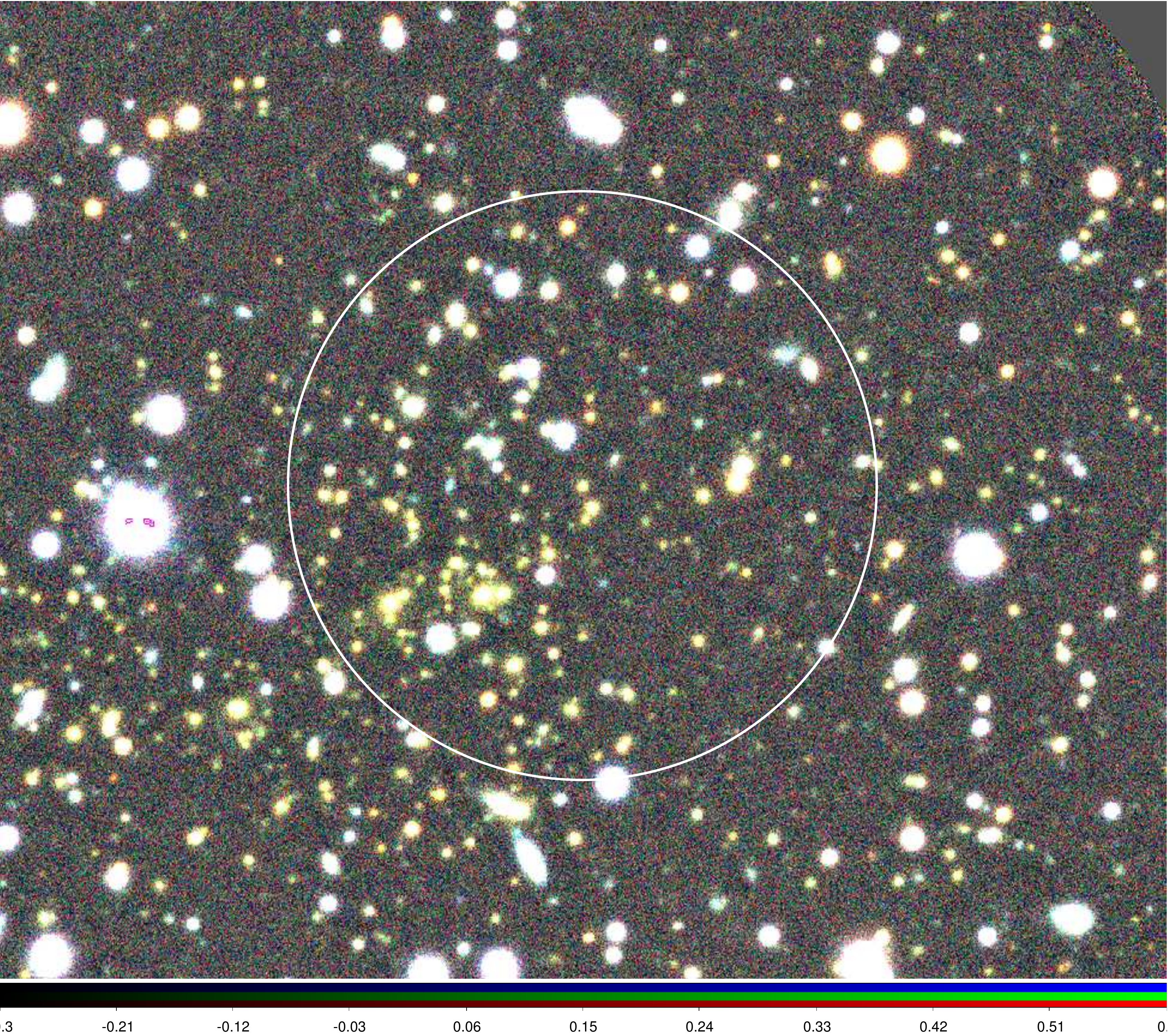}
		(b) PSZ2 G066.34+26.14
	\end{minipage}
	\hspace{0.1cm} % note: no blank line here
	\begin{minipage}{0.32\textwidth}
		\includegraphics[width=\linewidth]{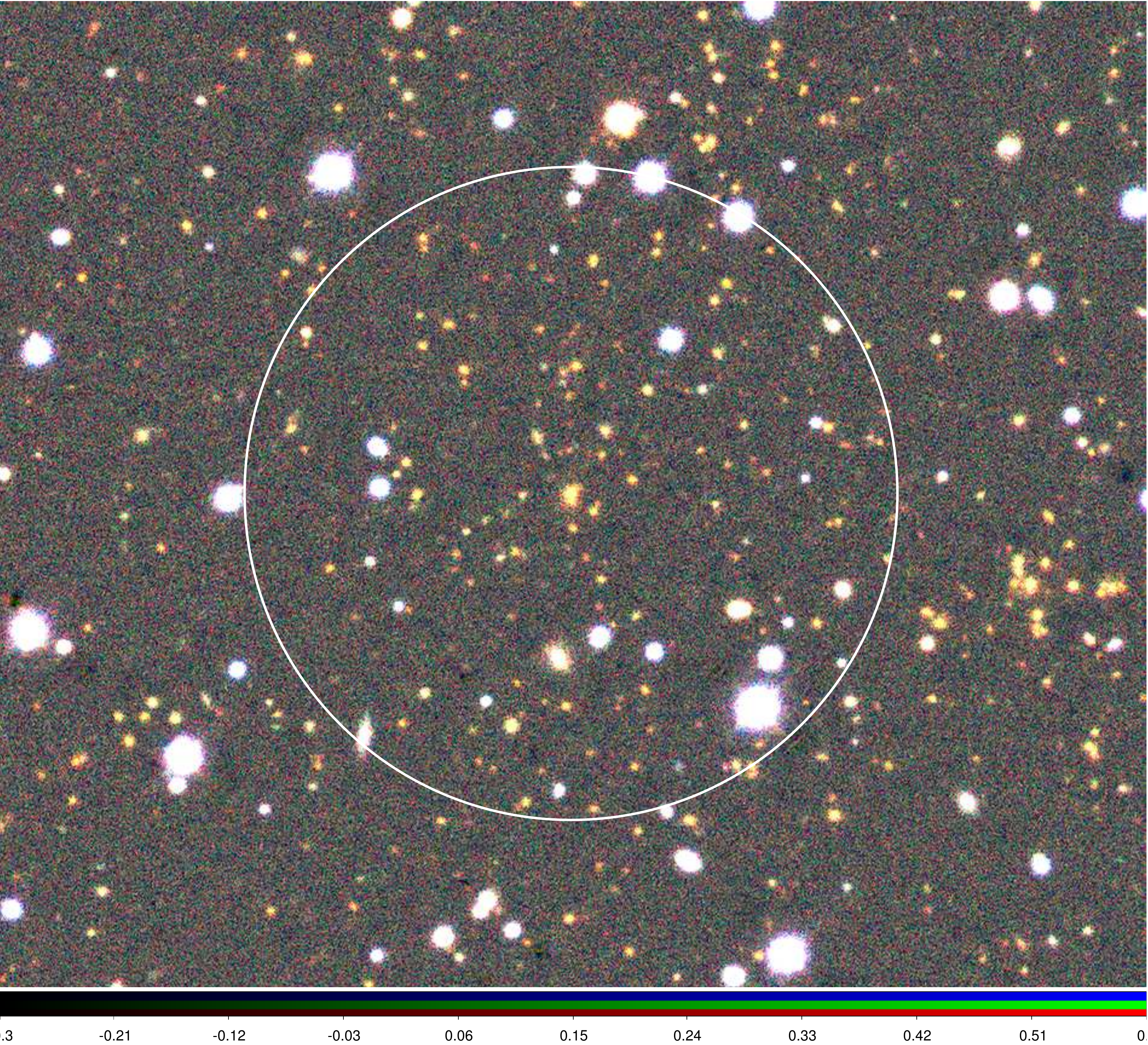}
		(c) PSZ2 G085.95+25.23
	\end{minipage}
	
	%	\vspace*{1cm} % vertical separation
	
	\begin{minipage}{0.32\textwidth}
		\includegraphics[width=\linewidth]{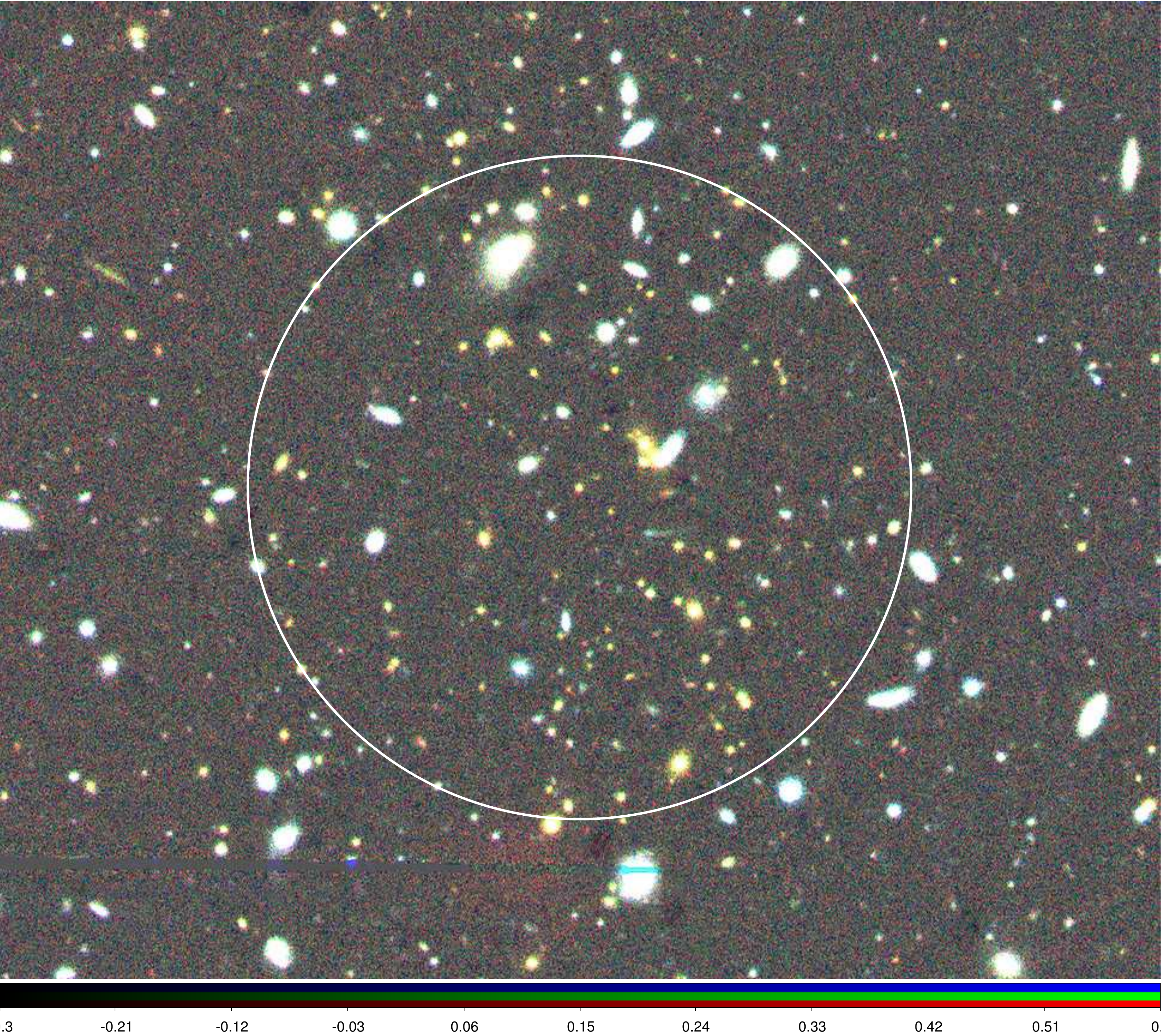}
		(d) PSZ2 G086.28+74.76
	\end{minipage}
	\hspace{\fill} % note: no blank line here
	\begin{minipage}{0.32\textwidth}
		\includegraphics[width=\linewidth]{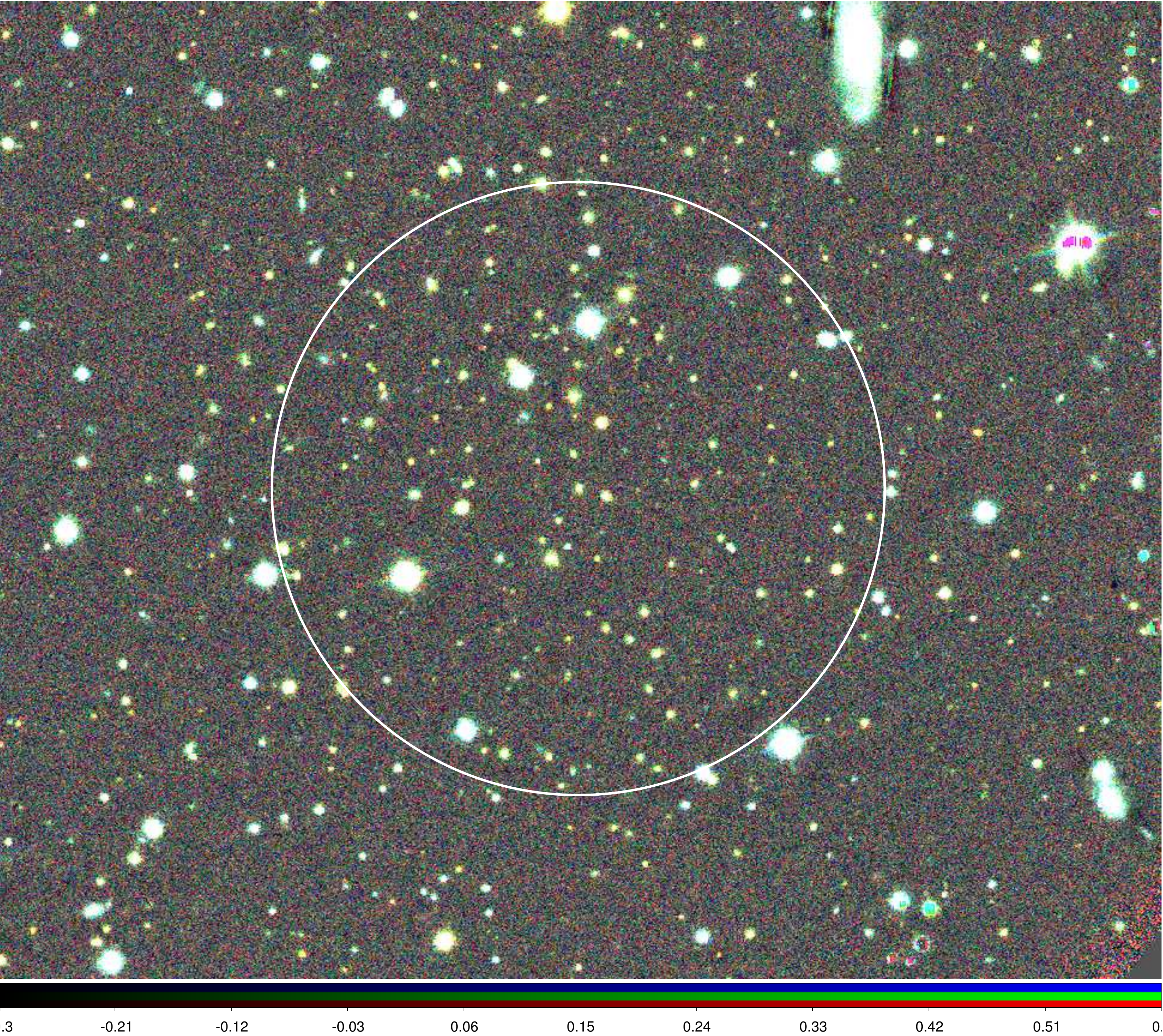}
		(e) PSZ2 G092.64+20.78
	\end{minipage}
	\hspace{\fill} % note: no blank line here
	\begin{minipage}{0.32\textwidth}
		\includegraphics[width=\linewidth]{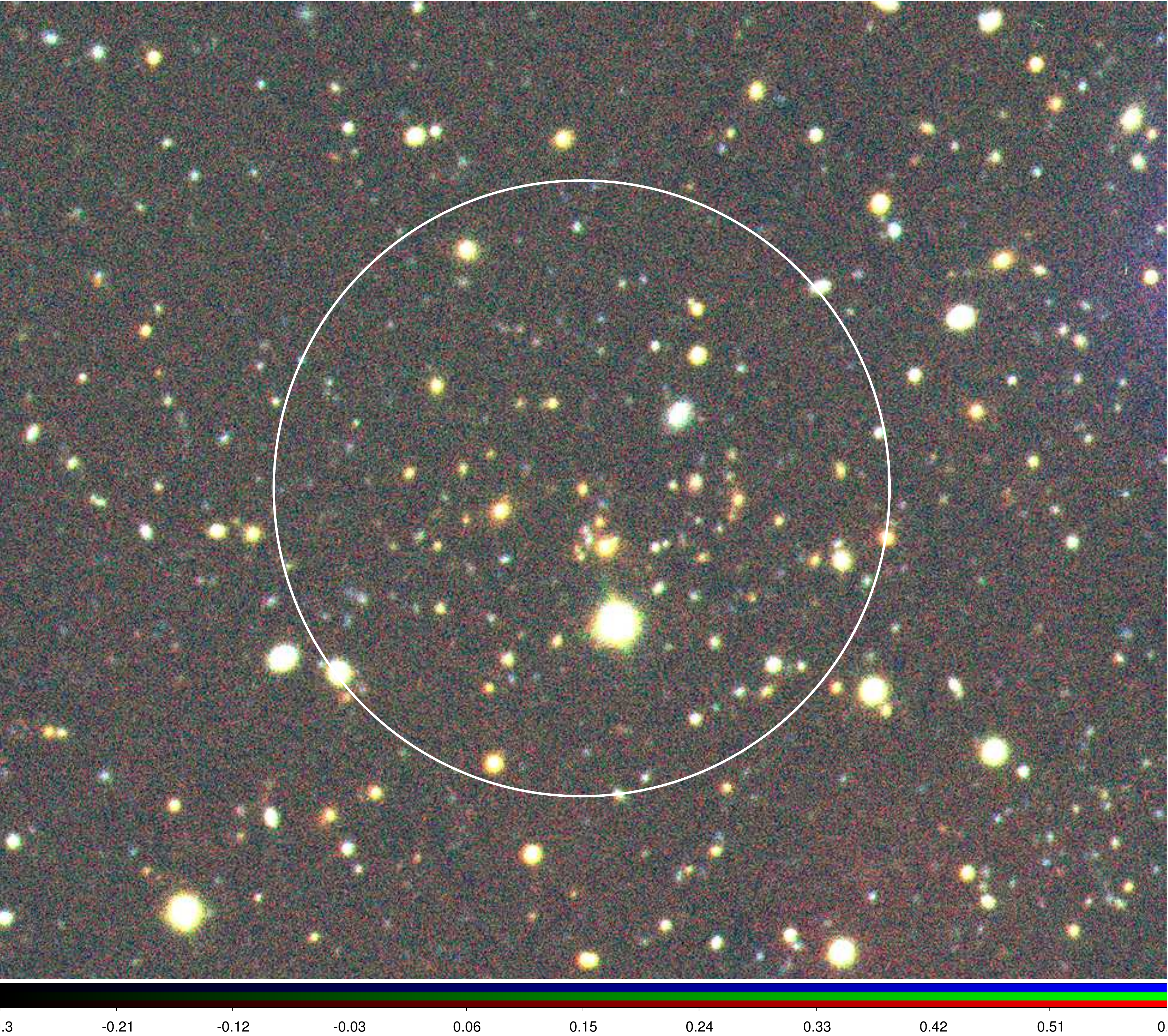}
		(f) PSZ2 G092.69+59.92
	\end{minipage}
	
	%	\vspace*{1cm} % vertical separation
	
	\begin{minipage}{0.32\textwidth}
		\includegraphics[width=\linewidth]{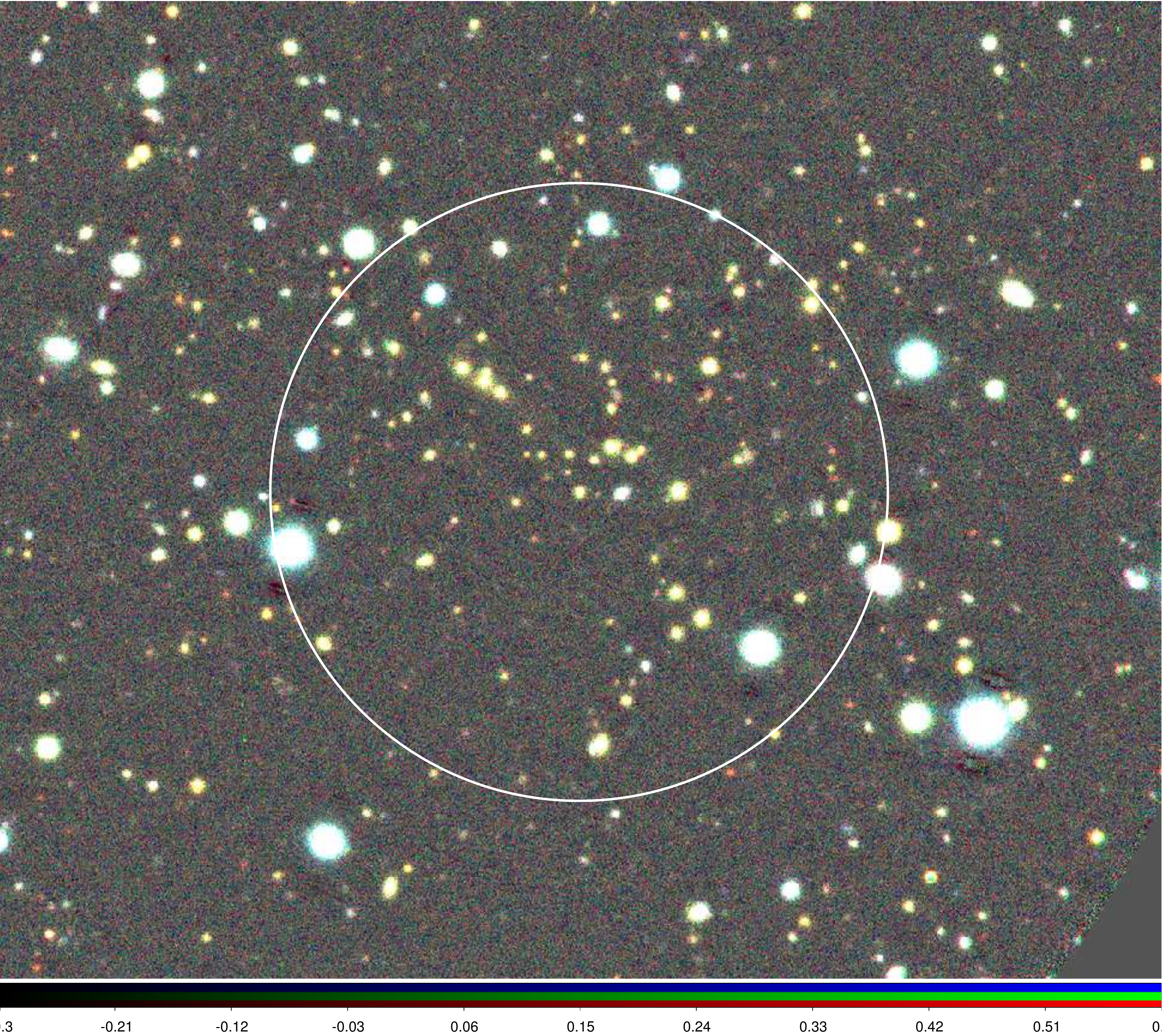}
		(g) PSZ2 G100.22+33.81
	\end{minipage}
	\hspace{\fill} % note: no blank line here	
	\begin{minipage}{0.32\textwidth}
		\includegraphics[width=\linewidth]{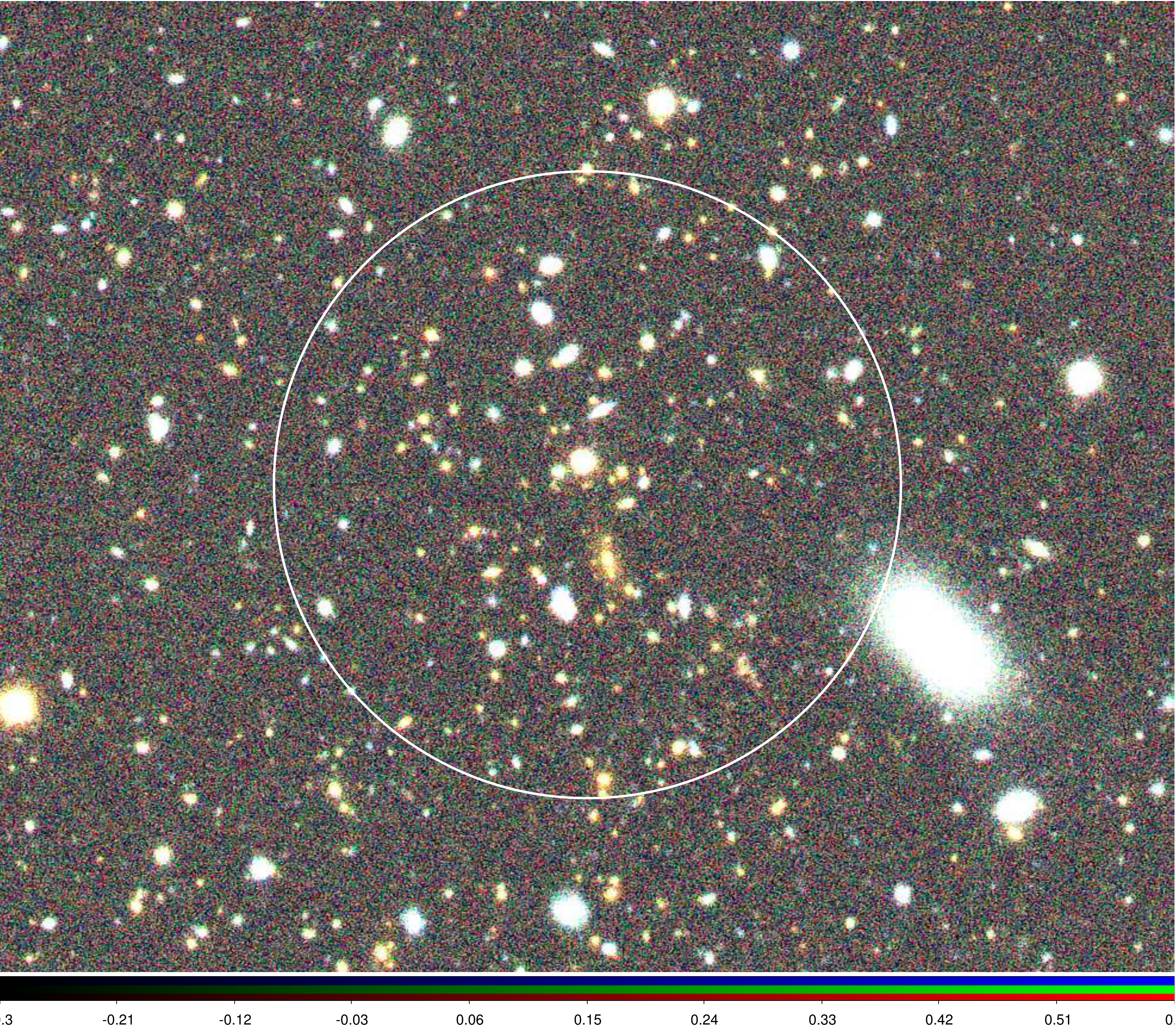}
		(h) PSZ2 G126.28+65.62
	\end{minipage}
	\hspace{\fill} % note: no blank line here
	\begin{minipage}{0.32\textwidth}						      \includegraphics[width=\linewidth]{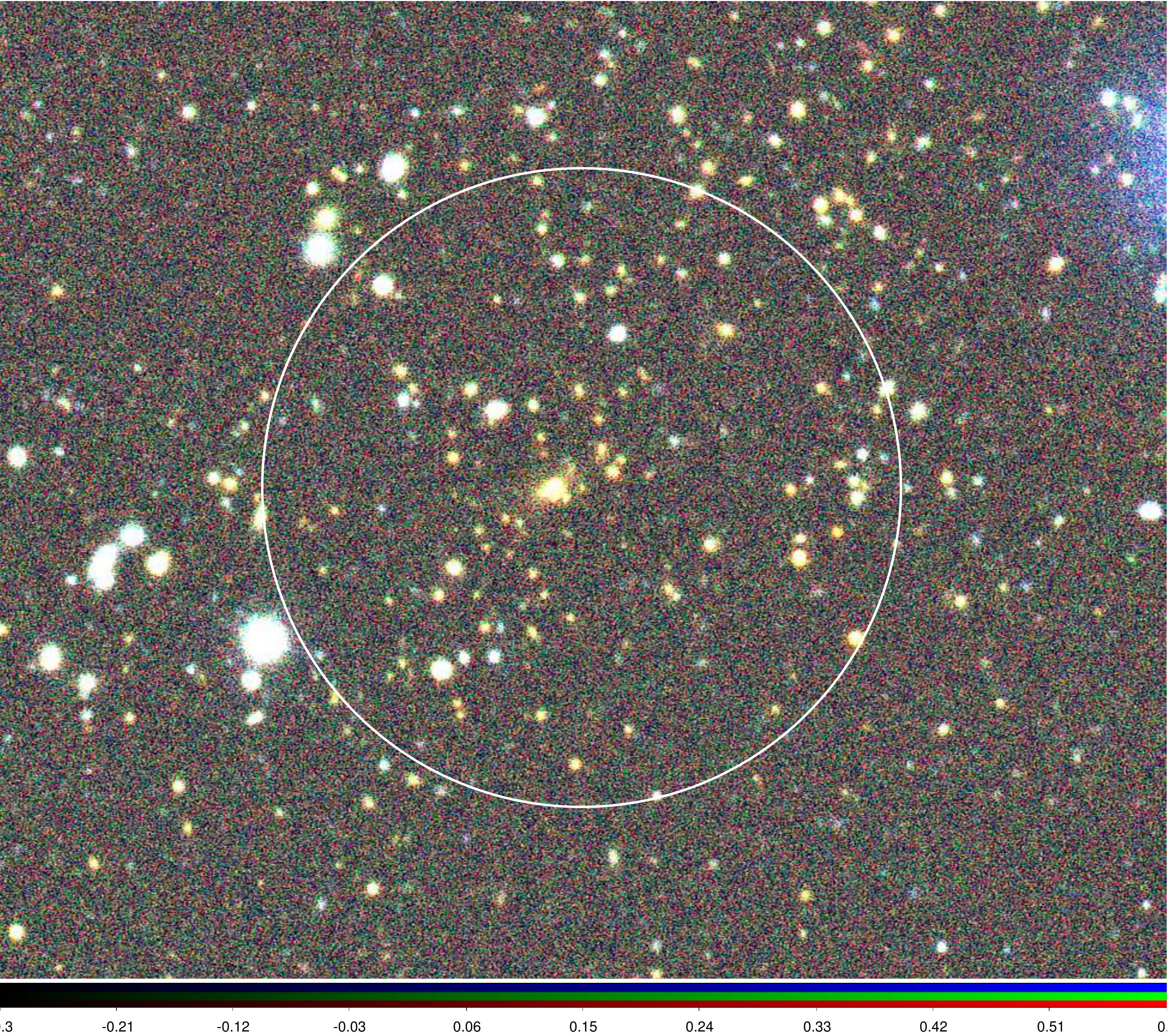}
		(i) PSZ2 G126.57+51.61
	\end{minipage}
	
	%	\vspace*{1cm} % vertical separation
	
	\begin{minipage}{0.32\textwidth}
		\includegraphics[width=\linewidth]{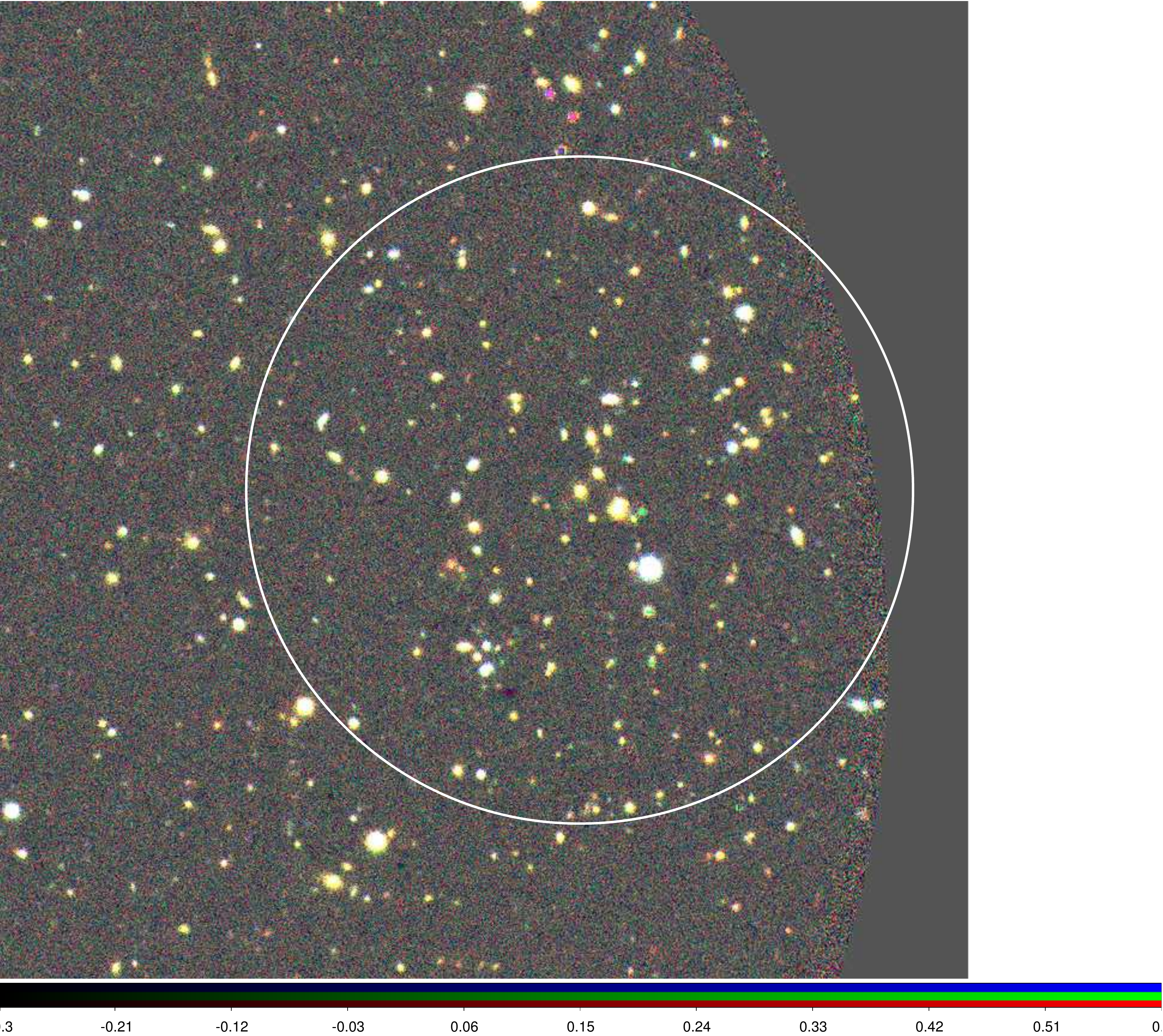}
		(j) PSZ2 G141.98+69.31
	\end{minipage}
	\hspace{\fill} % note: no blank line here
	\begin{minipage}{0.32\textwidth}						      \includegraphics[width=\linewidth]{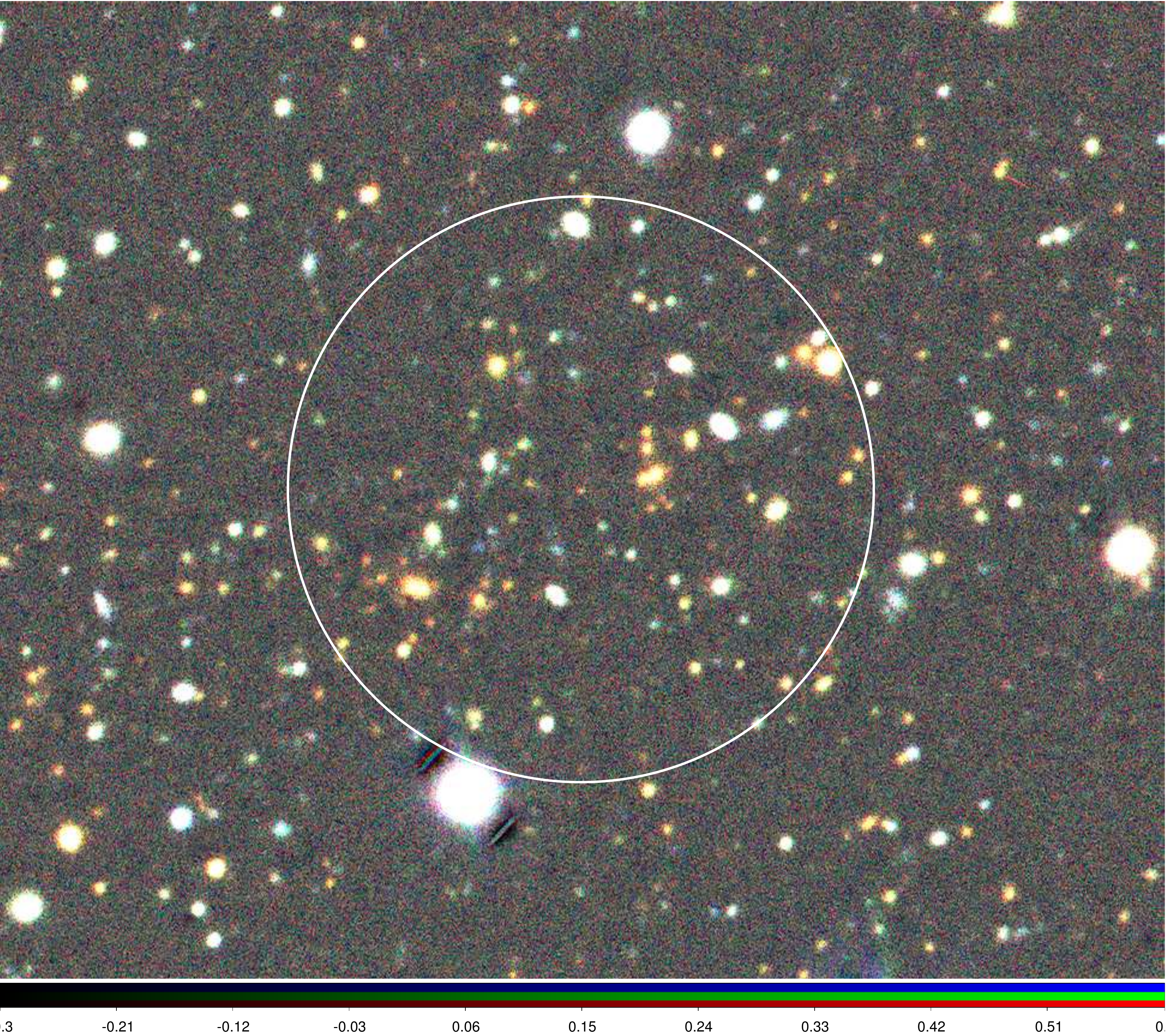}
		(k) PSZ2 G237.68+57.83
	\end{minipage}
	\hspace{\fill} % note: no blank line here
	\begin{minipage}{0.32\textwidth}
		\includegraphics[width=\linewidth]{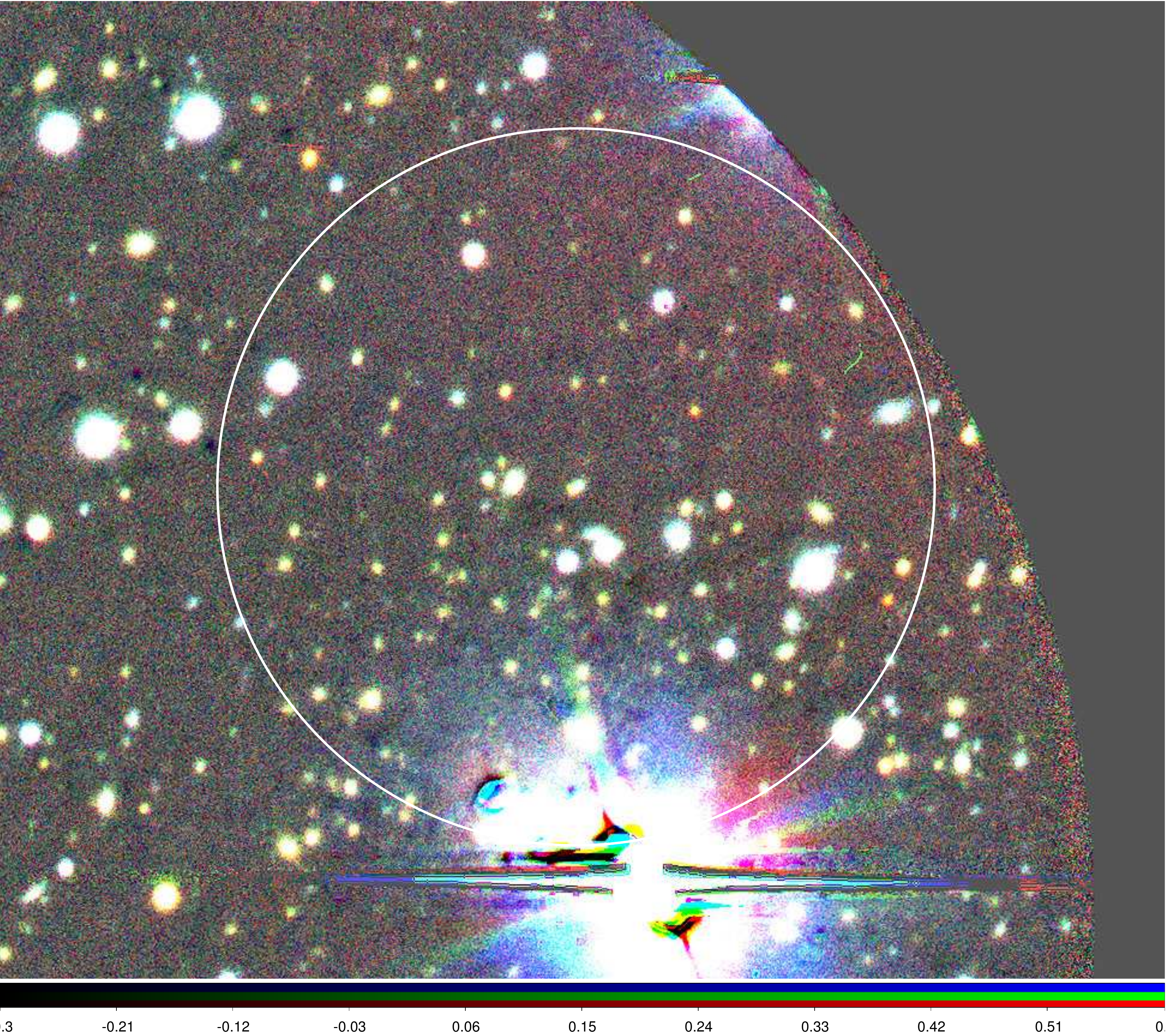}
		(l) PSZ2 G246.91+24.65
	\end{minipage}

	\caption{Clusters with $M_{\mathrm{500c,}\lambda}/M_{\mathrm{500c,SZ}}>0.25$. The white circles equal 0.5 Mpc in radius around the cluster centre at the estimated redshift.} \label{fig:4pics}
\end{figure*}

\begin{figure*}
	\begin{minipage}{0.32\textwidth}
		\includegraphics[width=\linewidth]{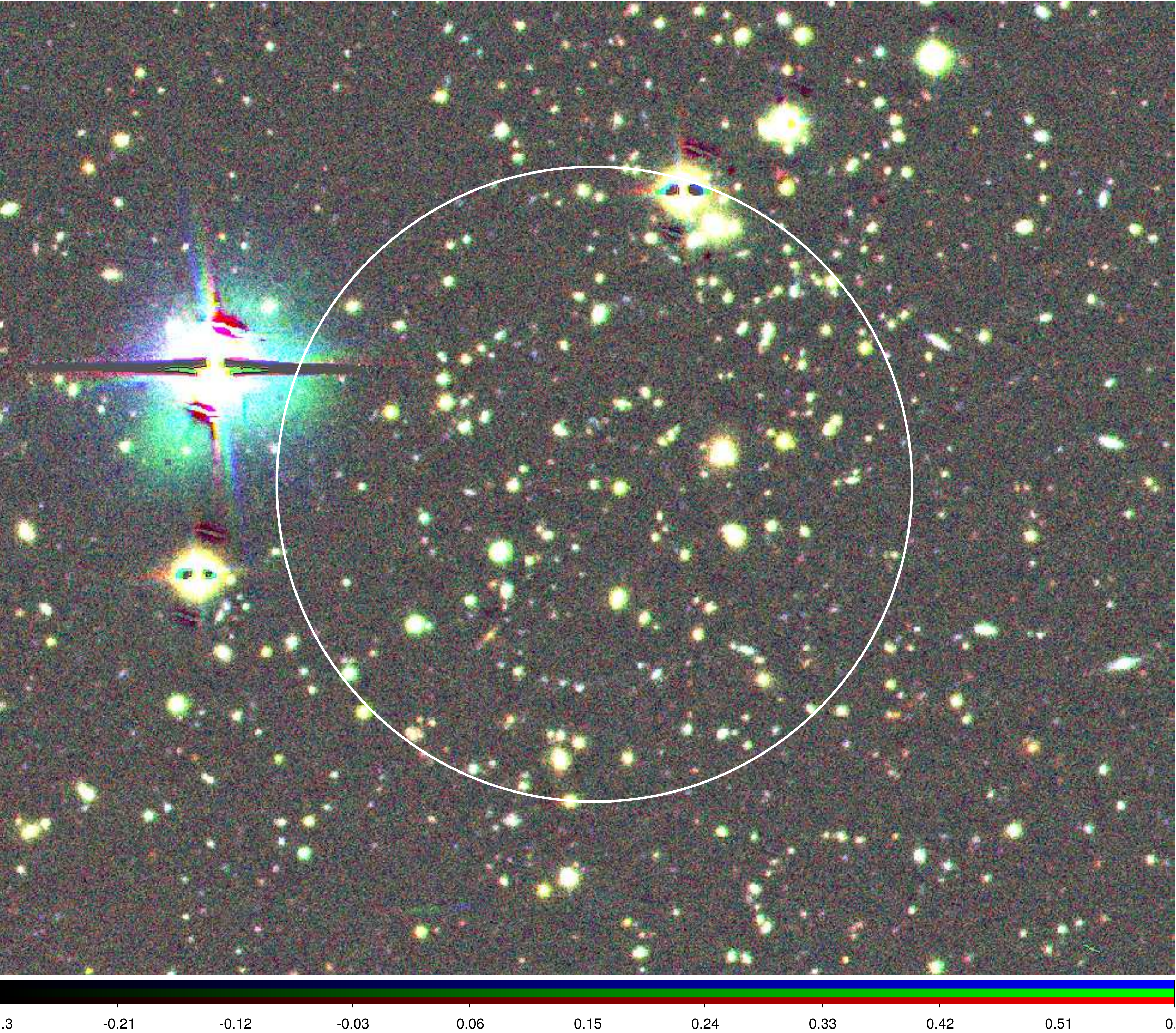}
		(a) PSZ2 G316.43+54.02
	\end{minipage}
	\hspace{0.1cm} % note: no blank line here
	\begin{minipage}{0.57\textwidth}
		\includegraphics[width=\linewidth]{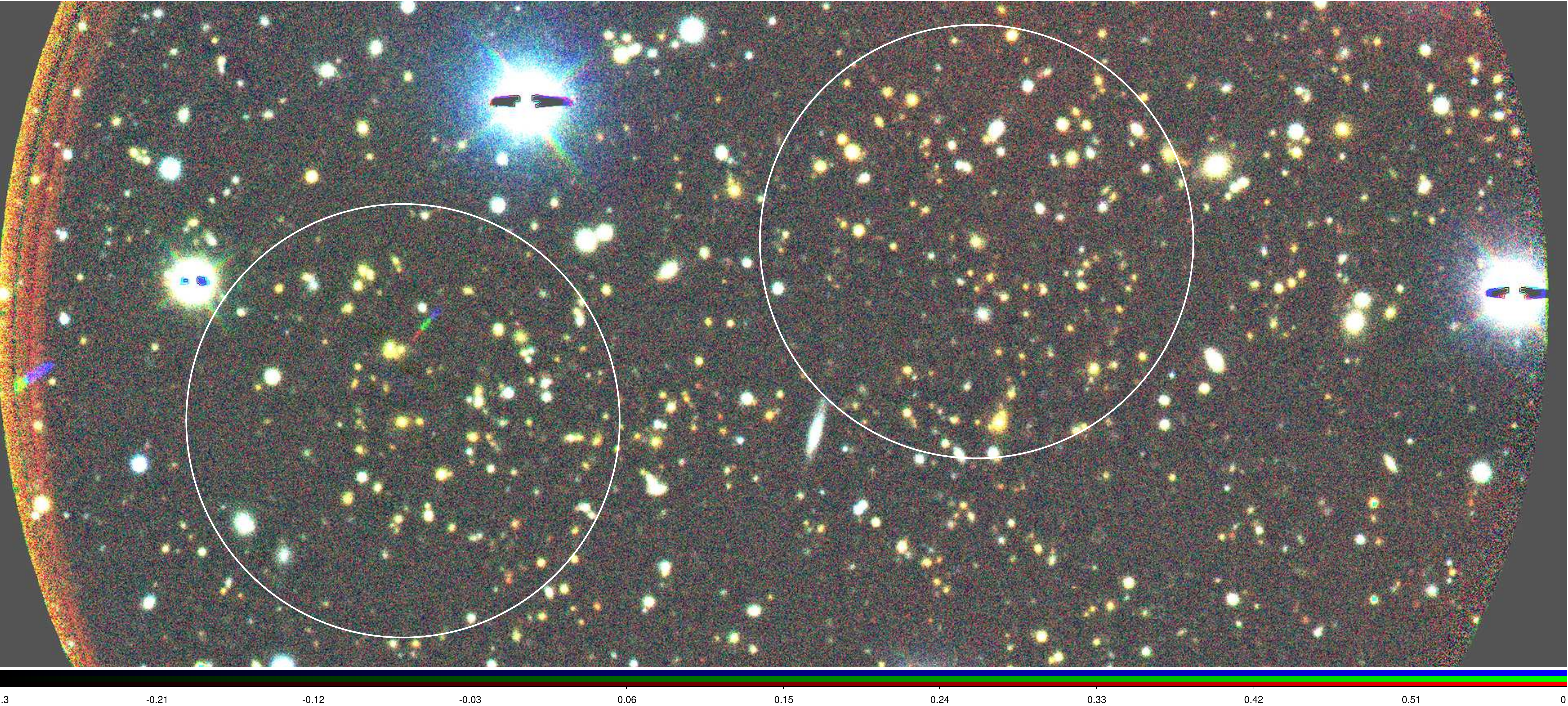}
		(b) PSZ2 G321.30+50.63
	\end{minipage}
	
	%	\vspace*{1cm} % vertical separation

	\begin{minipage}{0.32\textwidth}
		\includegraphics[width=\linewidth]{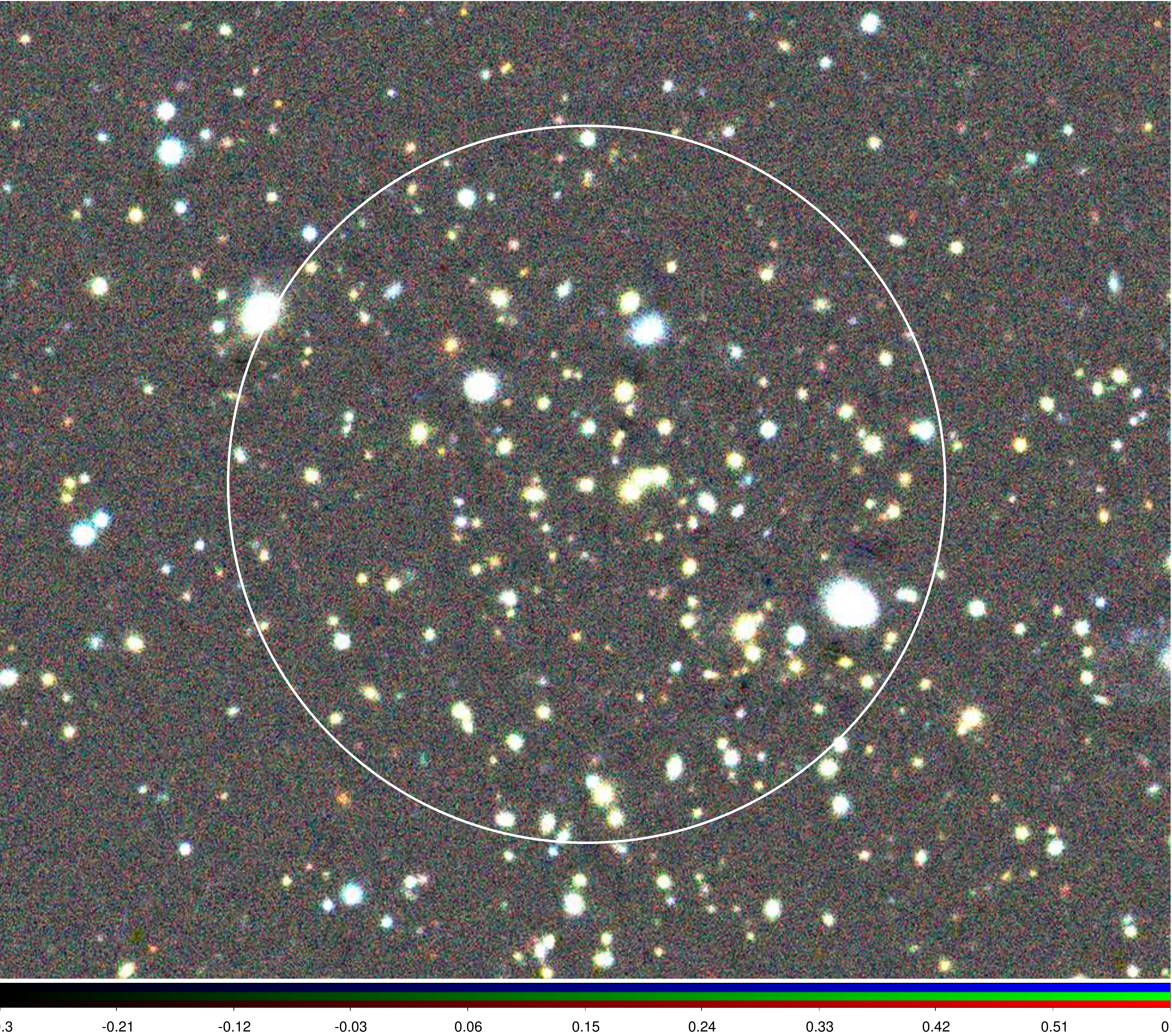}
		(c) PSZ2 G326.73+54.80
	\end{minipage}
	\hspace{\fill} % note: no blank line here
	\begin{minipage}{0.32\textwidth}
		\includegraphics[width=\linewidth]{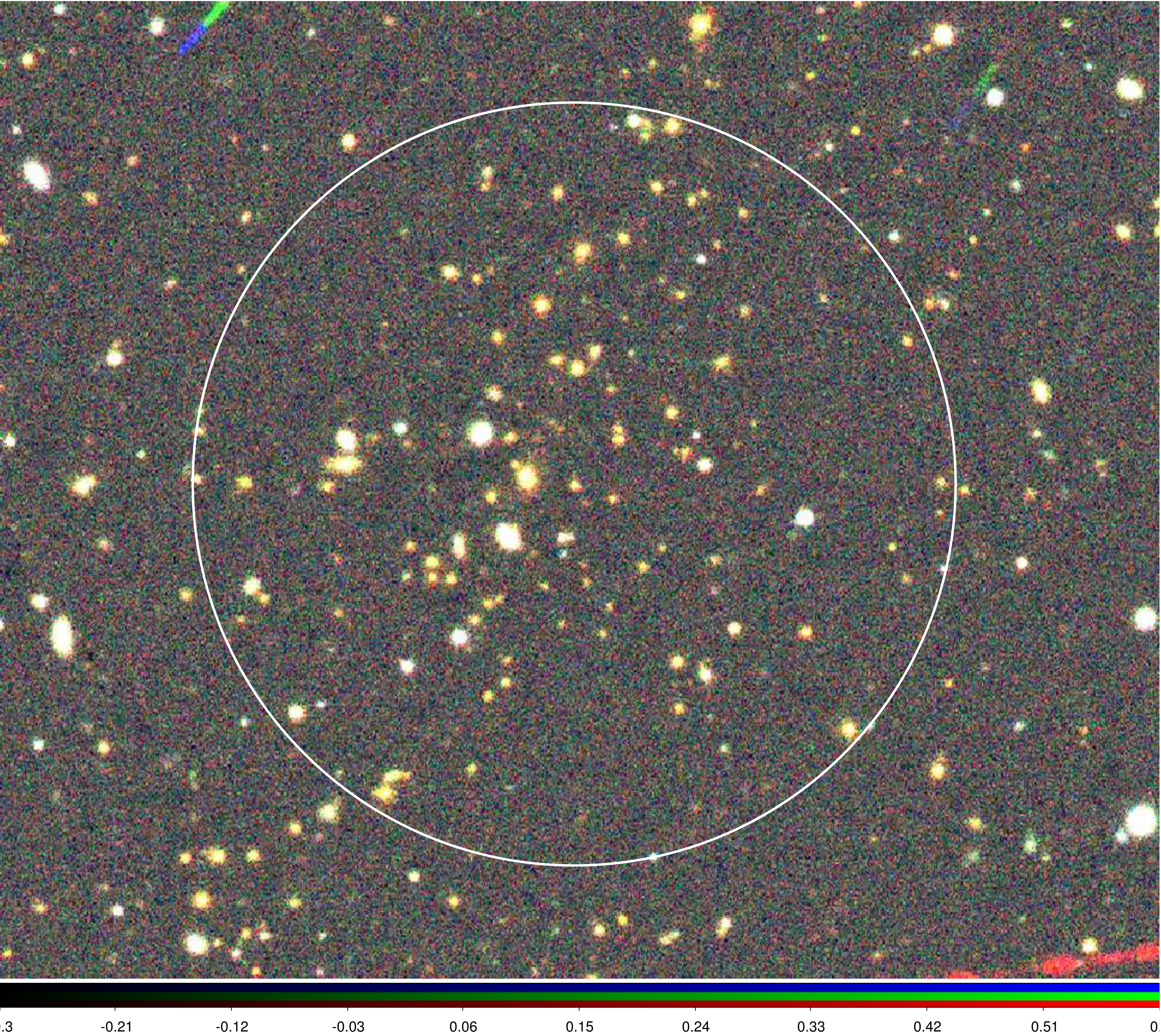}
		(d) PSZ2 G343.46+52.65
	\end{minipage}
	\hspace{0.1cm} % note: no blank line here	
	%	\vspace*{1cm} % vertical separation
	\begin{minipage}{0.32\textwidth}						      \includegraphics[width=\linewidth]{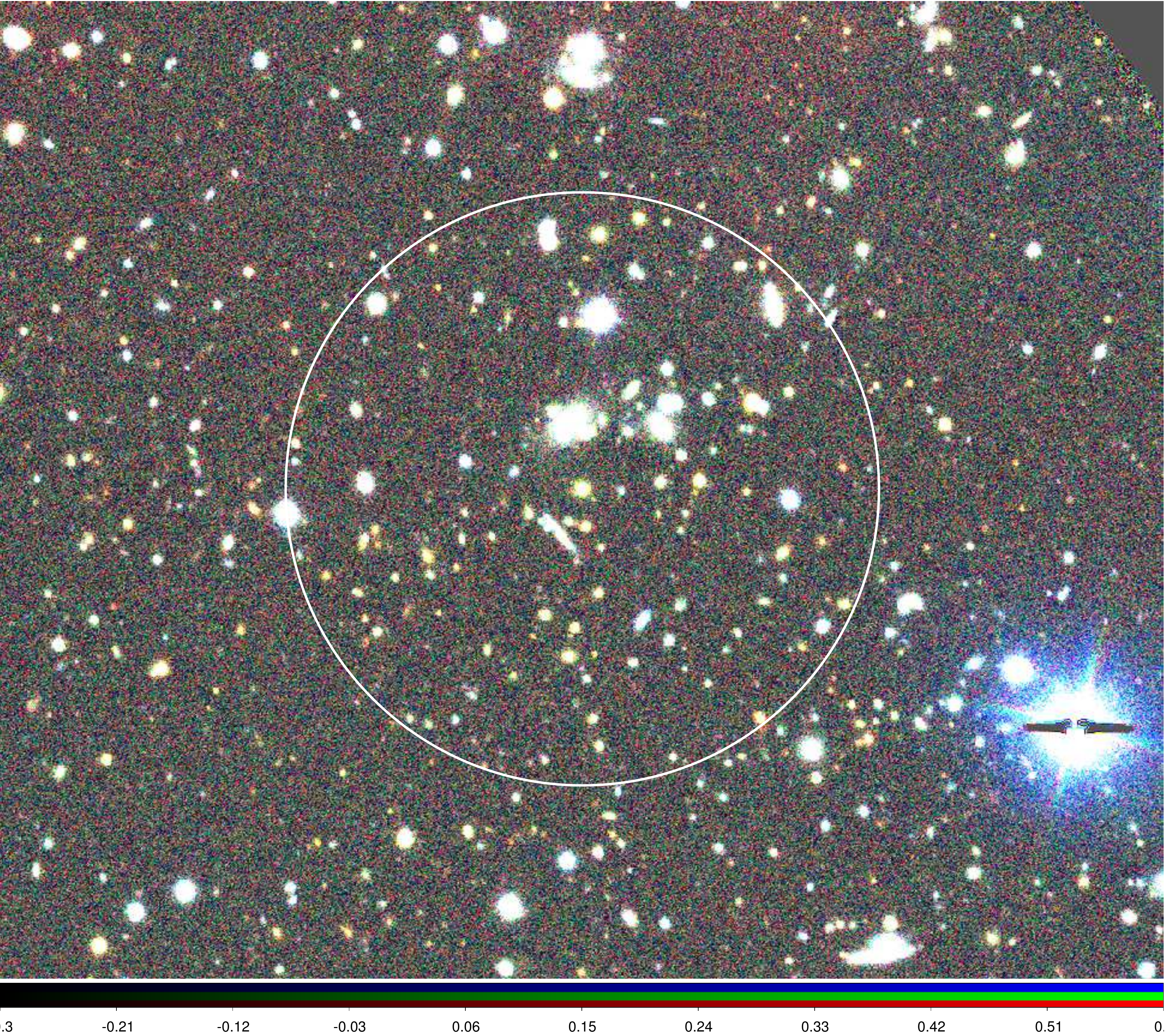}
		(e) PLCK G55.00-37.0
	\end{minipage}	
	
	\begin{minipage}{0.32\textwidth}
		\includegraphics[width=\linewidth]{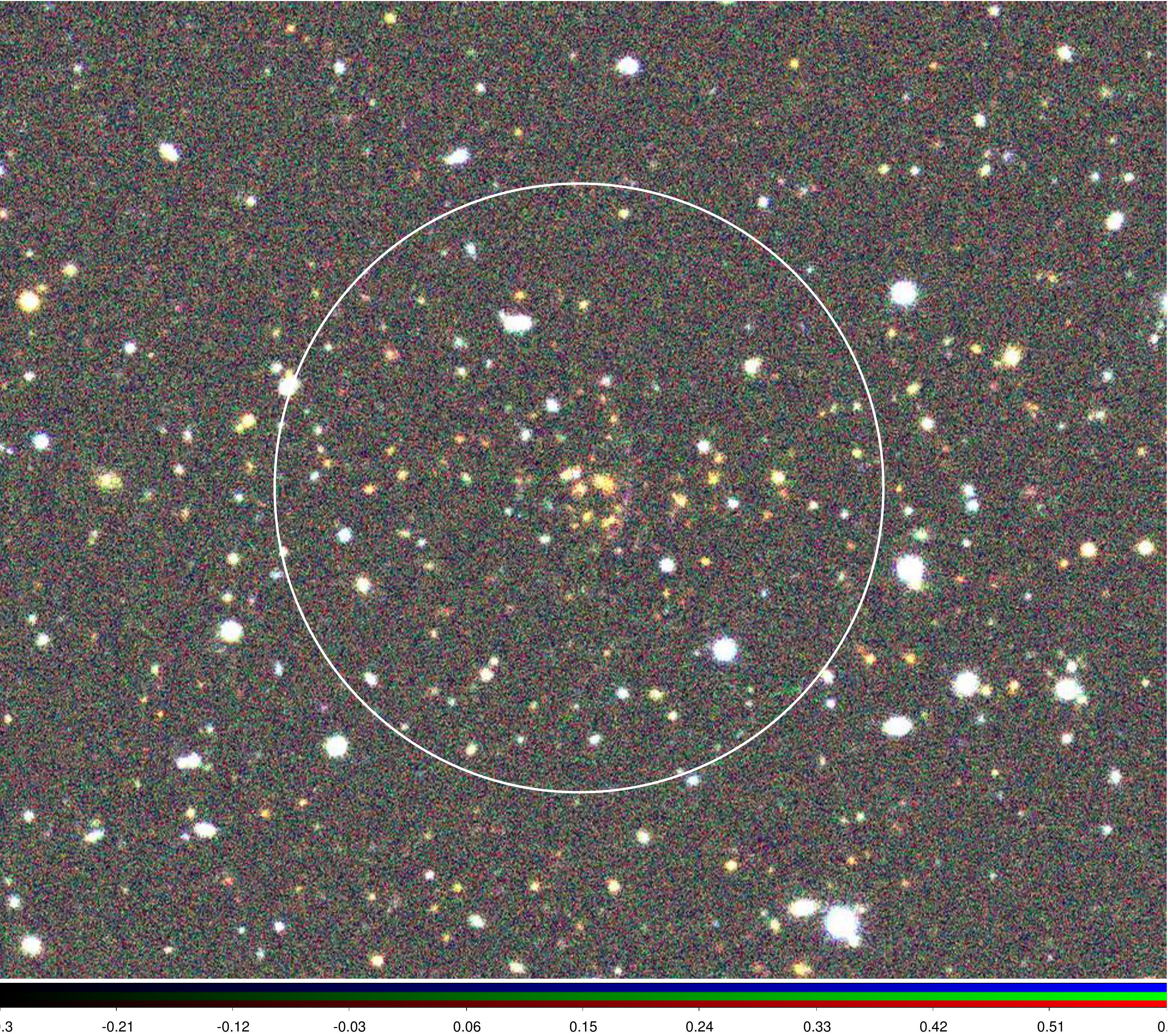}
		(f) PLCK G58.14-72.7
	\end{minipage}	
	\hspace{\fill} % note: no blank line here
	\begin{minipage}{0.32\textwidth}						      \includegraphics[width=\linewidth]{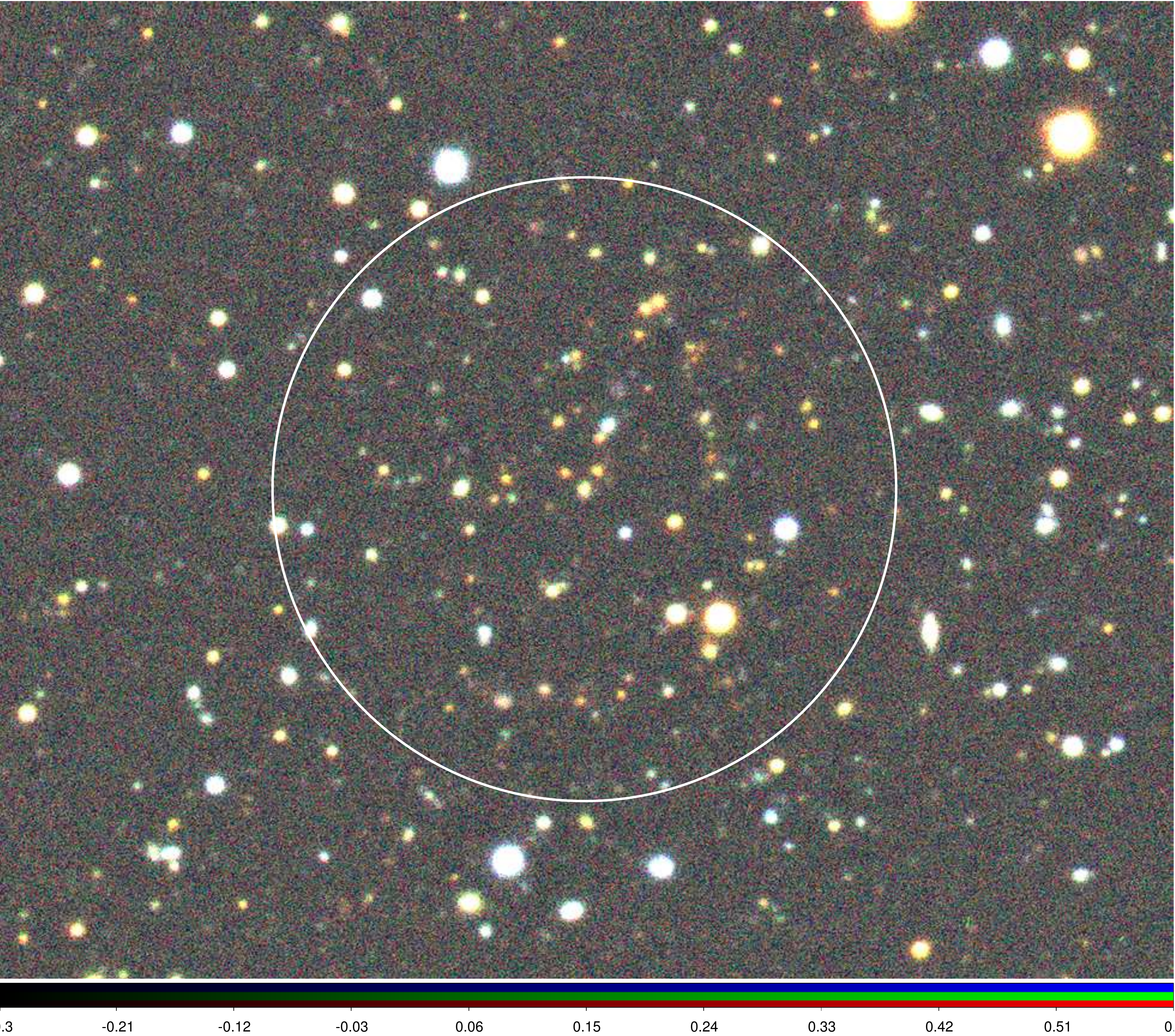}
		(g) PLCK G82.51+29.8
	\end{minipage}
	\hspace{\fill} % note: no blank line here	
	\begin{minipage}{0.32\textwidth}
		\includegraphics[width=\linewidth]{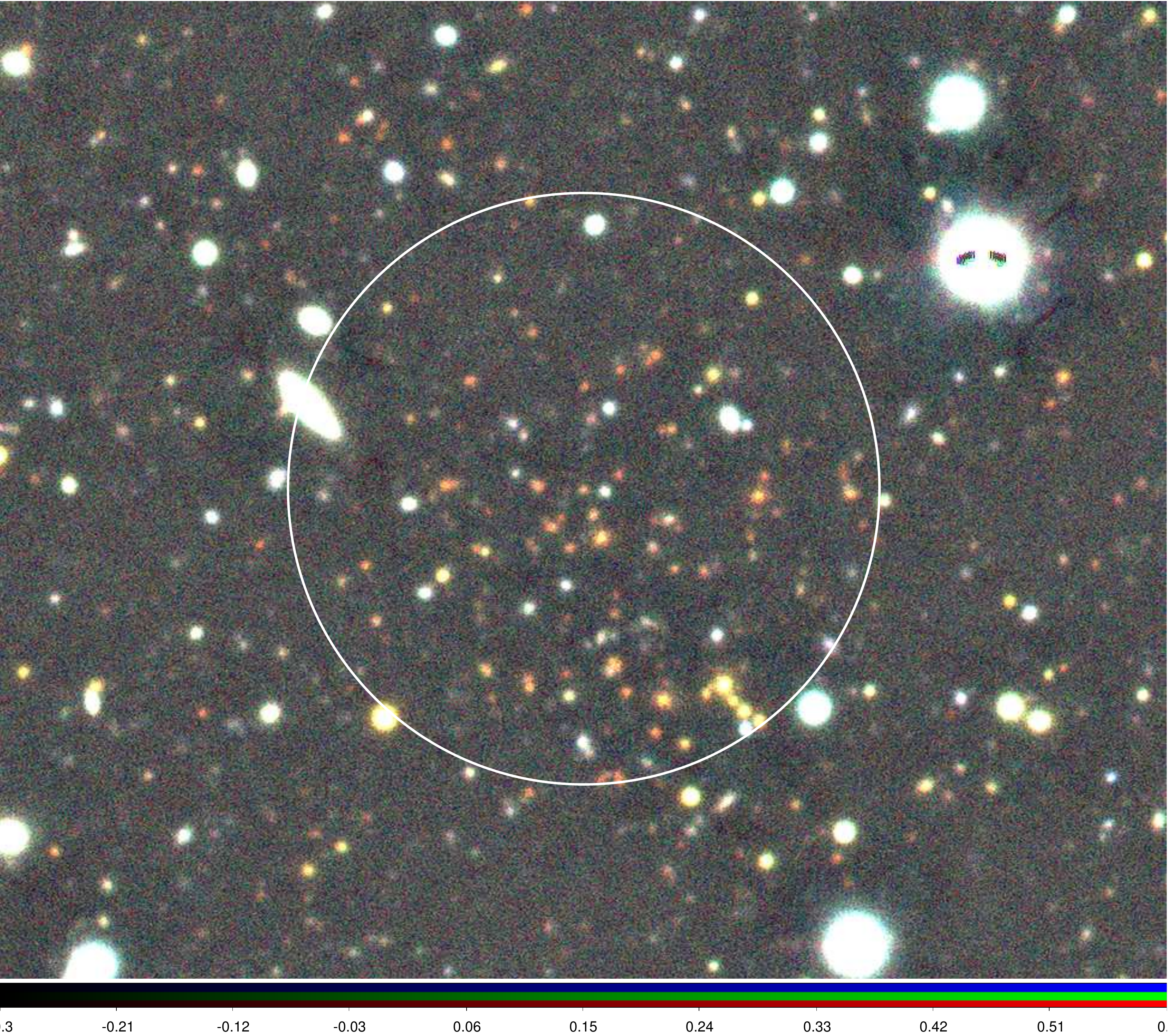}
		(h) PLCK G98.08-46.4
	\end{minipage}
	
	\caption{Clusters with $M_{\mathrm{500c,}\lambda}/M_{\mathrm{500c,SZ}}>0.25$ (continued). The white circles equal 0.5 Mpc in radius around the cluster centre at the estimated redshift. (b) We report the results for the cluster detection to the right of the image in Table \ref{tab:ResultsPSZ2}. More comments on candidate PSZ2 G321.30+50.63 can be found in Section \ref{Notes on Individual Cluster Candidates}.} \label{fig:4pics}
\end{figure*}	

\begin{figure*}		
	\begin{minipage}{0.32\textwidth}
		\includegraphics[width=\linewidth]{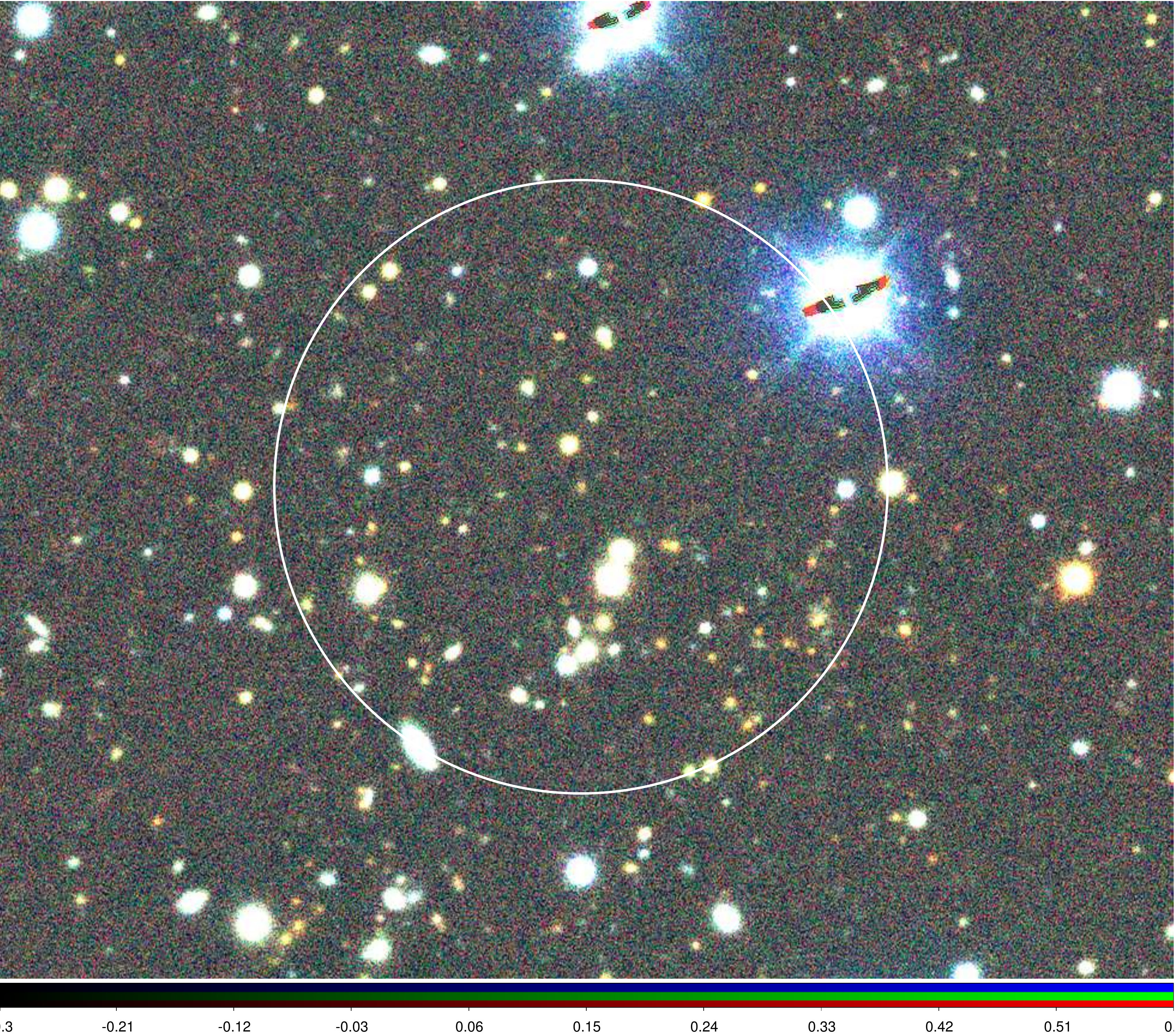}
		(a) PLCK G122.62-31.9
	\end{minipage}	
	\hspace{\fill} % note: no blank line here	
	\begin{minipage}{0.32\textwidth}
		\includegraphics[width=\linewidth]{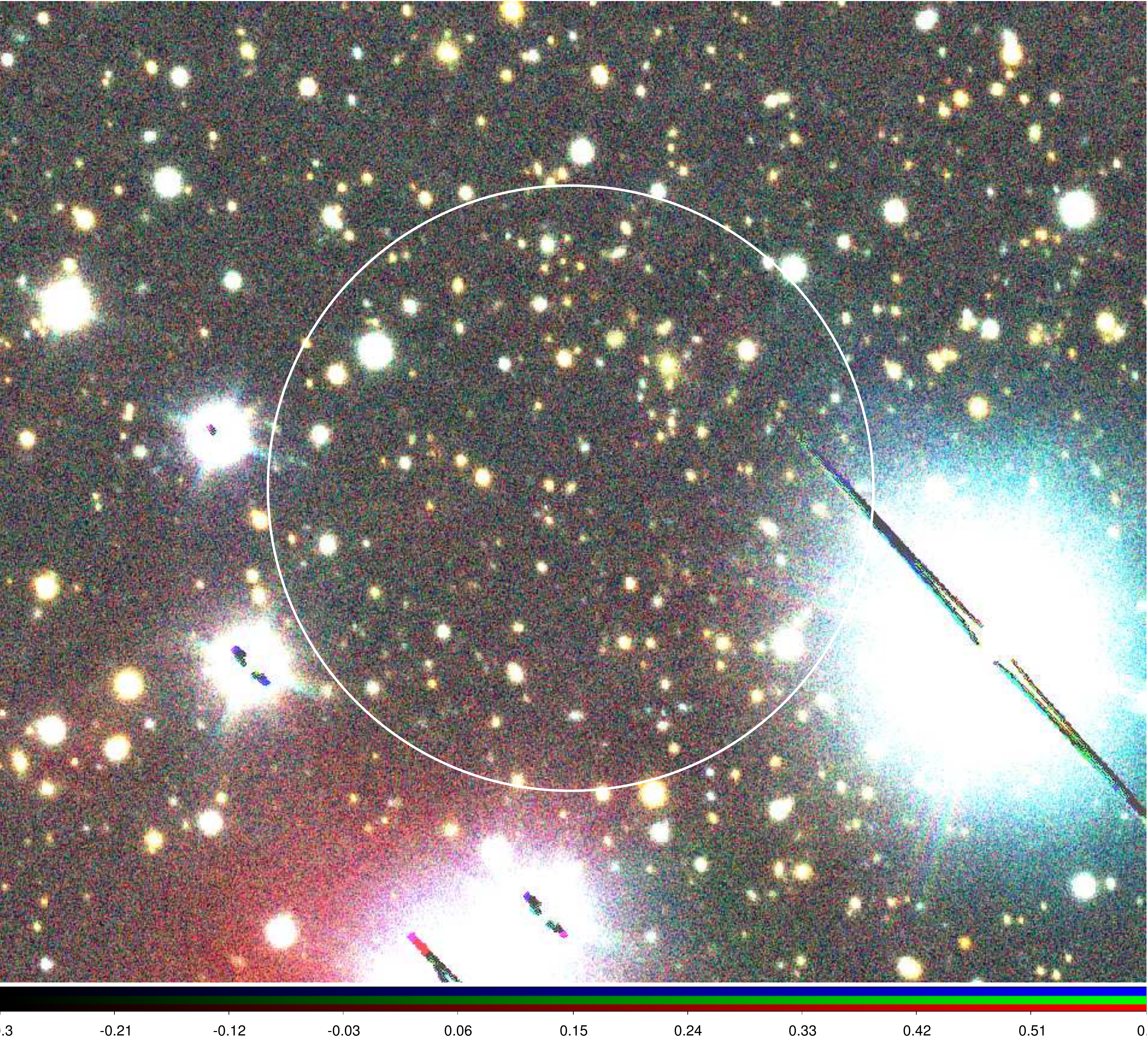}
		(b) PLCK G150.77+17.1
	\end{minipage}
	\hspace{\fill} % note: no blank line here		
	\begin{minipage}{0.32\textwidth}
		\includegraphics[width=\linewidth]{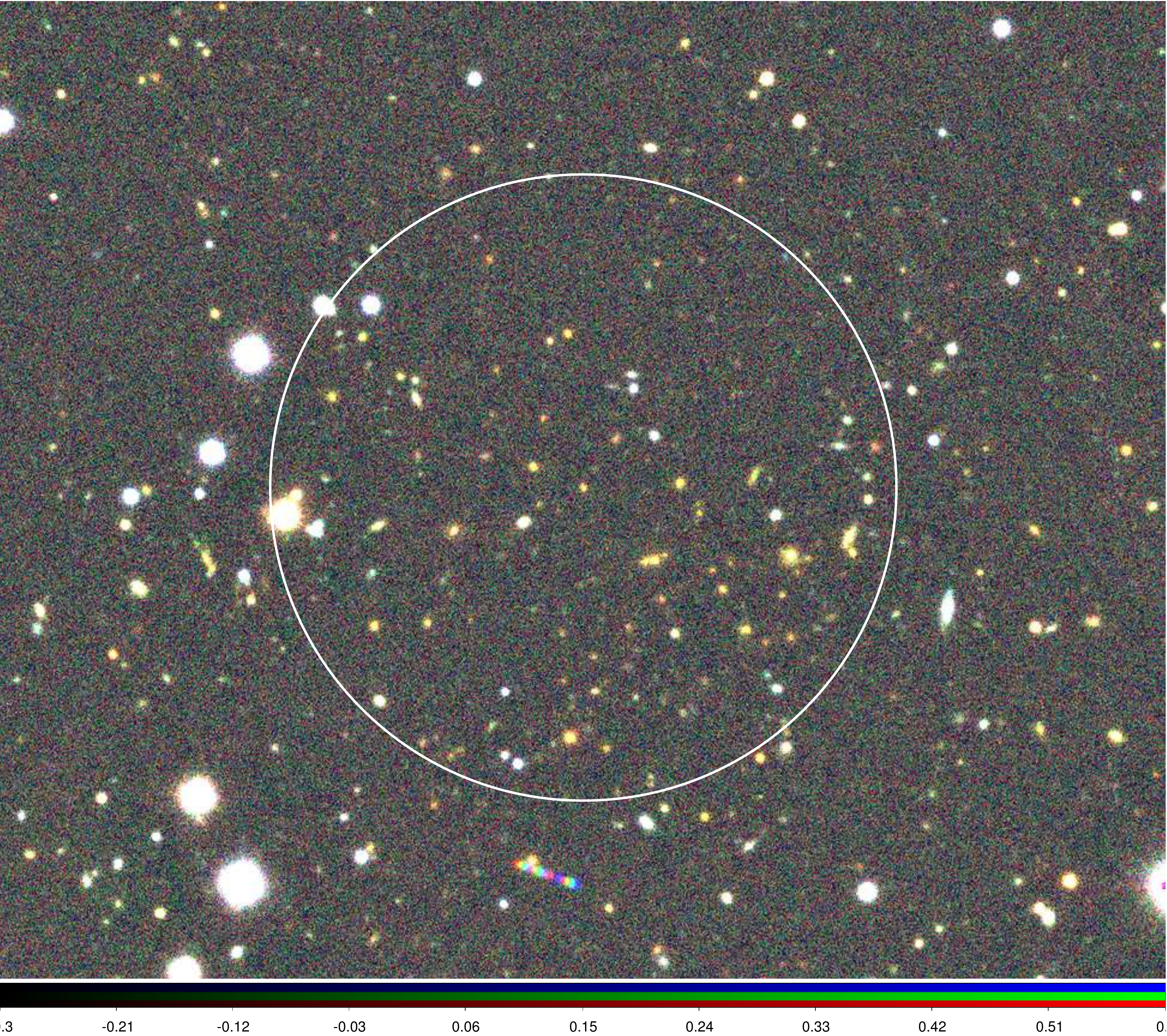}
		(c) PLCK G174.14-27.5
	\end{minipage}
	\hspace{\fill} % note: no blank line here
	\begin{minipage}{0.32\textwidth}
		\includegraphics[width=\linewidth]{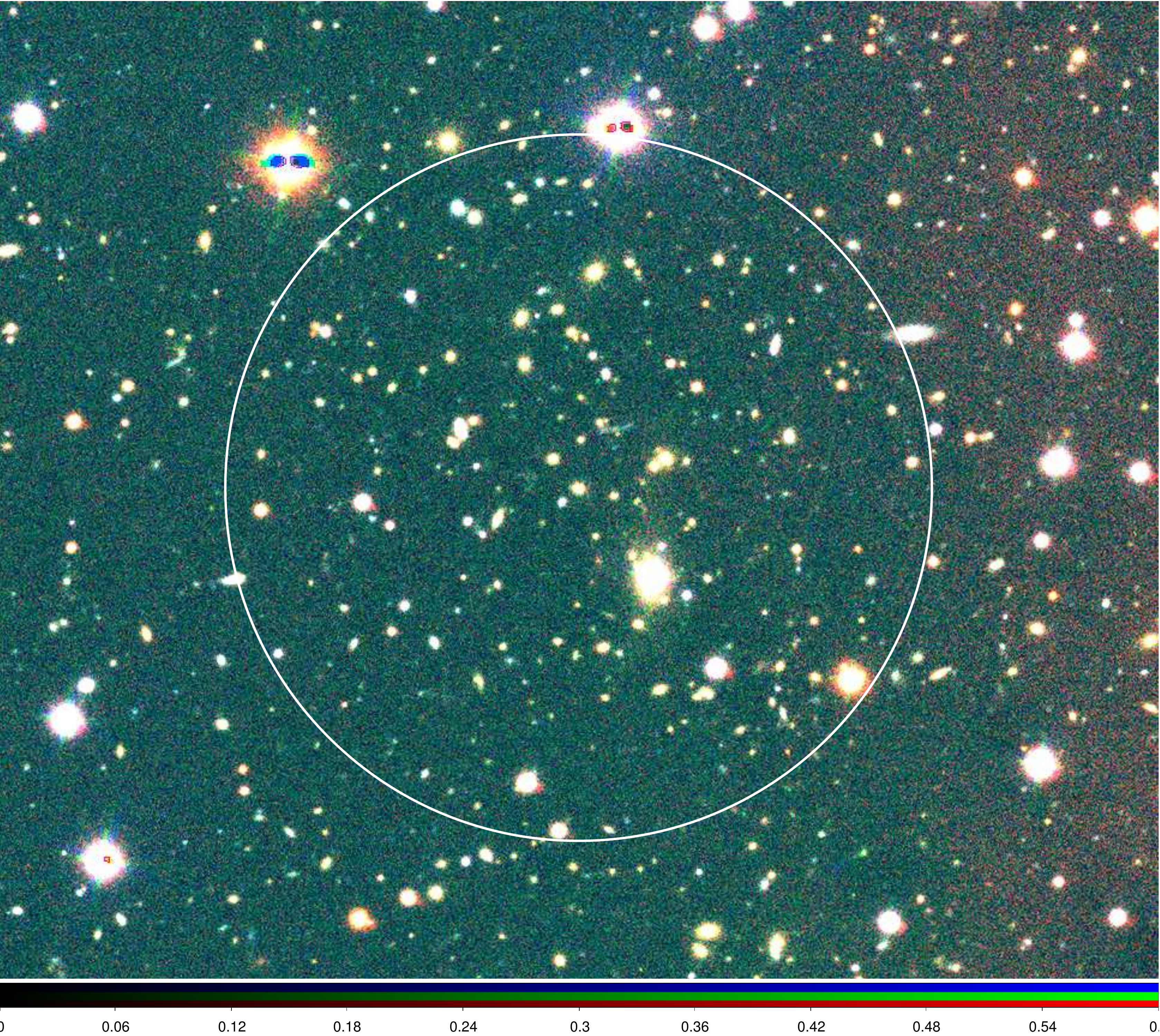}
		(d) PLCK G184.49+21.1
	\end{minipage}

	\caption{Clusters with $M_{\mathrm{500c,}\lambda}/M_{\mathrm{500c,SZ}}>0.25$ (continued). The white circles equal 0.5 Mpc in radius around the cluster centre at the estimated redshift.} \label{fig:4pics}
\end{figure*}

\begin{figure*}

	\begin{minipage}{0.32\textwidth}
		\includegraphics[width=\linewidth]{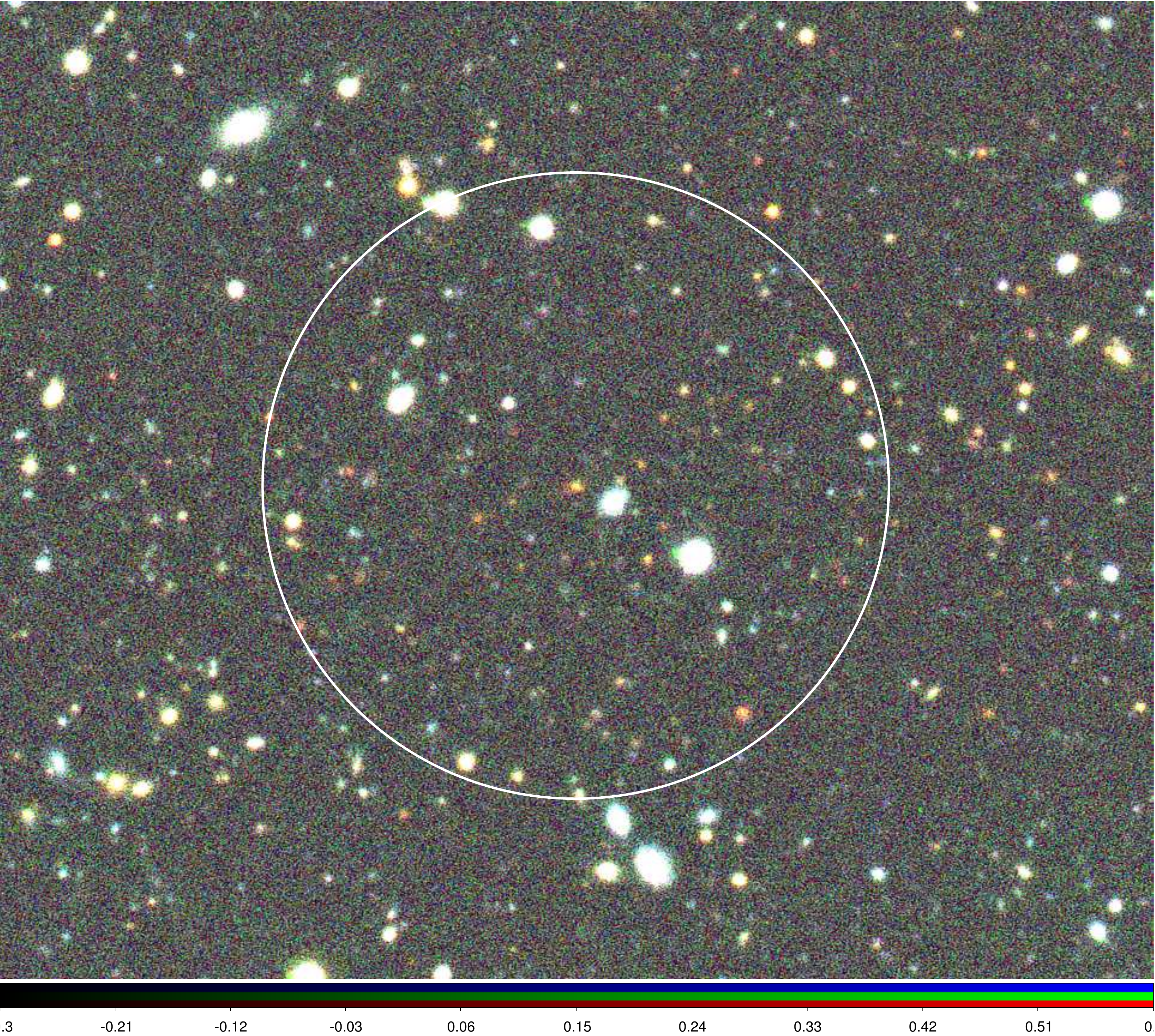}
		(a)PSZ2 G112.54+59.53
	\end{minipage}
	\hspace{0.1cm} % note: no blank line here
	\begin{minipage}{0.32\textwidth}
		\includegraphics[width=\linewidth]{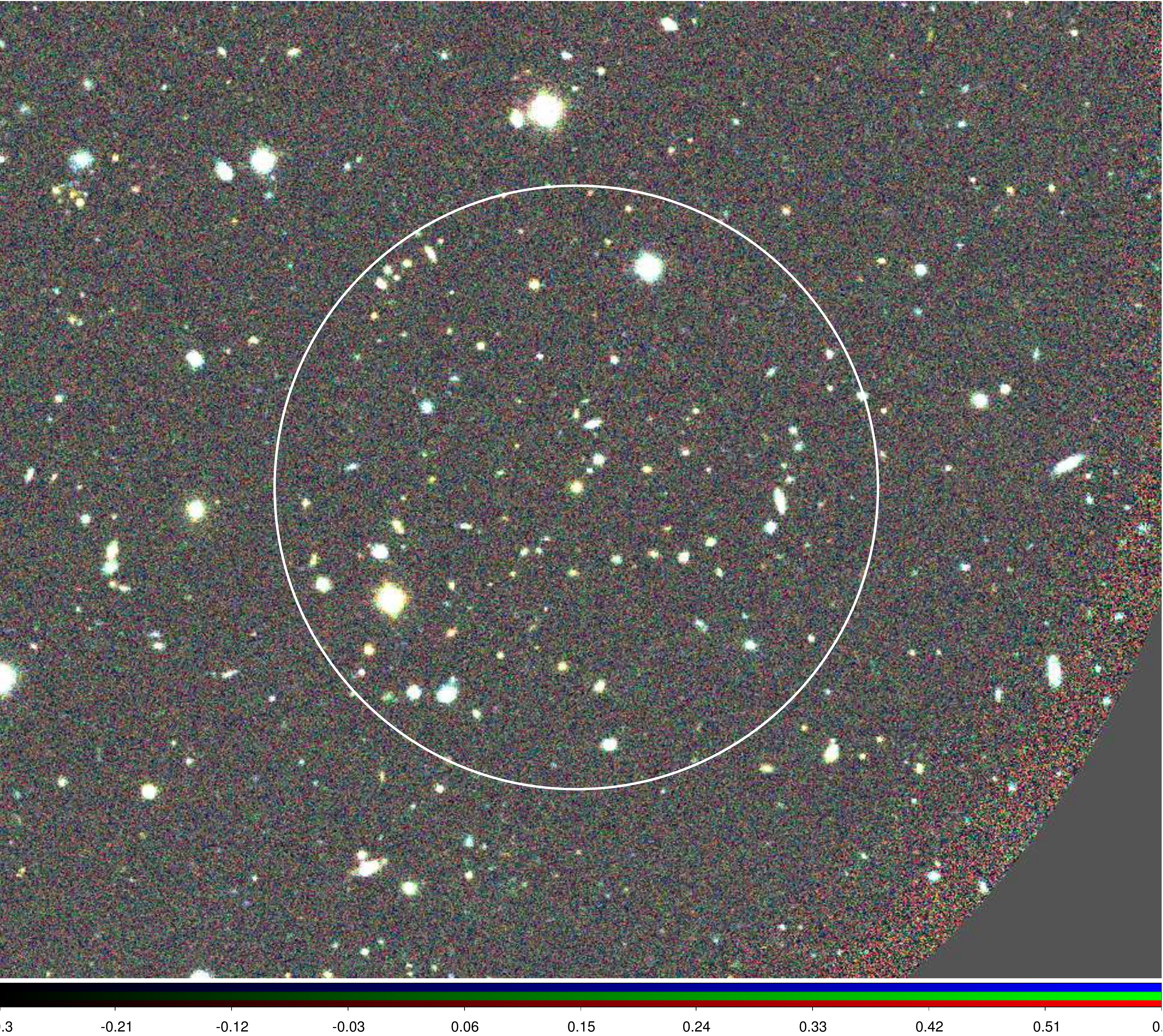}
		(b) PSZ2 G136.02-47.15
	\end{minipage}
	\hspace{0.1cm} % note: no blank line here
	\begin{minipage}{0.32\textwidth}
		\includegraphics[width=\linewidth]{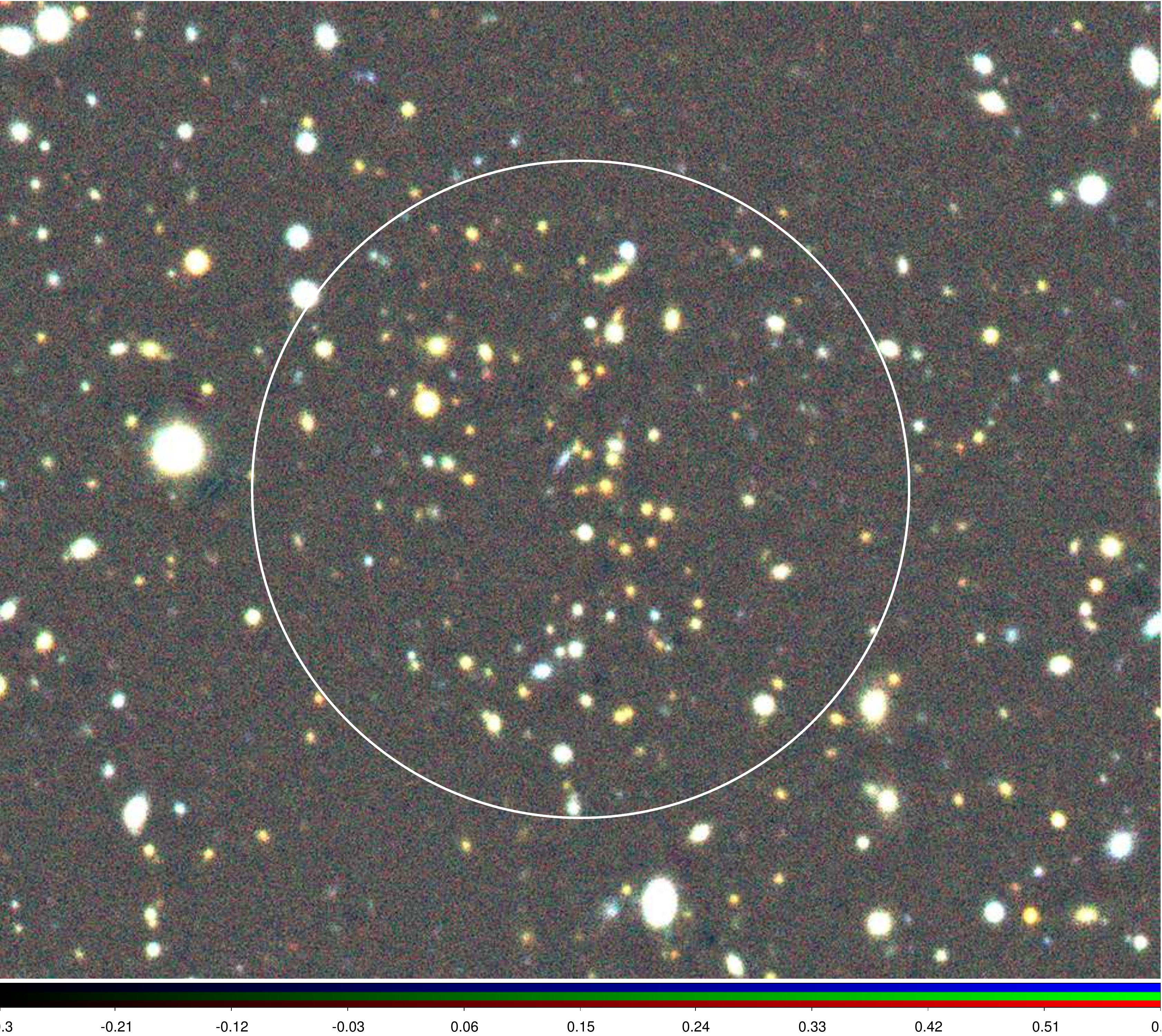}
		(c) PSZ2 G139.00+50.92
	\end{minipage}

	\begin{minipage}{0.32\textwidth}
		\includegraphics[width=\linewidth]{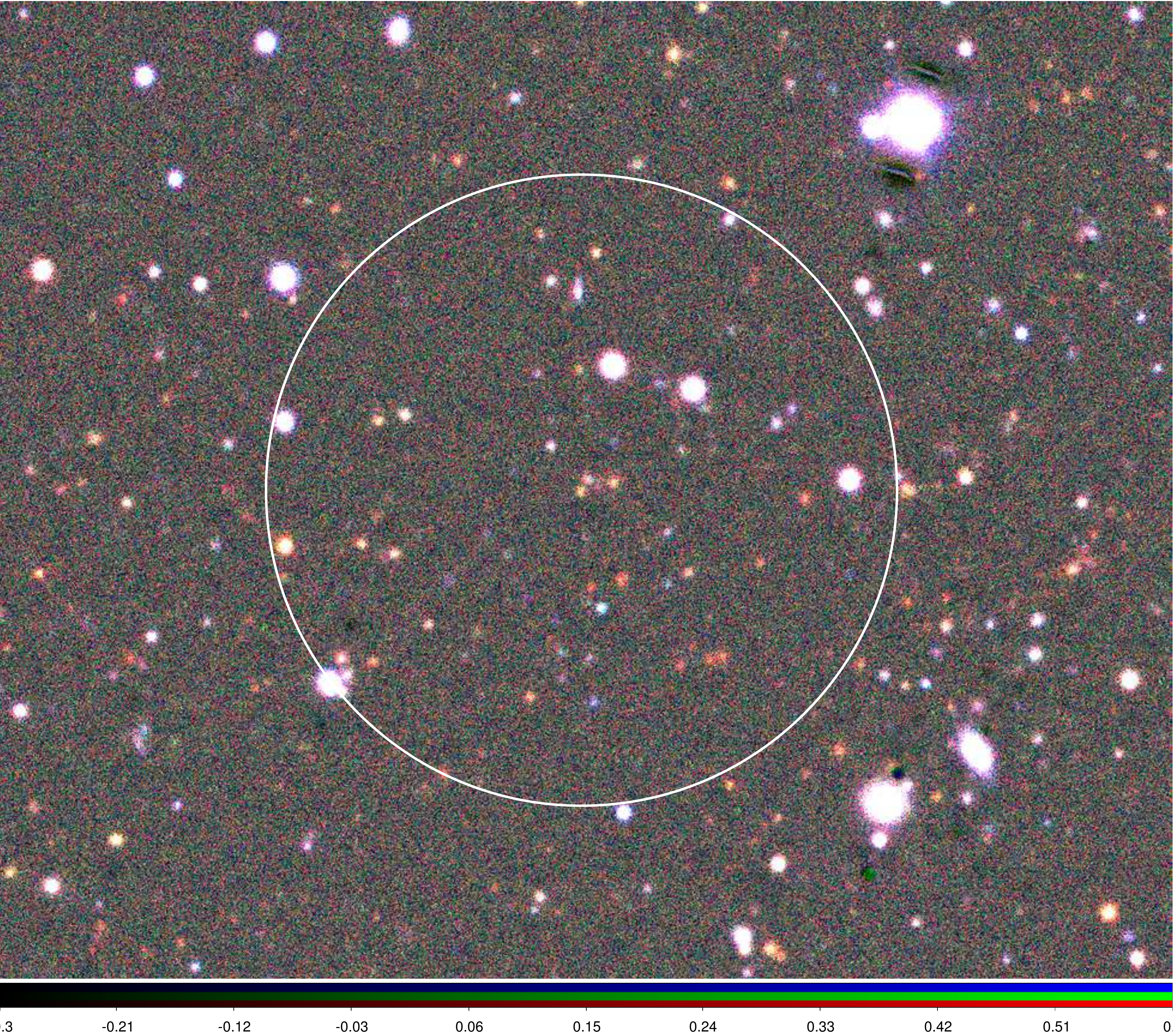}
		(d) PSZ2 G305.76+44.79
	\end{minipage}
	
	\hspace{0.1cm} % note: no blank line here	

	\caption{Clusters with $M_{\mathrm{500c,}\lambda}/M_{\mathrm{500c,SZ}}<0.25$. The white circles equal 0.5 Mpc in radius around the cluster centre at the estimated redshift.} \label{fig:4pics}
\end{figure*}

\begin{figure*}

	\begin{minipage}{0.32\textwidth}
		\includegraphics[width=\linewidth]{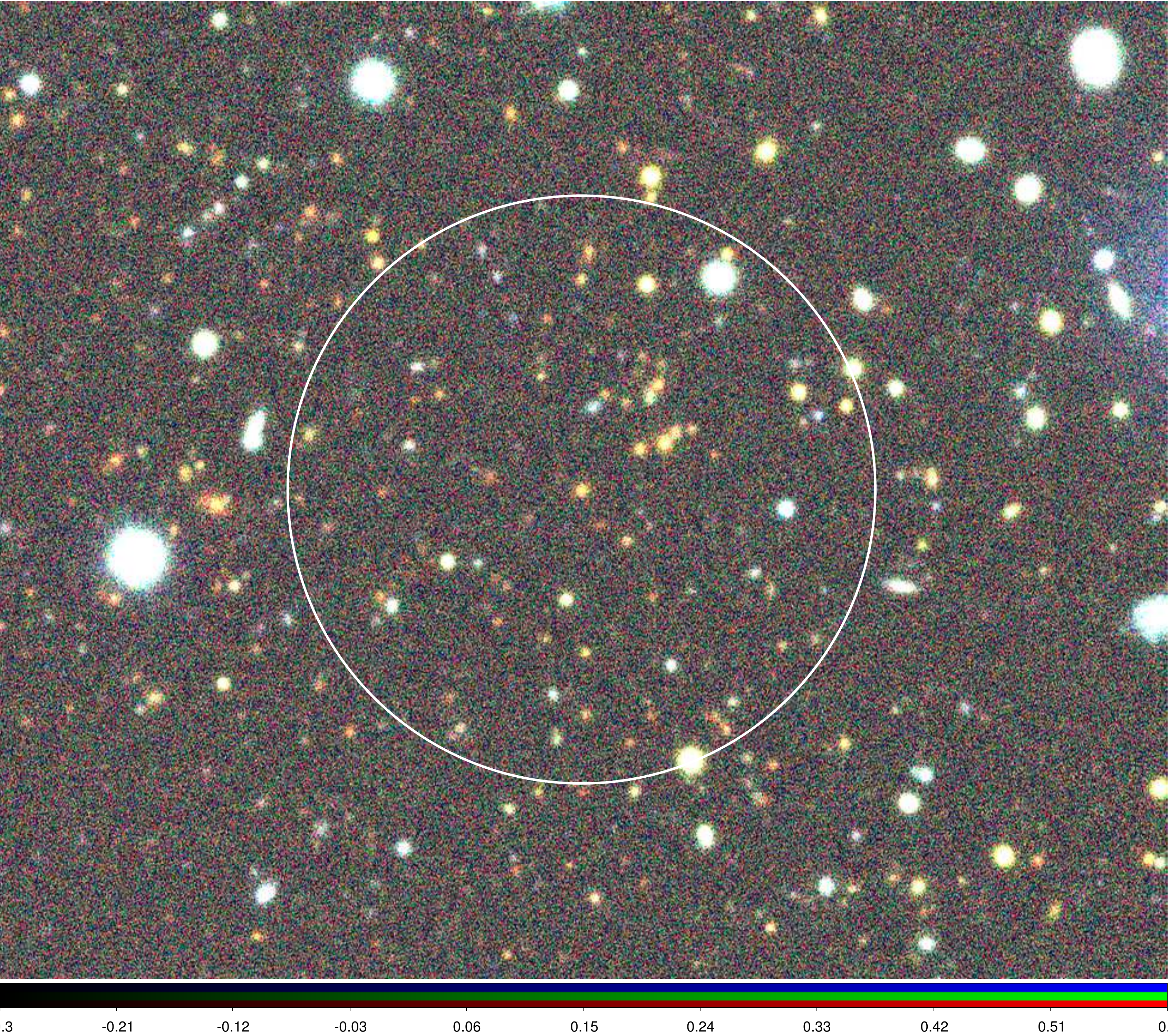}
		(a) PSZ2 G152.47+42.11
	\end{minipage}
	\hspace{0.1cm} % note: no blank line here	
	\begin{minipage}{0.32\textwidth}
		\includegraphics[width=\linewidth]{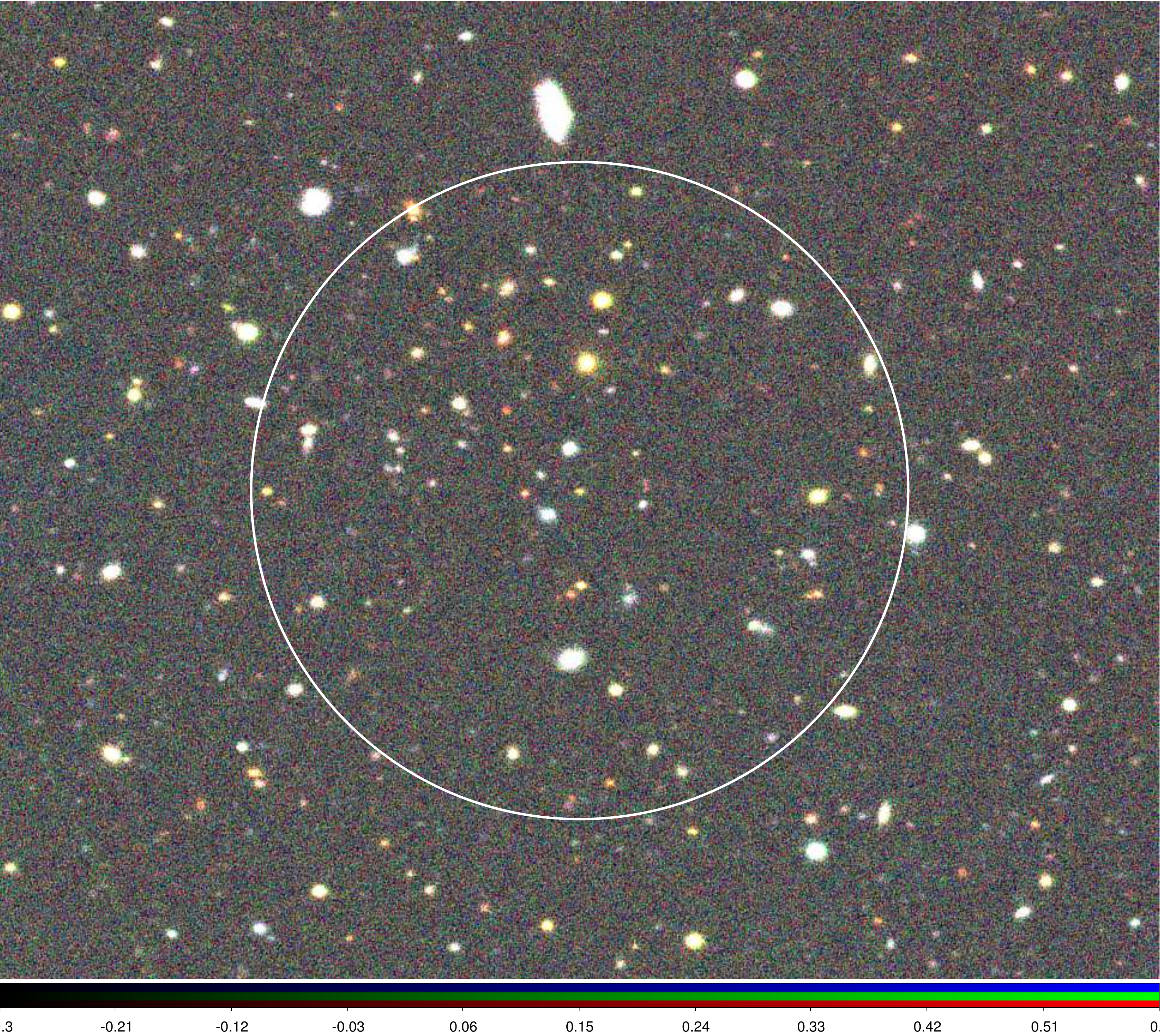}
		(b) PSZ2 G160.94+44.85
	\end{minipage}
	\hspace{0.1cm} % note: no blank line here
	\begin{minipage}{0.32\textwidth}
		\includegraphics[width=\linewidth]{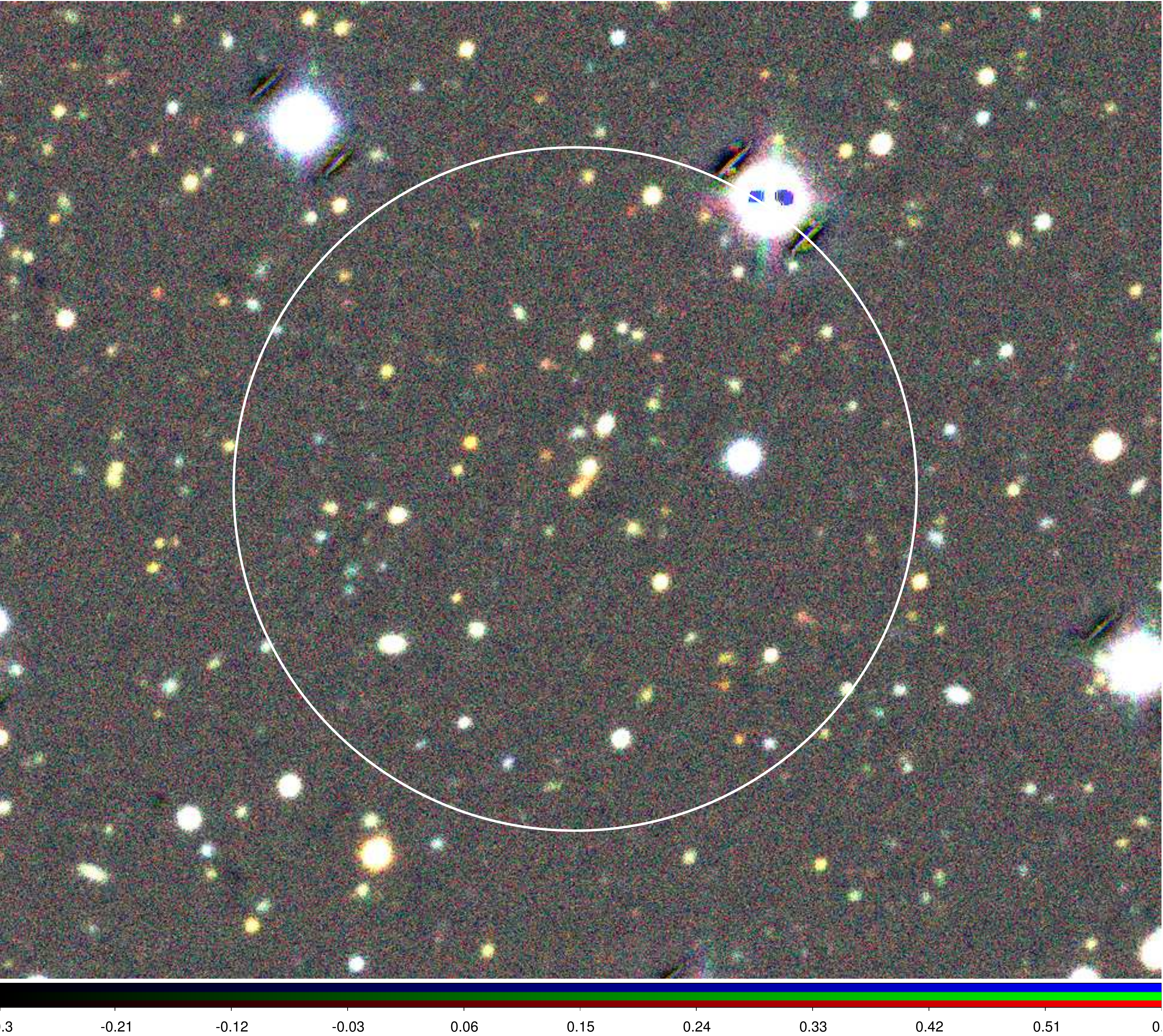}
		(c) PSZ2 G165.41+25.93
	\end{minipage}

	\begin{minipage}{0.32\textwidth}
		\includegraphics[width=\linewidth]{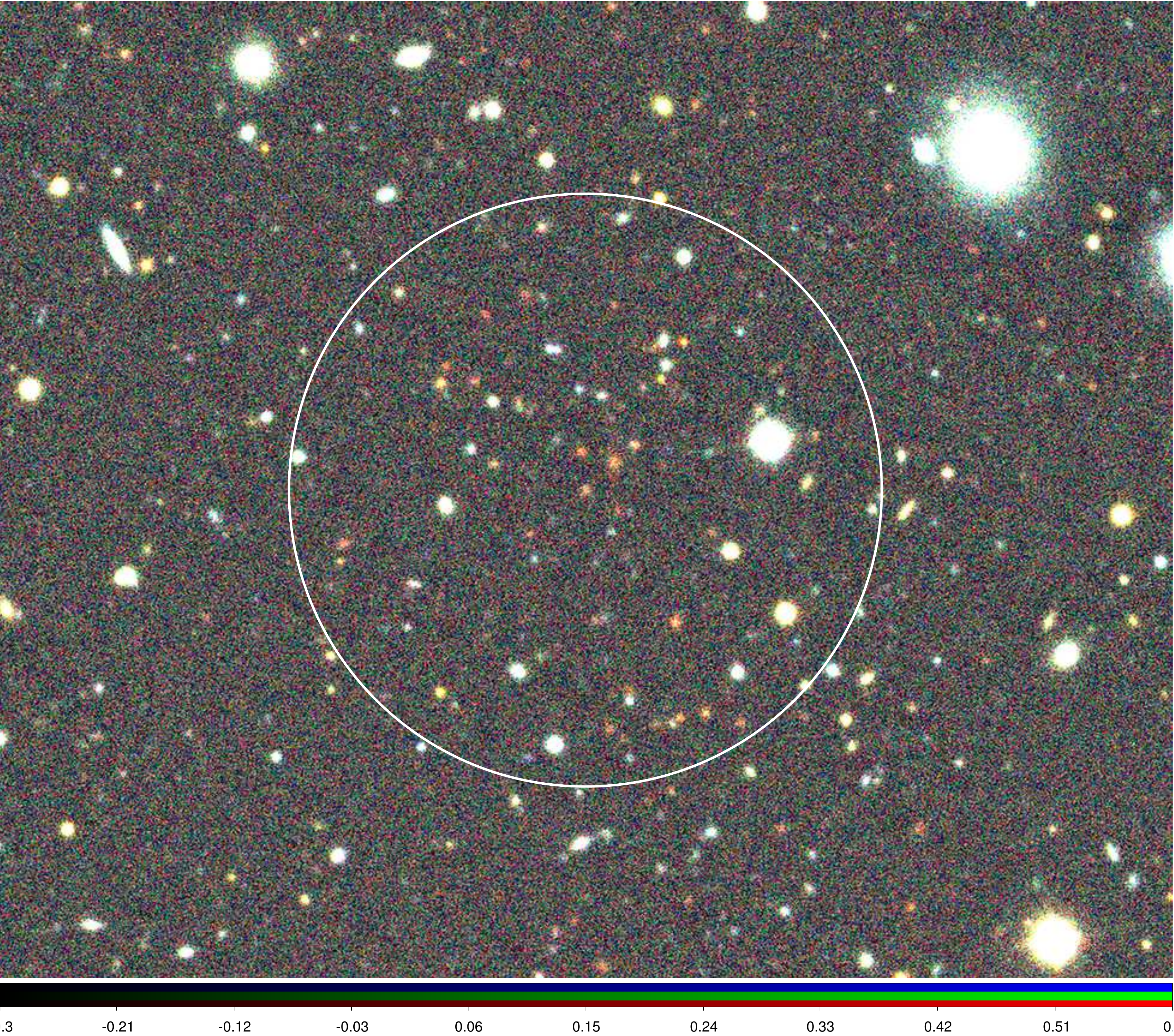}
		(d) PLCK G164.82-47.4
	\end{minipage}

	\caption{Clusters with $M_{\mathrm{500c,}\lambda}$ or $M_{\mathrm{500c,SZ}}$ unknown. The white circles equal 0.5 Mpc in radius around the cluster centre at the estimated redshift.} \label{fig:4pics}
\end{figure*}

%%%%%%%%%%%%%%%%%%%%%%%%%%%%%%%%%%%%%%%%%%%%%%%%%%

% Don't change these lines
\bsp	% typesetting comment
\label{lastpage}
\end{document}